\documentclass[aps,reprint]{revtex4-2}
\usepackage{etoolbox} %
\usepackage{blindtext}
\usepackage{graphicx}
\usepackage{subcaption}
\usepackage{amssymb}
\usepackage{xcolor}
\usepackage{graphicx}
\usepackage{floatrow}
\usepackage[fleqn]{amsmath}
\usepackage[font=small,skip=3pt]{caption}
\usepackage{makecell}
\usepackage{hyperref} 
\usepackage{cleveref}
\usepackage{url}
\makeatletter
\appto{\appendix}{%
  \@ifstar{\def\theequation@prefix{A.}}%
          {}%
}
\makeatother

\begin{document}
\title{Simultaneous Enhancement of Tritium Burn Efficiency and Fusion Power with Low-Tritium Spin-Polarized Fuel}
\author{J. F. Parisi$^{1}$}
\email{jparisi@pppl.gov}
\author{A. Diallo$^1$}
\author{J. A. Schwartz$^{1}$}
\affiliation{$^1$Princeton Plasma Physics Laboratory, Princeton University, Princeton, NJ, USA}
\begin{abstract}
This study demonstrates that using spin-polarized deuterium-tritium (D-T) fuel with more deuterium than tritium can increase tritium burn efficiency (TBE) by at least an order of magnitude without compromising fusion power output, compared to unpolarized fuel. Although previous studies show that a low tritium fraction can enhance TBE, this strategy resulted in reduced fusion power density. The surprising improvement in TBE at fixed power reported here is due to the TBE increasing nonlinearly with decreasing tritium fraction but the fusion power density increasing roughly linearly with D-T cross section. A study is performed for an ARC-like tokamak producing 482 MW of fusion power with unpolarized 51:49 D-T fuel, finding the minimum startup tritium inventory ($I_{\mathrm{startup,min}}$) is 0.69 kg. By spin-polarizing half of the fuel and using a 57:43 D-T mix, $I_{\mathrm{startup,min}}$ is reduced to 0.08 kg, and fully spin-polarizing the fuel with a 61:39 D-T mix further reduces $I_{\mathrm{startup,min}}$ to 0.03 kg. Some ARC-like scenarios could achieve plasma ignition with relatively modest spin polarization. These findings indicate that, with advancements in helium divertor pumping efficiency, TBE values of approximately 10-40\% could be achieved using low-tritium-fraction and spin-polarized fuel with minimal power loss. This would dramatically lower tritium startup inventory requirements and reduce the amount of on-site tritium. More generally than just for spin-polarized fuels, high plasma performance can be used to increase TBE. This strongly motivates the development of spin-polarized fuels and low-tritium-fraction operation for burning plasmas.
\end{abstract}

\maketitle

\setlength{\parskip}{0mm}
\setlength{\textfloatsep}{5pt}

\setlength{\belowdisplayskip}{6pt} \setlength{\belowdisplayshortskip}{6pt}
\setlength{\abovedisplayskip}{6pt} \setlength{\abovedisplayshortskip}{6pt}

\section{Introduction}

Deuterium-tritium (D-T) is widely considered the most feasible fuel for first-generation fusion power plants due to its high reactivity at experimentally realizable temperatures and the large energy release per fusion reaction \cite{Strachan1994short,Keilhacker1999,Wurzel2022}. However, tritium is scarce because it has a half-life of 12.3 years and because it is hard to produce in large quantities with current technology \cite{Kaufman1954,Kovari2018}. Due to tritium scarcity, D-T plants are designed to be tritium self-sufficient. This is achieved by breeding tritium on-site by neutron capture reactions with lithium in the blanket surrounding the core.

In order for a power plant to be tritium self-sufficient, the tritium-breeding-ratio (TBR), which measures the ratio of tritium production to burn-up, must exceed a minimum value \cite{El-Guebaly2009}. It has been reported that the tritium fractional burn-up is among the most important, if not the most important, variable for achieving a high TBR \cite{Jung1984,Abdou1986,Jackson2013,Abdou2021,Meschini2023}. A closely related quantity used in this work is the tritium burn efficiency (TBE) \cite{Whyte2023}, the ratio of the tritium burn rate to tritium injection rate. Improvements in the TBE can lessen requirements for other key tritium self-sufficiency parameters such as the startup inventory, the tritium doubling time, and tritium loss fractions \cite{Abdou1986}. A high TBE could significantly lower the cost and regulation complexity of a fusion plant \cite{Roth2008}.

Tritium supply for future power plants is particularly challenging \cite{Kovari2018}. D-T plants require a startup inventory because it takes time for tritium to be produced after the beginning of operations. During the time that on-site tritium production is ramping up to full capacity, the plant draws from a startup tritium inventory. The size of the tritium inventory can be considerable relative to global tritium supply. A `baseline' ARC \cite{Sorbom2015} device design has a tritium startup inventory of 1.1 kg and a `baseline' STEP \cite{Fradera2021,Schoofs2022} device design has a tritium startup inventory of 8.9 kg \cite{Meschini2023}. However, recent estimates have only $\approx$ 15 kg of theoretically available tritium for non-ITER fusion pilot plants after 2050 \cite{Kovari2018,Pearson2018,Meschini2023}, meaning that there could only be enough tritium to supply startup inventories for a handful of power plants. Fortunately, there are ways to significantly reduce the startup inventory requirement, with an increased TBE being found as by far the most important variable \cite{Meschini2023}. Tritium modeling of an ARC-class device showed that increasing the TBE from 0.5\% to 5\% could decrease the startup tritium inventory by roughly a factor of ten \cite{Meschini2023}.

While increasing the TBE is beneficial for tritium self-sufficiency, it also is deleterious for plant economics because the fusion power is significantly lower \cite{Kovari2018,Meschini2023}. One potential solution to reduce the tradeoff between fusion power and TBE is to increase the ratio of helium to hydrogen divertor pumping efficiency $\Sigma$ \cite{Whyte2023}, which could significantly increase the TBE (reducing the tritium startup inventory requirement) with a much smaller reduction in fusion power.

It has been suggested that the TBE can also be increased by decreasing the tritium fraction \cite{Boozer2021}. However, it was also noted that decreasing the tritium fraction also decreased the fusion power density, indicating a tradeoff between fusion power and TBE. With unpolarized D-T fuel, it is likely economically prohibitive to operate at lower tritium fraction and therefore high TBE without significant progress in fusion science and technology \cite{Kovari2018,Whyte2023}.

\begin{figure}[bt]
    \centering
    \includegraphics[width=1.0\textwidth]{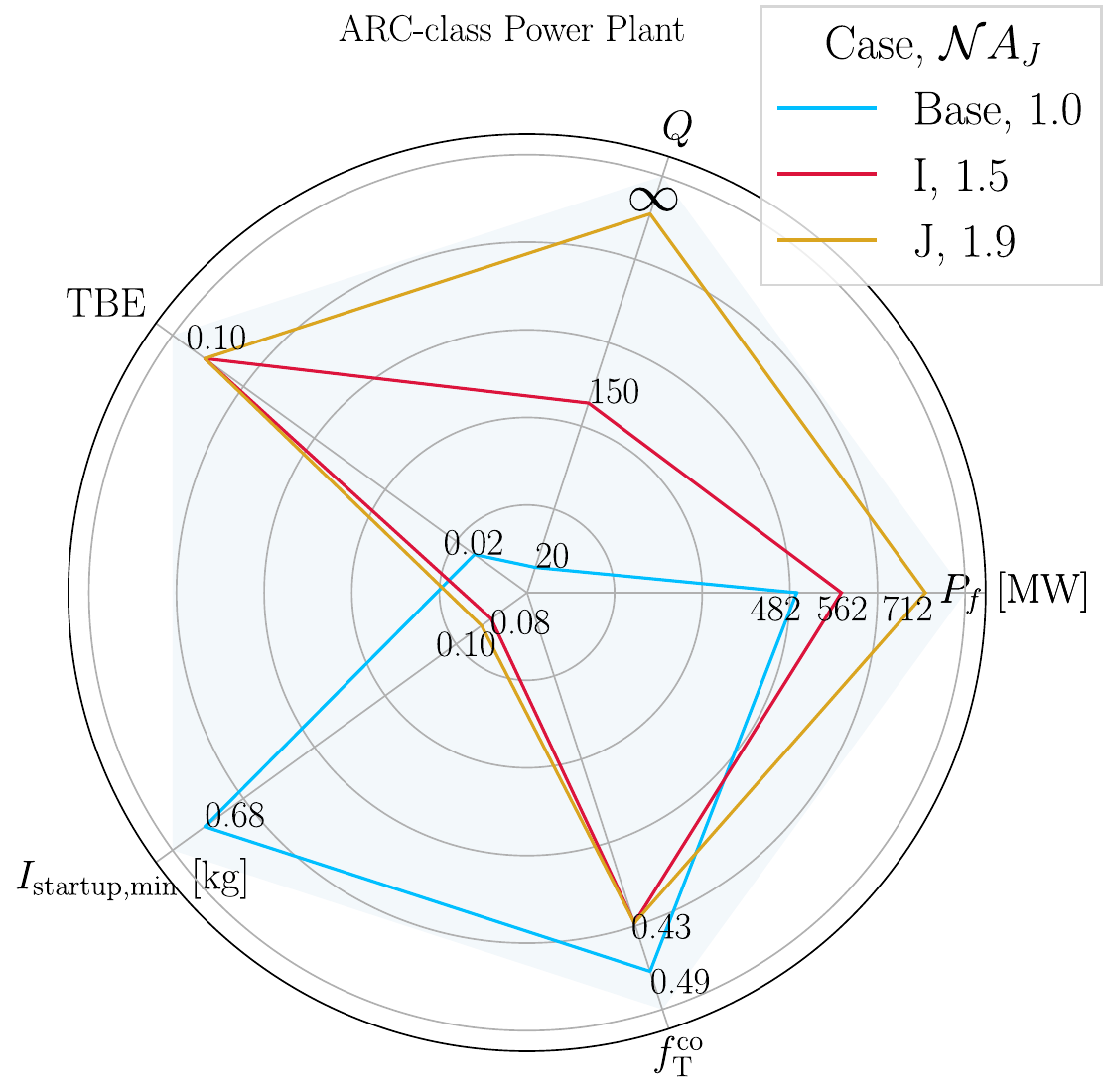}
    \caption{Tritium self-sufficiency and fusion power parameters of an ARC-like device for three cases: Base, with $\mathcal{N} A_J = 1.0$, case I with $\mathcal{N} A_J = 1.5$, and case J with $\mathcal{N} A_J = 1.9$. The spin-polarization power multiplier is $\mathcal{N} A_J$, the fusion power is $P_f$, the plasma gain is $Q$, the tritium burn efficiency is TBE, the tritium startup inventory is  $I_{\mathrm{startup,min}}$, and the tritium fraction is $f_{\mathrm{T}}^{\mathrm{co}}$. All five axes have a value of zero at the origin and are linearly scaled.}
    \label{fig:arcclass_spider}
\end{figure}

In this work, we show that operating with lower tritium fraction spin-polarized D-T fuel can simultaneously achieve high TBE and high fusion power. We show an example of a plant operating with a reduced tritium fraction and spin-polarized (SP) fuel that achieves a 15 times greater TBE than a plant operating with 50:50 D-T and unpolarized fuel, without any degradation in fusion power. A significant result of this approach is that the initial tritium inventory could be decreased to such minimal levels that shortages of tritium supply would likely be eliminated. Such a scheme could reduce tritium startup inventory requirements by an order of magnitude and could allow plants to breed tritium more quickly, increasing the global tritium inventory. A power plant designed to operate at a range of tritium fractions would provide operating flexibility and might stabilize tritium prices: if the marginal profit of additional electricity generation is higher than selling additional tritium, the plant could change its configuration to breed less tritium and generate more electricity. It would almost always be desirable to operate at higher polarization fraction.

While it is well known that polarizing the deuterium and tritium nuclear spins increases the cross section by up to 50\% \cite{Kulsrud1986}, SP fuels have not yet been tested in fusion plasmas. However, the first SP fusion experiments to test the polarization lifetime are planned for 2025 on the DIII-D tokamak using deuterium helium-3 fuel \cite{Garcia2023,Baylor2023,Heidbrink2024}. Recent advances have now made it possible to polarize deuterium and helium-3 gas at $\sim60-70\%$, to produce SP fuel at sufficiently large quantities for experiments, and to keep the fuel polarized during the injection process \cite{Garcia2023,Baylor2023,Heidbrink2024}. Due to nonlinear effects in the plasma, the total fusion power increase with SP fuels can be even higher than the 50\% cross-section enhancement, reportedly 80\% \cite{Smith2018IAEA} and 90\% \cite{Heidbrink2024}. Such benefits would dramatically improve the economics of fusion power plants. However, there are major obstacles to overcome before SP fuels could be used to fuel power plants. Ensuring that fuel remains polarized sufficiently long is particularly challenging, with ion-cyclotron-frequency resonances and metallic-wall interactions in high recycling regimes particularly worrisome \cite{Kulsrud1986,Garcia2023,Baylor2023,Heidbrink2024}. Additionally, is it technologically hard to simultaneously achieve a high polarization fraction and produce sufficient fuel in a power plant fueling scheme, although there are recent promising advances \cite{Sofikitis2017,Sofikitis2018,Kannis2021,Baylor2023}.

The core insight of this paper is as follows: while tritium self-sufficiency challenges such as supply shortages for startup inventory are concerning, many of these challenges directly result from insufficient fusion plasma performance. If fusion power density can be improved through plasma physics advances, it can then be exchanged for higher tritium burn efficiency. The result is a comparable or even higher total fusion power with tritium burn efficiency at least an order of magnitude higher. Spin-polarized fusion ($\to$ higher power density) with lower tritium fraction ($\to$ higher TBE), which we study in this work, is just one example of increasing the fusion plasma performance and TBE.

For the reader seeking a quick summary, we plot the main results in \Cref{fig:arcclass_spider}. \Cref{fig:arcclass_spider} has five axes: plasma gain $Q$, fusion power $P_f$, core tritium fraction $f_{\mathrm{T} }^{\mathrm{co} }$, minimum tritium startup inventory $I_{\mathrm{startup,min}}$, and the TBE. Each line color corresponds to an effective cross-section enhancement $\mathcal{N} A_J$ due to spin polarization. $\mathcal{N} A_J = 1.0$ is for unpolarized fuel and $\mathcal{N} A_J = 1.9$ represents the 90\% enhancement found in \cite{Heidbrink2024}. \Cref{fig:arcclass_spider} shows results for the following line of inquiry: for an ARC-like power plant, what is the effect of specifying three different spin-polarization values $\mathcal{N} A_J$ and requiring a given TBE? The outputs are $Q$, $P_f$, $I_{\mathrm{startup,min}}$, and $f_{\mathrm{T} }^{\mathrm{co} }$. With a 50\% effective cross-section enhancement due to spin polarization requiring TBE = 0.10, plasma Q increases 650\% from 20 to 150, the fusion power increases 17\% from 482 MW to 562 MW, and the tritium startup inventory decreases 89\% from 0.68 kg to 0.08 kg. To accomplish this, the core tritium fraction $f_{\mathrm{T} }^{\mathrm{co} }$ must decrease from 49\% to 43\%. If one could achieve even higher effective cross section multiplier $\mathcal{N} A_J = 1.9$, the plasma ignites and the fusion power increases to $P_f = 712$ MW. More details of these and other cases are provided in \Cref{sec:casestudy}.

This work is structured as follows: in \Cref{sec:spin_polarization}, we introduce spin polarization. Notation for variable-tritium-fraction plasmas is introduced in \Cref{sec:var_trit_fraction}, and we study its combined effects with spin-polarized fuel on tritium burn efficiency, fusion power density, and fusion gain. In \Cref{sec:min_startup_inventory}, we show the effect of spin polarization and tritium fraction on tritium startup inventory. We apply these results to an ARC-like device in \Cref{sec:casestudy}. We conclude in \Cref{sec:discussion}. 

We report further results in the appendices. We discuss the assumptions made of deuterium and tritium particle transport in \Cref{sec:core_trit_frac}. In \Cref{sec:tritfueling_exhaust} we study tritium injection and divertor fractions, in \Cref{sec:divertor_fusionpower} we constrain fusion power, tritium enrichment, tritium fraction, and divertor pumping. The effect of different deuterium and tritium pumping speeds is briefly discussed in \Cref{sec:trit_deut_pump_efficiency}. The workflow for calculating the ARC-like device parameters is described in \Cref{sec:workflow}. We discuss some of the limitations of our work in \Cref{sec:limitations}. The helium particle and energy confinement times are studied in \Cref{sec:heliumparticleconfinement}. An argument for why the tritium burn efficiency benefits so strongly with spin polarization is presented in \Cref{sec:intuition}. Potential applications of spin polarization and variable tritium fraction for power control of ignited plasmas is discussed in \Cref{sec:ignitionstab}. We plot the plasma gain and ignition condition against tritium burn efficiency and spin polarization for two ARC-like power plants in \Cref{sec:Q10_scan}. In \Cref{sec:tritium_enrichment} we show the effect of tritium enrichment on plasma gain.

The main parameters that we use in this work are listed in \Cref{tab:tab0}.

\section{Spin Polarization} \label{sec:spin_polarization}

\begin{table*}
\caption{Key quantities used in this work.}
\begin{ruledtabular}
\centering
  \begin{tabular}{ cccc  }
   Name & Quantity & Units & Equation  \\
    \hline
    Spin-polarization cross-section multiplier & $A_J$ &&   \cref{eq:AJ2} \\
    Nonlinear power enhancement factor & $\mathcal{N} $ &&   \cref{eq:NLenhancement} \\
    Fusion power density & $p_f$ & W m$^{-3}$&   \cref{eq:pfform2_init} \\
    Deuterium, tritium, hydrogen density & $n_D, n_{\mathrm{T}}, n_Q$ &m$^{-3}$&  \cref{eq:pfform2_init,eq:pfform0} \\
    D-T fusion reactivity & $\langle v \overline{ \sigma} \rangle$ & m$^{3}$s$^{-1}$&   \cref{eq:pfform2_init} \\
    D-T fusion energy release & $E$ & J &  \cref{eq:pfform2_init} \\
    Tritium injection flow rate fraction & $F_{\mathrm{T} }^{\mathrm{in} }$ &&  \cref{eq:fueling_ratio} \\
    Tritium divertor removal flow rate fraction & $F_{\mathrm{T} }^{\mathrm{div} }$ &&  \cref{eq:divertor_ratio} \\
    Deuterium, tritium, hydrogen flow rate & $\dot{N}_\mathrm{D}, \dot{N}_\mathrm{T}, \dot{N}_\mathrm{Q}$ &s$^{-1}$&  \cref{eq:fueling_ratio} \\
    Tritium flow rate enrichment & $H_{\mathrm{T}}$ && \cref{eq:HTfirst} \\
    Core tritium density fraction & $f_{\mathrm{T} }^{\mathrm{co} }$ &&  \cref{eq:fTco_def} \\
    Fusion power & $P_f$ &W& \cref{eq:totalfusion} \\
    Tritium burn rate & $\dot{N}_{\mathrm{T},\mathrm{burn}}$ &s$^{-1}$&  \cref{eq:tritium_burn} \\
    Helium ash removal rate & $\dot{N}_{\mathrm{He},\mathrm{div}}$ &s$^{-1}$&  \cref{eq:tritium_burn} \\
    Divertor neutral pumping speed for species $x$ & $S_x$ &m$^{3}$s$^{-1}$& \cref{eq:neutralpump} \\
    Tritium burn efficiency & TBE && \cref{eq:TBEfirst} \\
    Helium ash to hydrogen pumping speed ratio & $\Sigma$ && \cref{eq:ashpumpspeed} \\
    Helium-to-fuel divertor density ratio & $f_{\mathrm{He,div} }$ && \cref{eq:ashtofuel} \\
    Helium-to-electron core density ratio & $f_{\mathrm{dil} }$ && \cref{eq:dil1} \\
    Power density multiplier & $p_{\Delta}$ && \cref{eq:pDeltaform0} \\
    Helium density enrichment & $\eta_{\mathrm{He} }$ && \cref{eq:etaHe} \\
    Plasma gain & $Q$ && \cref{eq:corepower} \\
    Thermal energy density & $w_{\mathrm{th} }$ &J m$^{-3}$ & \cref{eq:corepower} \\
    Energy confinement time & $\tau_E$ &s& \cref{eq:corepower} \\
    Constant plasma gain multiplication factor & $C$ && \cref{eq:C} \\
    Minimum tritium startup inventory & $I_\mathrm{startup,min}$ & kg & \cref{eq:tritium_storage} \\
    Helium particle confinement time & $\tau_{\mathrm{He} }^*$ & s & \cref{eq:tauHestardef} \\
  \end{tabular}
\end{ruledtabular}
\label{tab:tab0}
\end{table*}

\begin{figure}[b]
    \centering
    \includegraphics[width=0.9\textwidth]{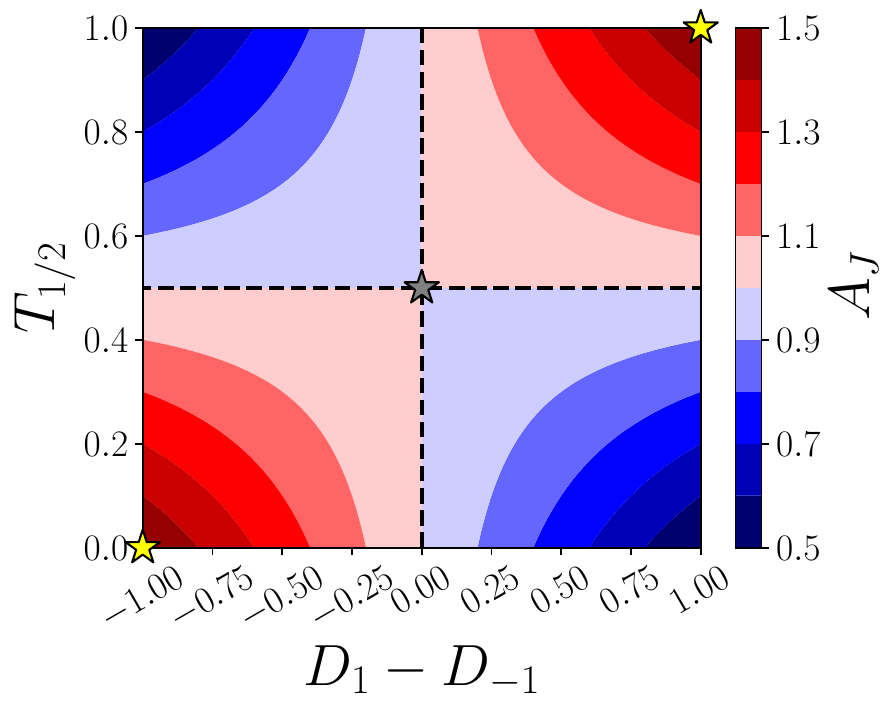}
    \caption{Polarization cross-section multiplier $A_J$ (\Cref{eq:AJ2}) versus deuterium spin difference $D_1 - D_{-1}$ and tritium spin probability $T_{1/2}$. The dashed contour indicates $A_J = 1.0$, the grey star is the operating point for unpolarized fuel, and the yellow stars are the maximum cross-section points.}
    \label{fig:AJ1}
\end{figure}

In this section we introduce the fuel polarization. The total D-T fusion cross section is
\begin{equation}
\sigma = \overline{\sigma} A_J,
\label{eq:AJ}
\end{equation}
where $\overline{\sigma}$ is the nominal unpolarized D-T cross section and $A_J$ is the polarization cross-section multiplier
\begin{equation}
A_J  \equiv \left( 1 + P_D P_T /2 \right),
\label{eq:AJ2}
\end{equation}
which generally satisfies $A_J \in [0.5,1.5]$ \cite{Kulsrud1986}. For unpolarized fusion, the nuclear spins are randomly oriented and $A_J$ = 1.0. Here, $P_D, P_T$ are the vector polarizations of deuterium and tritium, where $P_D = D_{1} - D_{-1}$ and $P_T = T_{1/2} - T_{-1/2}$. Here, $D_m$ and $T_m$ are the probabilities of being in a nuclear spin state $m$, where $m = 1,0,-1$ for deuterium and $m = 1/2,-1/2$ for tritium, satisfying $\sum D_m = \sum T_m = 1$. By choosing $P_D P_T = 1$, the cross section is enhanced by 50\%. We refer to this as the `enhanced parallel polarization' (EPP) scheme. We indicate the EPP scheme with yellow stars in \Cref{fig:AJ1} and an unpolarized fuel with a grey star. Shown in \Cref{fig:AJ1}, both tritium and deuterium must have some polarization bias for the multiplier $A_J$ to change from 1. While polarizing just one of deuterium or tritium does not change the total cross section, it does change the differential cross section \cite{Kulsrud1986}.

It is important to note that the fusion power increase has been reported to be higher than the cross-section increase for SP fuel. Recent works have found that with $A_J = 1.5$, the total fusion power increased by 80\% and 90\% \cite{Smith2018IAEA,Heidbrink2024} (higher than 50\% from the increased cross section) due to increased alpha heating. To approximately capture this nonlinear effect in this work, we pre-multiply the temperature-dependent D-T fusion reactivity $\langle v \overline{ \sigma} \rangle$ by a nonlinear-enhancement factor $\mathcal{N}$,
\begin{equation}
\langle v \overline{ \sigma} \rangle \to \mathcal{N} \langle v \overline{ \sigma} \rangle.
\label{eq:NLenhancement}
\end{equation}
Physically, $\mathcal{N}$ results from modifications to the temperature and (possibly) density due to alpha heating. The fusion power density on a flux surface is
\begin{equation}
\begin{aligned}
& p_f = \mathcal{N} A_J n_D n_{\mathrm{T}} \langle v \overline{ \sigma} \rangle E,
\end{aligned}
\label{eq:pfform2_init}
\end{equation}
where $n_D$ and $n_{\mathrm{T}}$ are the deuterium and tritium densities, and $E = 17.6$ MeV is the energy released from the D-T fusion reaction. For power increases of 80\% and 90\% with $A_J = 1.5$ \cite{Smith2018IAEA,Heidbrink2024}, one would set $\mathcal{N} A_J = 1.8, 1.9$. When $\mathcal{N} A_J = 1$, \Cref{eq:pfform2_init} returns to the standard expression for fusion power density, $p_f = n_D n_{\mathrm{T}} \langle v \overline{ \sigma} \rangle E$. Throughout this work $\mathcal{N}$ and $A_J$ will always appear together as $\mathcal{N} A_J$. Additionally, $\langle v \overline{ \sigma} \rangle$ should be interpreted as being at constant temperature for $\mathcal{N} \neq 1$; all of the temperature dependence is carried by $\mathcal{N}$. For further discussion of the limitations of and potential solutions to this approach, see \Cref{sec:constantT}.

\section{Variable Tritium Fraction} \label{sec:var_trit_fraction}

In this section, we introduce the notation for D-T plasmas with a variable tritium fraction. Throughout this work, we draw extensively from the notation and methods used in \cite{Whyte2023}. Our analysis is confined to steady-state operation in magnetic confinement fusion power plants such as tokamaks, stellarators, and mirrors. 

We study plasmas where the densities and flow rates are not necessarily equal for deuterium and tritium. A power plant operator controls the tritium fraction $F_{\mathrm{T}}^{\mathrm{in}}$ of the total fuel injection rate
\begin{equation}
F_{\mathrm{T}}^{\mathrm{in}} \equiv \frac{\dot{N}_{\mathrm{T},\mathrm{in}}}{\dot{N}_{Q,\mathrm{in}}},
\label{eq:fueling_ratio}
\end{equation}
where $\dot{N}_{\mathrm{T},\mathrm{in}}$ and $\dot{N}_{Q,\mathrm{in}}$ are the number of tritium and total unburned fuel particles injected into the device chamber per second. In the divertor, the tritium fraction $F_{\mathrm{T}}^{\mathrm{div}}$ of the total unburned fuel removal rate is
\begin{equation}
F_{\mathrm{T}}^{\mathrm{div}} \equiv \frac{\dot{N}_{\mathrm{T},\mathrm{div}}}{\dot{N}_{Q,\mathrm{div}}},
\label{eq:divertor_ratio}
\end{equation}
where $\dot{N}_{\mathrm{T},\mathrm{div}}$ and $\dot{N}_{Q,\mathrm{div}}$ are the number of tritium and total unburned fuel particles removed from the divertor per second. We define the tritium enrichment $H_{\mathrm{T}}$ as the ratio of the divertor tritium fraction to the injection tritium fraction,
\begin{equation}
H_{\mathrm{T} } \equiv \frac{F_{\mathrm{T}}^{\mathrm{div}}}{F_{\mathrm{T}}^{\mathrm{in}}}.
\label{eq:HTfirst}
\end{equation}
We assume that the D-T core density mix is $1:a$, where $a \geq 0$ is a real number. The tritium and deuterium core densities satisfy
\begin{equation}
n_{\mathrm{T},\mathrm{co}} = a n_{\mathrm{D},\mathrm{co}}, \;\; n_{\mathrm{T},\mathrm{co}} + n_{\mathrm{D},\mathrm{co}} = (1+a) n_{\mathrm{D},\mathrm{co}}.
\end{equation}
Practically, a D-T mix that is not $1:1$ is maintained by differing injection and divertor removal rates for deuterium and tritium. In \Cref{sec:core_trit_frac}, we show that the core tritium flow rate fraction
\begin{equation}
F_{\mathrm{T}}^{\mathrm{co}} \equiv \frac{\dot{N}_{\mathrm{T},\mathrm{co}}}{\dot{N}_{Q,\mathrm{co}}},
\label{eq:core_ratio}
\end{equation}
is equal to the typical tritium density fraction $f_{\mathrm{T}  }^{\mathrm{co} }$, under certain assumptions. Therefore, for this work we will assume that
\begin{equation}
f_{\mathrm{T}  }^{\mathrm{co} } = F_{\mathrm{T}}^{\mathrm{co}},
\label{eq:ftco_FTco}
\end{equation}
where
\begin{equation}
f_{\mathrm{T}  }^{\mathrm{co} } \equiv \frac{n_{\mathrm{T},\mathrm{co}}}{n_{\mathrm{Q},\mathrm{co}}}. 
\end{equation}
In future work, it may be interesting to explore the consequences of $f_{\mathrm{T}  }^{\mathrm{co} } \neq F_{\mathrm{T}}^{\mathrm{co}}$ arising from effects such as differing deuterium and tritium particle transport \cite{Estrada-Mila:2005aa,Belli2021}. We assume that the tritium fractions satisfy,
\begin{equation}
F_{\mathrm{T}}^{\mathrm{div}} \leq f_{\mathrm{T}}^{\mathrm{co}} \leq F_{\mathrm{T}}^{\mathrm{in}}, \;\;\;\;\;\; (F_{\mathrm{T}}^{\mathrm{in}} < 1/2)
\label{eq:FTinboundlesshalf}
\end{equation}
for reduced tritium injection fraction $(F_{\mathrm{T}}^{\mathrm{in}} < 1/2)$ and
\begin{equation}
F_{\mathrm{T}}^{\mathrm{div}} \geq f_{\mathrm{T}}^{\mathrm{co}} \geq F_{\mathrm{T}}^{\mathrm{in}}, \;\;\;\;\;\; (F_{\mathrm{T}}^{\mathrm{in}} > 1/2)
\label{eq:FTinboundmorehalf}
\end{equation}
for enhanced tritium injection fraction $(F_{\mathrm{T}}^{\mathrm{in}} > 1/2)$. For $H_{\mathrm{T}} \neq 1$, the tritium density and flow rate fractions $F_{\mathrm{T}}^{\mathrm{co}}$ and $f_{\mathrm{T}  }^{\mathrm{co} }$ have a radial dependence. Therefore, we interpret $f_{\mathrm{T}  }^{\mathrm{co} }$ as an `average' core tritium density fraction rather than the exact value on a given flux surface. Thus,
\begin{equation}
\begin{aligned}
f_{\mathrm{T}  }^{\mathrm{co}} & = \frac{F_{\mathrm{T}}^{\mathrm{div}} + F_{\mathrm{T}}^{\mathrm{in}}}{2} \\
& = F_{\mathrm{T}}^{\mathrm{in}} \frac{1 + H_{\mathrm{T}}}{2} = F_{\mathrm{T}}^{\mathrm{div}} \frac{1 + 1/H_{\mathrm{T}}}{2},
\end{aligned}
\label{eq:fTco_def}
\end{equation}
where we used $F_{\mathrm{T}}^{\mathrm{div}} = H_{\mathrm{T}} F_{\mathrm{T}}^{\mathrm{in}}$. Given that $f_{\mathrm{T}  }^{\mathrm{co} }$ must always satisfy $f_{\mathrm{T}  }^{\mathrm{co} } \leq 1$, this constrains $F_{\mathrm{T}}^{\mathrm{in}}$ and $H_{T}$,
\begin{equation}
F_{\mathrm{T}}^{\mathrm{in}} \leq \frac{2}{1 + H_{\mathrm{T}}}. 
\label{eq:FTinconstraint}
\end{equation}
Furthermore, for reduced tritium fraction $(F_{\mathrm{T}}^{\mathrm{in}} < 1/2)$, \Cref{eq:FTinboundlesshalf} and requiring $F_{\mathrm{T}}^{\mathrm{div}} \geq 0$ constrains $H_{\mathrm{T}}$,
\begin{equation}
0 \leq H_{\mathrm{T}} < 1,
\end{equation}
and for enhanced tritium fraction $(F_{\mathrm{T}}^{\mathrm{in}} > 1/2)$, \Cref{eq:FTinboundmorehalf} constrains $H_{\mathrm{T}}$,
\begin{equation}
1< H_{\mathrm{T}} < \frac{1}{F_{\mathrm{T}}^{\mathrm{in}}},
\label{eq:HT_bounds2}
\end{equation}
where the upper bound is derived in \Cref{sec:enrichmentbounds}, and corresponds to $F_{\mathrm{T}}^{\mathrm{div}} = 1$.

The total fusion power $P_f$ is 
\begin{equation}
P_f \equiv \int p_f d V = E \dot{N}_{\alpha},
\label{eq:totalfusion}
\end{equation}
where the integral is evaluated over the total plasma volume $V$ and $\dot{N}_{\alpha}$ is the alpha production rate in the whole plasma. The power density on a flux surface $p_f$ is
\begin{equation}
\begin{aligned}
& p_f = f_{\mathrm{T}}^{\mathrm{co}} (1-f_{\mathrm{T}}^{\mathrm{co}})  \mathcal{N} A_J n_{Q,\mathrm{co}}^2 \langle v \overline{ \sigma} \rangle E,
\end{aligned}
\label{eq:pfform0}
\end{equation}
where $n_{Q,\mathrm{co}} = n_{D,\mathrm{co}} + n_{T,\mathrm{co}}$ is the unburned fuel density. We remind the reader that $f_{\mathrm{T}}^{\mathrm{co}}$, defined in \Cref{eq:fTco_def}, is not the exact tritium density fraction on a flux surface; when $|1 - H_{\mathrm{T}}| = \mathcal{O}\left( 1 \right)$, the error in the power density expression in \Cref{eq:pfform0} can become large because $f_{\mathrm{T}}^{\mathrm{co}}$ will have a strong radial dependence.

By particle conservation, the alpha production rate is equal to the tritium burn rate $\dot{N}_{\mathrm{T},\mathrm{burn}}$ and the helium ash removal in the divertor $\dot{N}_{\mathrm{He},\mathrm{div}}$,
\begin{equation}
\dot{N}_{\alpha} = \dot{N}_{\mathrm{T},\mathrm{burn}} = \dot{N}_{\mathrm{He},\mathrm{div}}.
\label{eq:tritium_burn}
\end{equation}
Conservation of particles requires that
\begin{equation}
\dot{N}_{Q,\mathrm{in}} = 2 \dot{N}_{\mathrm{He},\mathrm{div}} + \dot{N}_{Q,\mathrm{div}}.
\label{eq:particle_conservation}
\end{equation}
Using \Cref{eq:tritium_burn,eq:particle_conservation}, \Cref{eq:divertor_ratio} becomes
\begin{equation}
\dot{N}_{\mathrm{T},\mathrm{div}} = F_{\mathrm{T}}^{\mathrm{div}} \left( \frac{\dot{N}_{\mathrm{T},\mathrm{in}}}{F_{\mathrm{T}}^{\mathrm{in}}} - 2 \dot{N}_{\alpha} \right).
\label{eq:tritiumdivertor}
\end{equation}
\Cref{eq:tritiumdivertor} shows that the tritium divertor flow rate is proportional to the difference between injected hydrogen and alpha particle production. The individual tritium and deuterium flow rates satisfy
\begin{equation}
\begin{aligned}
& \dot{N}_{\mathrm{T},\mathrm{in}} = \dot{N}_{\mathrm{T},\mathrm{div}} + \dot{N}_{\mathrm{He},\mathrm{div}} = \frac{F_{\mathrm{T}}^{\mathrm{in}}}{1-F_{\mathrm{T}}^{\mathrm{in}}}  \dot{N}_{\mathrm{D},\mathrm{in}}.
\label{eq:individualtritium}
\end{aligned}
\end{equation}

The new contributions of this work are the effects of the spin-polarization multiplier $\mathcal{N} A_J$ and tritium injection fraction $F_{\mathrm{T}}^{\mathrm{in}}$. A fusion power plant operator could control $A_J$ and $F_{\mathrm{T}}^{\mathrm{in}}$ (but not necessarily $\mathcal{N}$) with a fueling scheme where the polarization and tritium fraction are adjustable.

\subsection{Tritium Burn Efficiency} \label{sec:tritium_burn_efficiency}

In order to measure the efficiency of tritium burn, we use tritium burn efficiency (TBE) \cite{Whyte2023}, defined as
\begin{equation}
\mathrm{TBE} \equiv \frac{\dot{N}_{\mathrm{T},\mathrm{burn}}}{\dot{N}_{\mathrm{T},\mathrm{in}}} =\left( \frac{\dot{N}_{\mathrm{T},\mathrm{div}}}{\dot{N}_{\mathrm{He},\mathrm{div}}} +1 \right)^{-1}.
\label{eq:TBEfirst}
\end{equation}
Physically, the TBE measures the probability of a tritium particle undergoing a fusion reaction from the moment it is injected into the chamber to the moment it leaves the chamber through the divertor. Note that the TBE is different to the more frequently used burn fraction \cite{Abdou2021}, which measures the fraction of tritium burned in a single pass through the plasma. For a more extensive discussion of TBE versus other tritium burn metrics, see \cite{Abdou2021,Whyte2023,Meschini2023}. For an analytic study of tritium burn fraction, see \cite{Jackson2013}.

\begin{figure}[!tb]
    \centering
    \begin{subfigure}[t]{0.88\textwidth}
    \centering
    \includegraphics[width=1.0\textwidth]{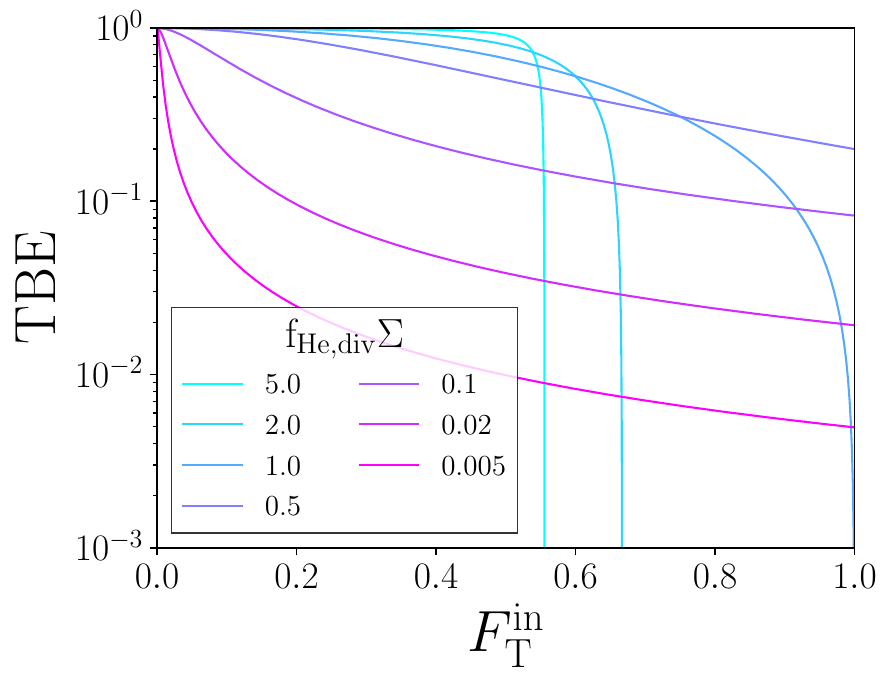}
    \caption{TBE (\Cref{eq:TBEform3_c}) versus $F_{\mathrm{T}}^{\mathrm{in}}$.}
    \end{subfigure}
     ~
     ~
    \begin{subfigure}[t]{0.88\textwidth}
    \centering
    \includegraphics[width=1.0\textwidth]{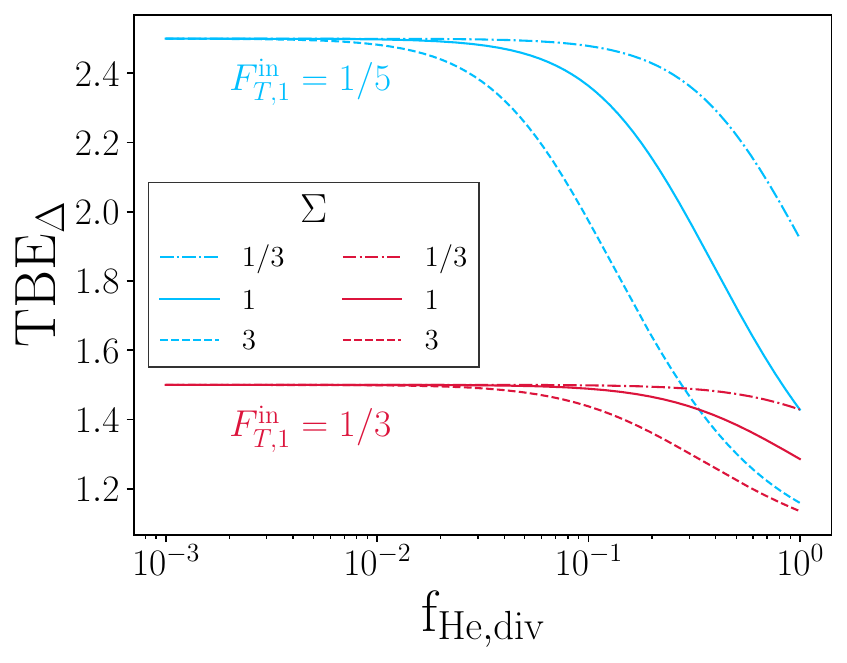}
    \caption{TBE enhancement (\Cref{eq:TBEenhancement}) versus $f_{\mathrm{He},\mathrm{div}}$.}
    \end{subfigure}
     ~
    \begin{subfigure}[t]{0.88\textwidth}
    \centering
    \includegraphics[width=1.0\textwidth]{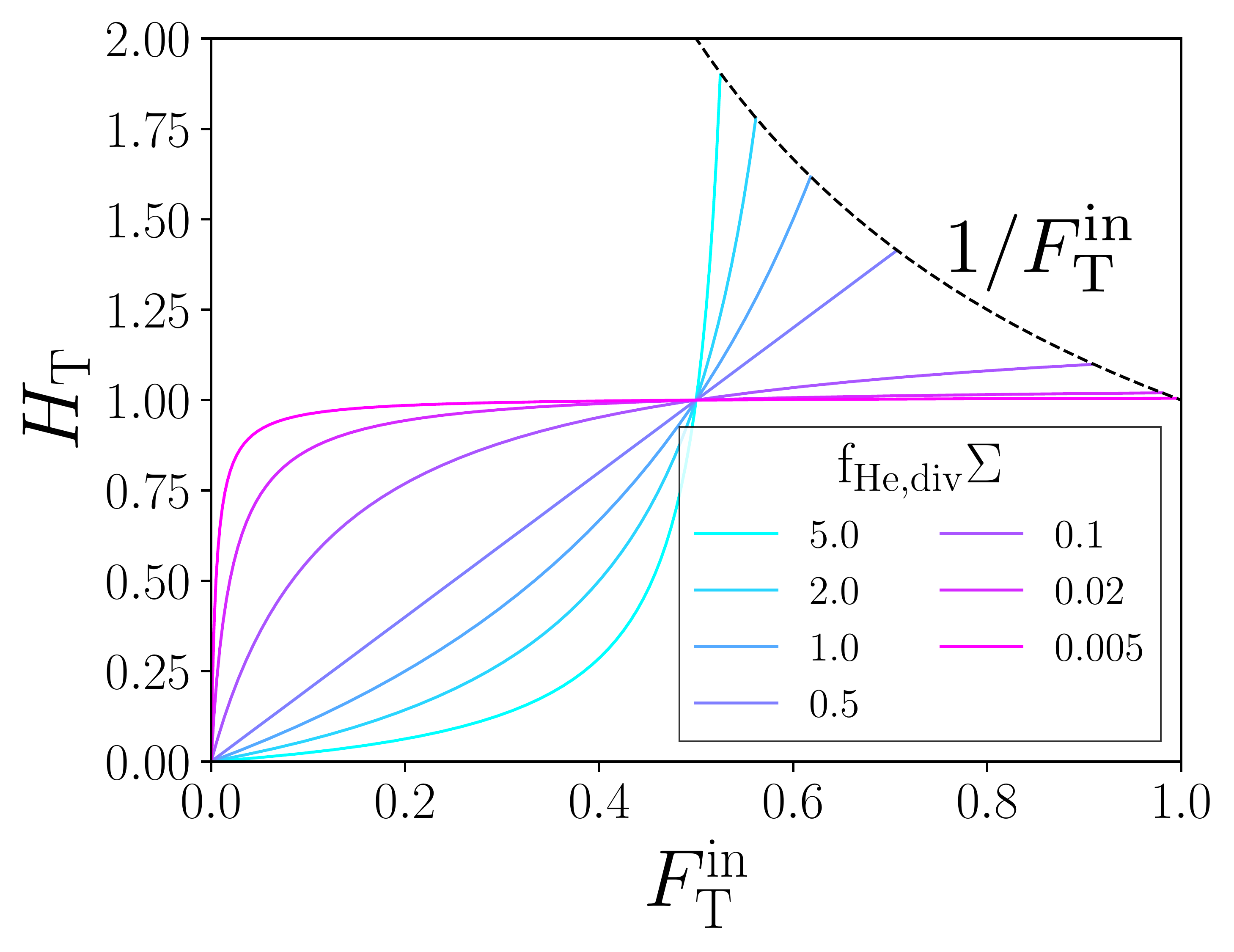}
    \caption{Tritium enrichment (\Cref{eq:FTdiv_new}) versus $F_{\mathrm{T}}^{\mathrm{in}}$, including $1/F_{\mathrm{T}}^{\mathrm{in}}$ scaling for forbidden region (see \Cref{eq:HT_bounds2}).}
    \end{subfigure}
    \caption{Tritium burn efficiency (TBE) (a) and tritium enrichment $H_{\mathrm{T}}$ (c) as a function of tritium injection fraction $F_{\mathrm{T}}^{\mathrm{in}}$ for different $f_{\mathrm{He},\mathrm{div}} \Sigma$ values. In (b), we plot the ratio of TBE for two tritium input fractions $F_{\mathrm{T}}^{\mathrm{in}}$ as a function of the ratio of helium to unburned hydrogen fuel in the divertor $f_{\mathrm{He},\mathrm{div}}$.}
    \label{fig:TBE1}
\end{figure}

Substituting $\dot{N}_{\mathrm{T},\mathrm{div}} = F_{\mathrm{T}}^{\mathrm{div}} \dot{N}_{Q,\mathrm{div}}$ (see \Cref{eq:divertor_ratio}) into \Cref{eq:TBEfirst},
\begin{equation}
\mathrm{TBE} = \left( F_{\mathrm{T}}^{\mathrm{div}} \frac{\dot{N}_{Q,\mathrm{div}}}{\dot{N}_{\mathrm{He},\mathrm{div}}}  +1 \right) ^{-1}.
\label{eq:TBEform2}
\end{equation}
We wish to replace $\dot{N}_{Q,\mathrm{div}}/\dot{N}_{\mathrm{He},\mathrm{div}}$ in terms of dimensionless variables. To do this, we write the divertor flow rate for a species $x$ as given by the neutral gas density $n_{x,\mathrm{div}}$ and effective pumping speed $S_x$,
\begin{equation}
\dot{N}_{x,\mathrm{div}} = n_{x,\mathrm{div}} S_x.
\label{eq:neutralpump}
\end{equation}
The helium-to-fuel divertor pumping ratio is
\begin{equation}
\Sigma \equiv \frac{S_{\mathrm{He}}}{S_Q},
\label{eq:ashpumpspeed}
\end{equation}
and the helium-to-fuel divertor density ratio is
\begin{equation}
f_{\mathrm{He},\mathrm{div}} \equiv \frac{n_{\mathrm{He},\mathrm{div}}}{n_{Q,\mathrm{div}}}.
\label{eq:ashtofuel}
\end{equation}
Using \Cref{eq:neutralpump,eq:ashpumpspeed,eq:ashtofuel} in \Cref{eq:TBEform2} we find
\begin{equation}
\frac{\dot{N}_{\mathrm{He},\mathrm{div}}}{\dot{N}_{Q,\mathrm{div}}} = f_{\mathrm{He},\mathrm{div}} \Sigma.
\label{eq:helium_exhaust_here}
\end{equation}
Therefore, substituting \Cref{eq:helium_exhaust_here} into \Cref{eq:TBEform2},
\begin{equation}
\mathrm{TBE} = \left( F_{\mathrm{T}}^{\mathrm{div}} \frac{1}{f_{\mathrm{He},\mathrm{div}} \Sigma} +1 \right)^{-1}.
\label{eq:TBEform3}
\end{equation}

From the perspective of a plant operator, the tritium injection fraction $F_{\mathrm{T}}^{\mathrm{in}}$ is easier to control than $F_{\mathrm{T}}^{\mathrm{div}}$, and so we rewrite the TBE in terms of $F_{\mathrm{T}}^{\mathrm{in}}$. Using the expression for $F_{\mathrm{T}}^{\mathrm{div}}$ in \Cref{eq:FTdiv0}, the TBE expressed in terms of $F_{\mathrm{T}}^{\mathrm{in}}$ is
\begin{equation}
\mathrm{TBE} = \left( \frac{F_{\mathrm{T}}^{\mathrm{in}}}{ f_{\mathrm{He},\mathrm{div}} \Sigma \left( 1 +   f_{\mathrm{He},\mathrm{div}} \Sigma \left( \frac{1}{F_{\mathrm{T}}^{\mathrm{in}}} - 2 \right) \right)} +1 \right)^{-1},
\label{eq:TBEform3_c}
\end{equation}
which we plot in \Cref{fig:TBE1}(a): decreasing $F_{\mathrm{T}}^{\mathrm{in}}$ always increases the TBE, and increasing $f_{\mathrm{He},\mathrm{div}}$ increases the TBE for $F_{\mathrm{T}}^{\mathrm{in}}<1/2$ but decreases the TBE for $F_{\mathrm{T}}^{\mathrm{in}}>1/2$.

To measure the improvement in the TBE with lower $F_{\mathrm{T}}^{\mathrm{in}}$, we define the TBE enhancement,
\begin{equation}
\begin{aligned}
& \mathrm{TBE}_{\Delta} (F_{\mathrm{T}}^{\mathrm{in}}, f_{\mathrm{He},\mathrm{div}}, \Sigma) \\ & \equiv \frac{\mathrm{TBE} (F_{\mathrm{T}}^{\mathrm{in}}, f_{\mathrm{He},\mathrm{div}}, \Sigma) }{\mathrm{TBE} (F_{\mathrm{T}}^{\mathrm{in}}=1/2, f_{\mathrm{He},\mathrm{div}}, \Sigma) },
\end{aligned}
\label{eq:TBEenhancement}
\end{equation}
which measures the enhancement (or degradation) of the TBE relative to the TBE with an equal tritium and deuterium mix, $F_{\mathrm{T}}^{\mathrm{in}} = 1/2$. In \Cref{fig:TBE1}(b) we plot $\mathrm{TBE}_{\Delta}$ for $F_{\mathrm{T}}^{\mathrm{in}} = 1/5, 1/3$. In the region $F_{\mathrm{T}}^{\mathrm{div}} /f_{\mathrm{He},\mathrm{div}} \Sigma \gg 1$,
\begin{equation}
\mathrm{TBE}_{\Delta} \approx \frac{1}{F_{\mathrm{T}}^{\mathrm{div}}} \simeq \frac{1}{F_{\mathrm{T}}^{\mathrm{in}}}.
\label{eq:TBE_approx}
\end{equation}
\Cref{fig:TBE1}(b) shows that the relative improvement in the TBE scheme for a small tritium fraction is only significantly less than $1/F_{\mathrm{T}}^{\mathrm{in}}$ when $f_{\mathrm{He},\mathrm{div}} \Sigma$ is order unity, and even when $f_{\mathrm{He},\mathrm{div}} \Sigma$ is order unity, there is still a substantial improvement in the TBE.

Using \Cref{eq:helium_exhaust_here}, the tritium enrichment is
\begin{equation}
H_{\mathrm{T}} = \frac{1}{1 +   f_{\mathrm{He},\mathrm{div}} \Sigma \left( 1 /F_{\mathrm{T}}^{\mathrm{in}} - 2 \right)  }.
\label{eq:HT_expression}
\end{equation}
\Cref{eq:HT_expression} provides one way to conceptualize tritium enrichment: a plant operator controls $\Sigma$ and $F_{\mathrm{T}}^{\mathrm{in} }$ through divertor pumping and the fueling system. In \Cref{fig:TBE1}(c), we plot $H_{\mathrm{T}}$ versus $F_{\mathrm{T}}^{\mathrm{in}}$ for different $f_{\mathrm{He},\mathrm{div}} \Sigma$ values. At fixed $F_{\mathrm{T}}^{\mathrm{in}} < /2$, increasing $f_{\mathrm{He},\mathrm{div}} \Sigma$ decreases $H_{\mathrm{T}}$. This describes a higher TBE since less tritium is removed in the divertor. Through \Cref{eq:HT_expression}, $H_{\mathrm{T}} $ and $ f_{\mathrm{He},\mathrm{div}}$ are coupled: knowing $F_{\mathrm{T}}^{\mathrm{in}}$, $\Sigma$, and $H_{\mathrm{T}}$ uniquely specifies $f_{\mathrm{He},\mathrm{div}}$.

\subsection{Fusion power density}

\begin{figure}[tb]
    \centering
    \includegraphics[width=0.9\textwidth]{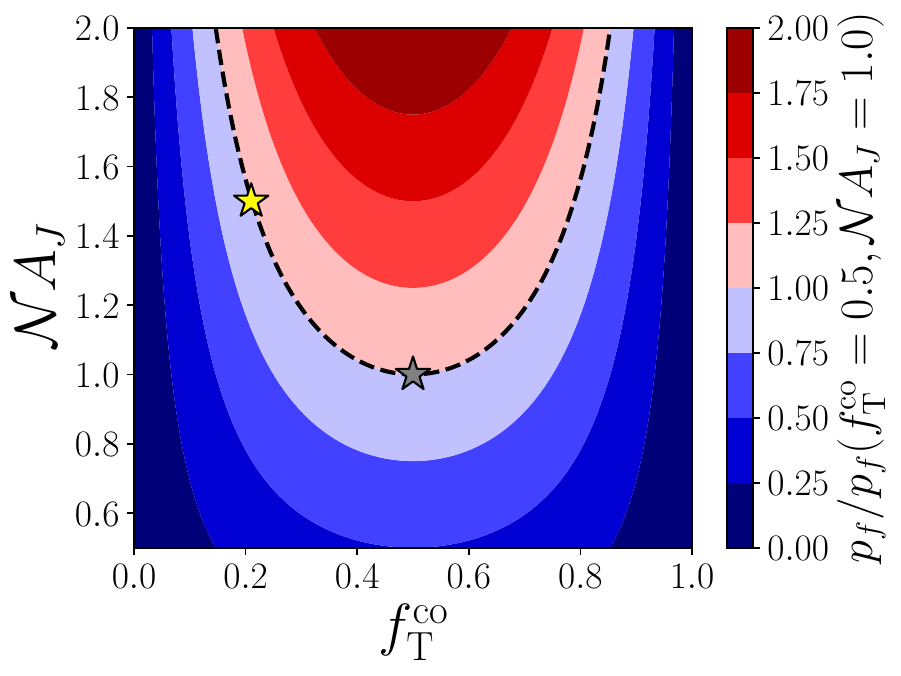}
    \caption{Neglecting helium dilution effects, fusion power density $p_f$ in \Cref{eq:pfform2} normalized to the nominal power density with zero polarization $\mathcal{N} A_J = 1.0$ and a 50:50 D-T mix ($f_{\mathrm{T}}^{\mathrm{co}} = 0.5$). The dashed contour indicates a value of 1.0, the grey star indicates $\mathcal{N} A_J = 1.0$, $f_{\mathrm{T}}^{\mathrm{co}} = 0.5$, and the yellow star indicates $\mathcal{N} A_J = 1.5$, $f_{\mathrm{T}}^{\mathrm{co}} = 0.21$.}
    \label{fig:powerdensity1}
\end{figure}

We now consider the effect of tritium fraction and spin polarization on the power density. The helium-to-electron core density ratio is,
\begin{equation}
f_{\mathrm{dil}} \equiv \frac{n_{\alpha}}{n_e},
\label{eq:dil1}
\end{equation}
also known as the ash dilution fraction \cite{Whyte2023}. Using quasineutrality
\begin{equation}
n_e = n_{Q,\mathrm{co}} + 2 n_{\alpha},
\end{equation}
and ignoring impurities, the fusion power density as a function of $f_{\mathrm{T}}^{\mathrm{co}}$, $f_{\mathrm{dil}}$, and $\mathcal{N} A_J$ is
\begin{equation}
\begin{aligned}
& p_f = 4 f_{\mathrm{T}}^{\mathrm{co}} (1-f_{\mathrm{T}}^{\mathrm{co}}) (1-2 f_{\mathrm{dil}})^2 \mathcal{N} A_J n_e^2 \langle v \overline{ \sigma} \rangle \frac{E}{4},
\end{aligned}
\label{eq:pfform2}
\end{equation}
where $n_e$ is the electron density. Thus, at fixed $n_e$ and temperature, the fusion power density $p_f$ relative to its maximum value $p_{f,max}$ where $p_{f,max}$ has $f_{\mathrm{T}}^{\mathrm{co}} = 1/2, f_{\mathrm{dil}} = 0, \mathcal{N} A_J = 1$, is given by the power multiplier $p_{\Delta}$
\begin{equation}
p_{\Delta} \equiv \frac{p_f}{p_{f,max}} = 4 f_{\mathrm{T}}^{\mathrm{co}} (1-f_{\mathrm{T}}^{\mathrm{co}}) (1-2 f_{\mathrm{dil}})^2  \mathcal{N} A_J.
\label{eq:pDeltaform0}
\end{equation}
Importantly, in \Cref{eq:pDeltaform0} $f_{\mathrm{dil}}$ cannot be varied independently of $f_{\mathrm{T}}^{\mathrm{co}}$ and $\mathcal{N} A_J$, and hence, at fixed $\mathcal{N} A_J$ the power density $p_{\Delta}$ is not necessarily maximized for an equal D-T mixture, $f_{\mathrm{T}}^{\mathrm{co}} = 1/2$. This is unlike in \cite{Whyte2023} that used $p_{\Delta} =(1-2 f_{\mathrm{dil}})^2$ where it was assumed that $f_{\mathrm{T}}^{\mathrm{co}} = 1/2$ and $\mathcal{N} A_J = 1.0$. 

\begin{figure*}[!tb]
    \centering
    \begin{subfigure}[t]{0.47\textwidth}
    \includegraphics[width=1.0\textwidth]{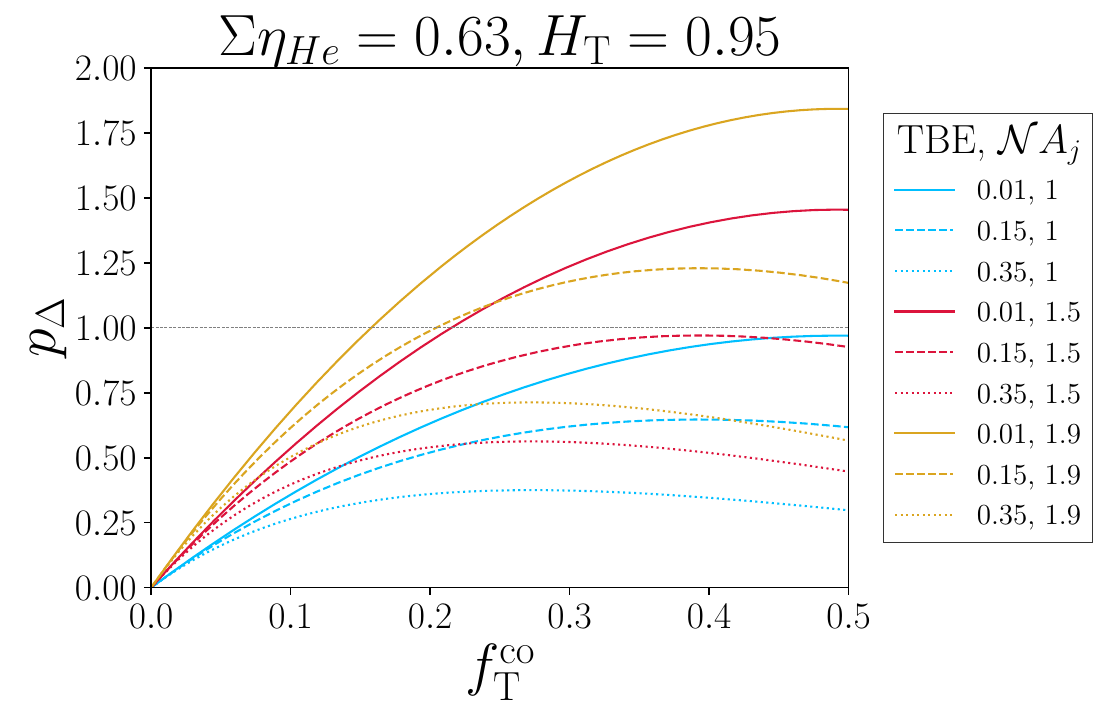}
    \caption{$p_{\Delta}$ versus $f_{\mathrm{T}}^{\mathrm{co}}$, fixed enrichment $H_\mathrm{T} = 0.95$.}
    \end{subfigure}
     ~
    \centering
    \begin{subfigure}[t]{0.47\textwidth}
    \includegraphics[width=1.0\textwidth]{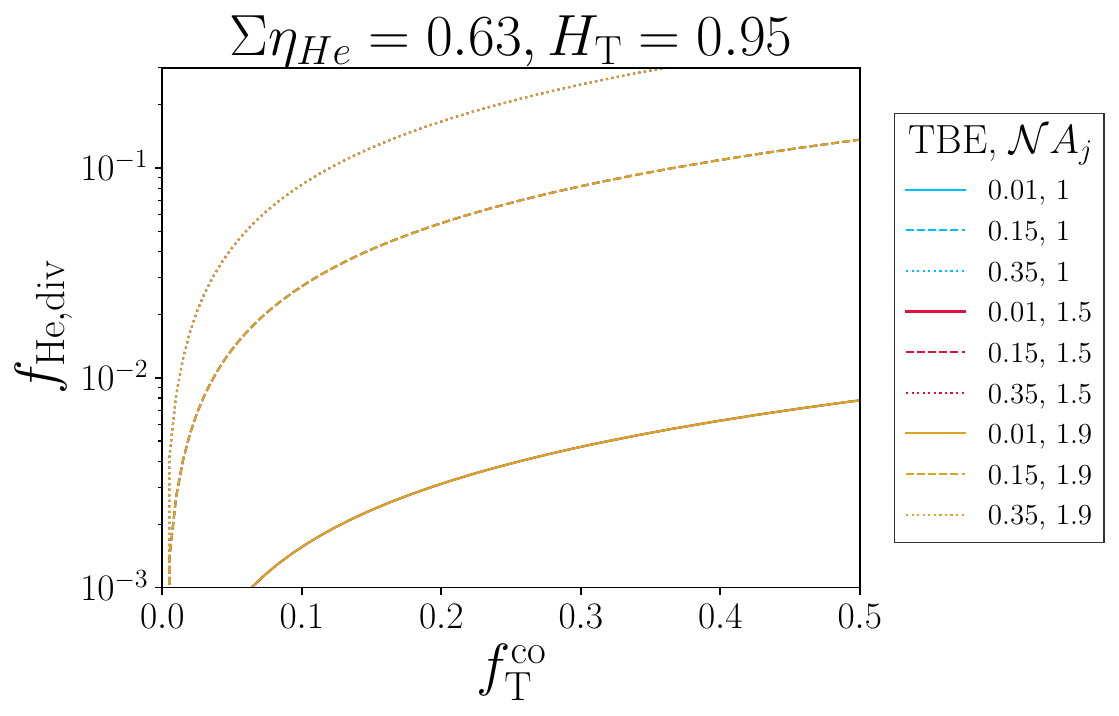}
    \caption{$f_{\mathrm{He,div}}$ versus $f_{\mathrm{T}}^{\mathrm{co}}$, fixed enrichment $H_\mathrm{T} = 0.95$.}
    \end{subfigure}
    \centering
    \begin{subfigure}[t]{0.47\textwidth}
    \includegraphics[width=1.0\textwidth]{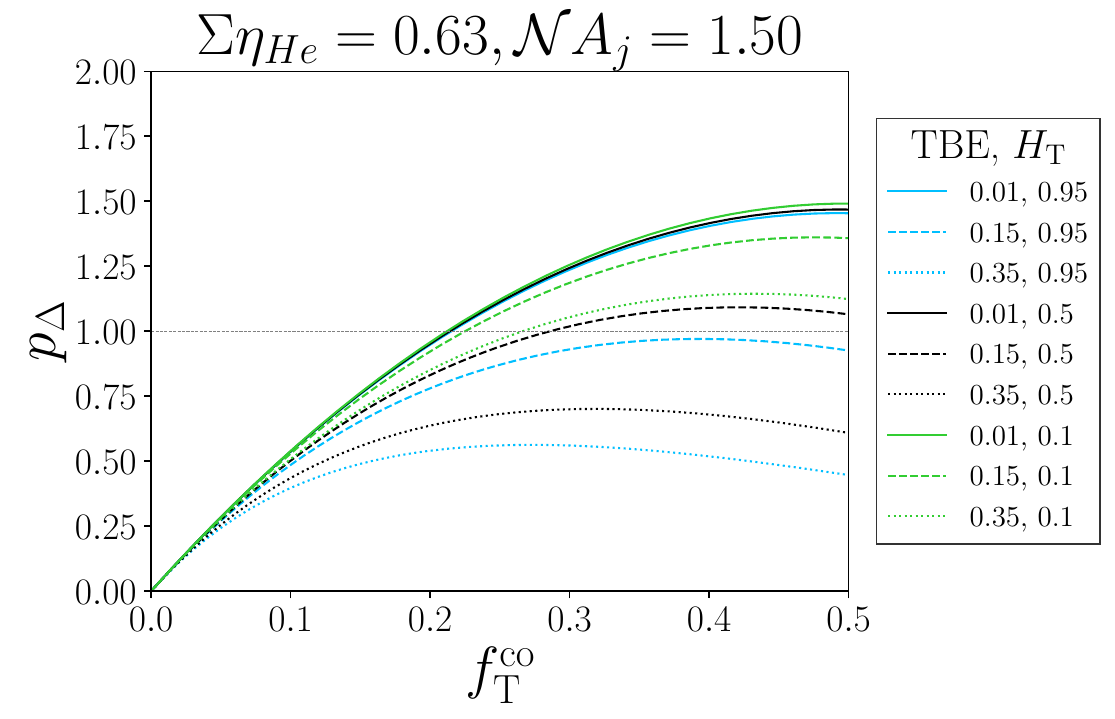}
    \caption{$p_{\Delta}$ versus $f_{\mathrm{T}}^{\mathrm{co}}$, fixed polarization multiplier $\mathcal{N} A_J = 1.5$.}
    \end{subfigure}
     ~
    \centering
    \begin{subfigure}[t]{0.47\textwidth}
    \includegraphics[width=1.0\textwidth]{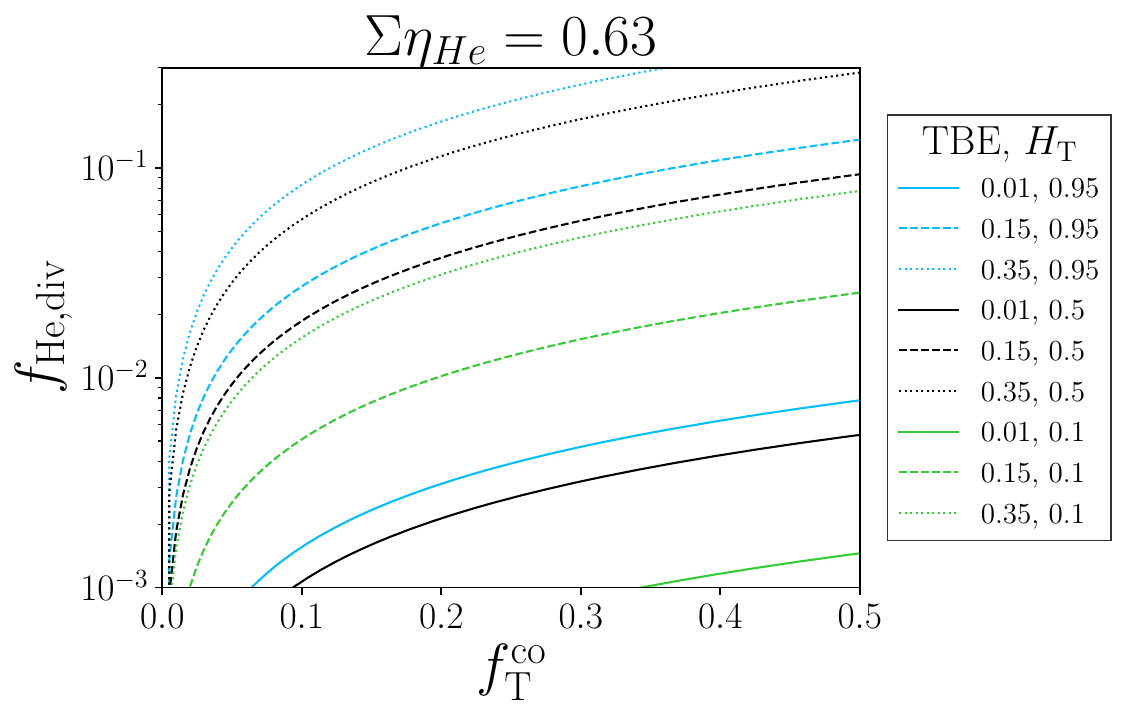}
    \caption{$f_{\mathrm{He,div}}$ versus $f_{\mathrm{T}}^{\mathrm{co}}$, fixed polarization multiplier $\mathcal{N} A_J = 1.5$.}
    \end{subfigure}
    \caption{Fusion power multiplier $p_{\Delta}$ in \Cref{eq:pDeltaform2_HT} versus core tritium fraction $f_{\mathrm{T}}^{\mathrm{co}}$ (a,c) and corresponding divertor helium-to-fuel ratio $f_{\mathrm{He,div}}$ in (b,d).}
    \label{fig:pDelta_fT}
\end{figure*}

Using a simple estimate for $p_{\Delta}$, we first show how a plasma with $F_{\mathrm{T}}^{\mathrm{in}} = 0.21$ and spin polarization $\mathcal{N} A_J = 1.5$ has the same power as an unpolarized plasma with $F_{\mathrm{T}}^{\mathrm{in}} = 0.5$. This simple estimate is obtained by neglecting helium dilution effects (setting $f_{\mathrm{dil}} = 0.0$ in \Cref{eq:pDeltaform0}). 
In \Cref{fig:powerdensity1}, we plot the power density enhancement factor $p_f / p_f (f_{\mathrm{T}}^{\mathrm{co}} = 0.5, \mathcal{N} A_J = 1.0)$ versus $f_{\mathrm{T}}^{\mathrm{co}}$ and $\mathcal{N} A_J$. The grey star indicates the unpolarized $50:50$ D-T mix $p_f (f_{\mathrm{T}}^{\mathrm{co}} = 0.5, \mathcal{N} A_J = 1.0)$ and the yellow star the EPP-enhanced $79:21$ D-T mix $p_f (f_{\mathrm{T}}^{\mathrm{co}} = 0.21, \mathcal{N} A_J = 1.5)$. \Cref{fig:powerdensity1} demonstrates that with 100\% core fuel polarization in the EPP scheme, at fixed temperature and $n_{Q,\mathrm{co}}$ the power density with $f_{\mathrm{T}}^{\mathrm{co}} = 0.21$ is equal to an unpolarized fuel with $f_{\mathrm{T}}^{\mathrm{co}} = 0.5$. That is,
\begin{equation}
\begin{aligned}
& p_f (f_{\mathrm{T}}^{\mathrm{co}} = 0.21, \mathcal{N} A_J = 1.5) \\
& = p_f (f_{\mathrm{T}}^{\mathrm{co}} = 0.5, \mathcal{N} A_J = 1.0).
\end{aligned}
\label{eq:pf_pola_runpolar}
\end{equation}
\Cref{eq:pf_pola_runpolar} means that 58\% less tritium (but 58\% more deuterium) needs to be injected in the EPP scheme to match the power density of the unpolarized scheme. Using the highest value of $\mathcal{N} A_J = 1.9$ reported in the literature \cite{Heidbrink2024} gives a 15\% tritium fraction to match the unpolarized fuel power density,
\begin{equation}
\begin{aligned}
& p_f (f_{\mathrm{T}}^{\mathrm{co}} = 0.15, \mathcal{N} A_J = 2.0) \\
& = p_f (f_{\mathrm{T}}^{\mathrm{co}} = 0.5, \mathcal{N} A_J = 1.0).
\end{aligned}
\label{eq:pf_pola_runpolar_extreme}
\end{equation}
We now proceed to study $p_{\Delta}$ including helium dilution effects.

An important quantity is the helium enrichment, the ratio of the helium-to-fuel density ratios in the divertor and the core,
\begin{equation}
\eta_{\mathrm{He}} \equiv \frac{f_{\mathrm{He},\mathrm{div}}}{f_{\alpha,\mathrm{co}}},
\label{eq:etaHe}
\end{equation}
where the helium-to-fuel density ratio in the core is
\begin{equation}
f_{\alpha,\mathrm{co}} \equiv \frac{n_{\alpha}}{n_{Q,\mathrm{co}}}. 
\end{equation}
Combining \Cref{eq:pDeltaform0,eq:etaHe} we obtain
\begin{equation}
p_{\Delta} = 4 f_{\mathrm{T}}^{\mathrm{co}} (1-f_{\mathrm{T}}^{\mathrm{co}}) \left[1 - \left(1 + \frac{\eta_{\mathrm{He}}}{ 2 f_{\mathrm{He},\mathrm{div}} }   \right)^{-1}  \right]^{2}  \mathcal{N} A_J,
\label{eq:pDeltaform1}
\end{equation}
where we used
\begin{equation}
f_{\mathrm{dil}} = \frac{f_{\mathrm{He},\mathrm{div}}/\eta_{\mathrm{He}}}{1 + 2 f_{\mathrm{He},\mathrm{div}}/\eta_{\mathrm{He}}}.
\label{eq:fdil}
\end{equation}
Rearranging the TBE in \Cref{eq:TBEform3} gives,
\begin{equation}
f_{\mathrm{He},\mathrm{div}} = \frac{F_{\mathrm{T}}^{\mathrm{div}}}{\Sigma \left( \frac{1}{\mathrm{TBE}} -1 \right) },
\label{eq:fHediv}
\end{equation}
and substituting in \Cref{eq:pDeltaform1} we find
\begin{equation}
p_{\Delta} = 4 f_{\mathrm{T}}^{\mathrm{co}} (1-f_{\mathrm{T}}^{\mathrm{co}})  \left(1-  \frac{2 F_{\mathrm{T}}^{\mathrm{div}}}{\Sigma \eta_{\mathrm{He}} \left(1 - \frac{1}{\mathrm{TBE} }  \right) } \right)^{-2}  \mathcal{N} A_J.
\label{eq:pDeltaform2}
\end{equation}
One must be careful in interpreting \Cref{eq:pDeltaform2}, since not all variables are independent. Furthermore, we prefer to work with $f_{\mathrm{T}}^{\mathrm{co}}$ than $F_{\mathrm{T}}^{\mathrm{div}}$. Therefore, we substitute $F_{\mathrm{T}}^{\mathrm{div}} = 2f_{\mathrm{T}}^{\mathrm{co}}/ (1+1/H_{\mathrm{T}})$ (see \Cref{eq:fTco_def}), giving
\begin{equation}
p_{\Delta} = \frac{ 4 f_{\mathrm{T}}^{\mathrm{co}}}{ (1-f_{\mathrm{T}}^{\mathrm{co}})  \left(1-  \frac{4 f_{\mathrm{T}}^{\mathrm{co}} / (1+1/H_{\mathrm{T}})}{\Sigma \eta_{\mathrm{He}} \left(1 - \frac{1}{\mathrm{TBE} }  \right) } \right)^{2}}  \mathcal{N} A_J,
\label{eq:pDeltaform2_HT}
\end{equation}

Plotting \Cref{eq:pDeltaform2_HT} in \Cref{fig:pDelta_fT} reveals several important trends. First, when increasing TBE at constant $\Sigma \eta_{\mathrm{He}}$ and $\mathcal{N} A_J$ values, the maximum power enhancement ${p}_{\Delta}$ always occurs when $f_{\mathrm{T}}^{\mathrm{co}} < 0.5$. This counterintuitive result is shown by the curves in \Cref{fig:pDelta_fT}(a) with different linestyles. The maximum power enhancement occurring at $f_{\mathrm{T}}^{\mathrm{co}} < 0.5$ is valid when $f_{\mathrm{He},\mathrm{div}}$ in \Cref{eq:fHediv} varies self-consistently with changes in $F_{\mathrm{T}}^{\mathrm{div}}$, $\Sigma$, TBE, $H_{\mathrm{T}}$, and $\eta_{\mathrm{He} }$.

\begin{figure}[bt!]
    \centering
    \includegraphics[width=0.99\textwidth]{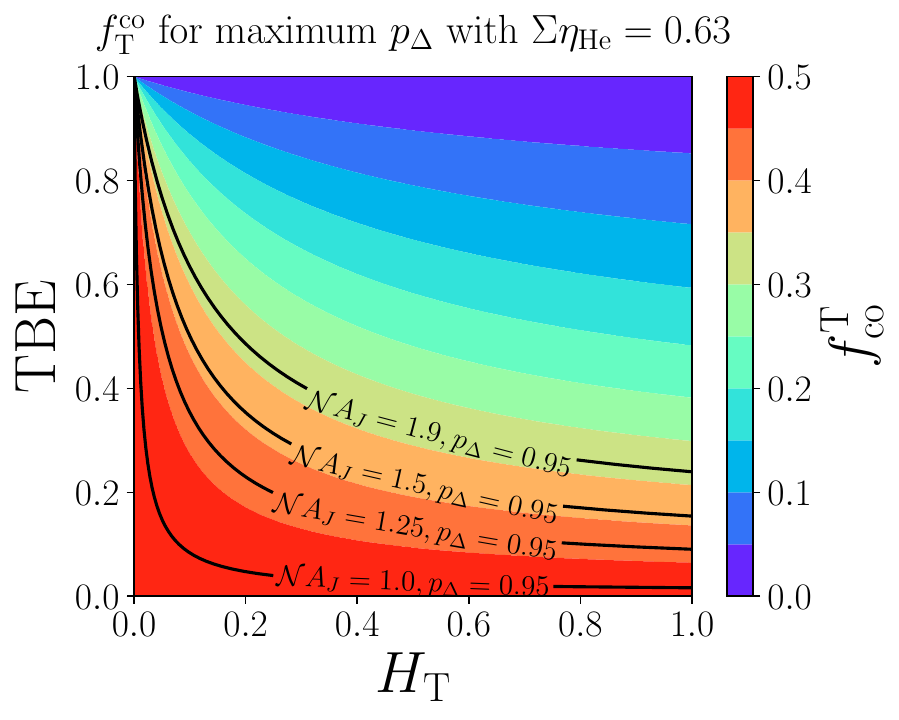}
    \caption{Tritium core fraction $f_{\mathrm{T}}^{\mathrm{co}}$ for which the power multiplier $p_{\Delta}$ is maximized plotted versus tritium burn efficiency (TBE) and tritium enrichment $H_{\mathrm{T} }$. Notably, at fixed $p_{\Delta}$, there is a one-to-one mapping of contours of constant $f_{\mathrm{T}}^{\mathrm{co}}$ and constant $\mathcal{N} A_J$.}
    \label{fig:TBE_HT_pdeltaconts}
\end{figure}

The second notable is polarizing the fuel increases TBE by an order of magnitude without a power decrease compared with the unpolarized fuel. For example, in \Cref{fig:pDelta_fT}(a), for unpolarized fuel with $f_{\mathrm{T}}^{\mathrm{co}} = 0.5$ to achieve $p_{\Delta} = 0.95$, it must have very low TBE -- for the $\Sigma \eta_{\mathrm{He}} = 0.63$ value used in \Cref{fig:pDelta_fT}(a), the TBE is $\sim$1\%. However, for a plasma with EPP fuel, $\mathcal{N} A_J = 1.5-1.9$, the TBE will be 15-20 times larger, TBE = 15-20\%. If a moderate power degradation is accepted, $p_{\Delta} = 0.8$, unpolarized fuel with $f_{\mathrm{T}}^{\mathrm{co}} = 0.5$ achieves a TBE of 7\% and EPP polarized fuel achieves a TBE of 23-32\%. Spin polarization allows for a much more favorable tradeoff between high power-density and higher TBE, especially at higher $p_{\Delta}$ values. \Cref{fig:pDelta_fT}(b) shows how $f_{\mathrm{He,div} }$ increases with increasing $f_{\mathrm{T}}^{\mathrm{co}}$ and with increasing TBE.  \Cref{fig:pDelta_fT}(b) also shows that $f_{\mathrm{He,div} }$ is independent of polarization at fixed $H_{\mathrm{T}}$ -- all curves with equal TBE lay on top of each other, so only $\mathcal{N} A_J = 1.9$ curves are visible.

\begin{figure}[!tb]
    \centering
    \begin{subfigure}[t]{0.97\textwidth}
    \includegraphics[width=1.0\textwidth]{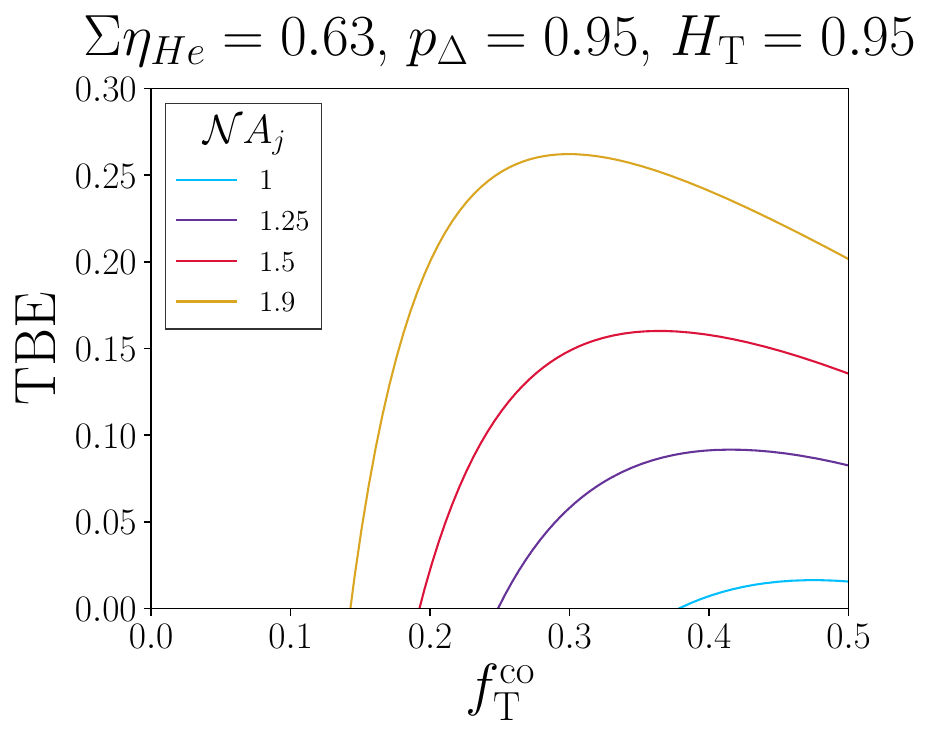}
    \caption{TBE versus $f_{\mathrm{T}}^{\mathrm{co}}$ at $\Sigma \eta_{\mathrm{He} = 0.63 }$, $p_{\Delta} = 0.95$, $H_{\mathrm{T}} = 0.95$.}
    \end{subfigure}
     ~
    \centering
    \begin{subfigure}[t]{0.97\textwidth}
    \includegraphics[width=1.0\textwidth]{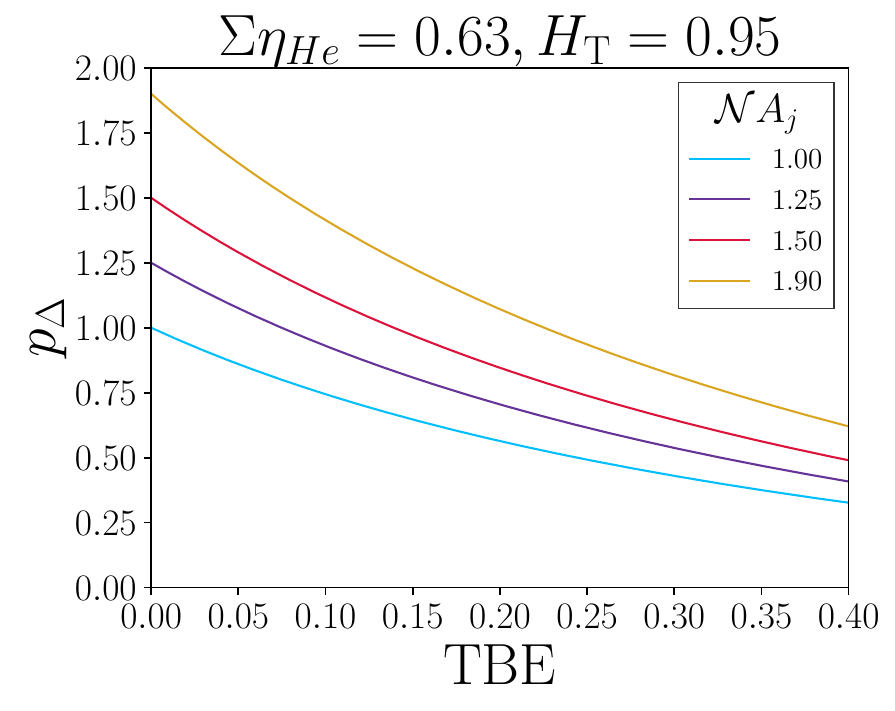}
    \caption{Maximum $p_{\Delta}$ (over all $f_{\mathrm{T}}^{\mathrm{co}}$ values) at $\Sigma \eta_{\mathrm{He} = 0.63 }$, $H_{\mathrm{T}} = 0.95$.}
    \end{subfigure}
    \caption{Using $p_{\Delta}$ in \Cref{eq:pDeltaform2_HT} to find the TBE (a) and $p_{\Delta}$ for different $\mathcal{N} A_J$ values. Minor variations in $\mathcal{N} A_J$ lead to substantial increases in the TBE.}
    \label{fig:TBEfull}
\end{figure}

The third notable is that decreasing $H_{\mathrm{T}}$ can significantly enhance the TBE at fixed power multiplier. This is shown in \Cref{fig:pDelta_fT}(c). The explanation is intuitive: at fixed power, if relatively less tritium is making it to the divertor, a larger tritium fraction must have been burned in the core. In \Cref{fig:TBE_HT_pdeltaconts}, we contour plot the $f_{\mathrm{T}}^{\mathrm{co}}$ that gives the maximum power against the TBE and $H_{\mathrm{T}}$. At fixed power, $p_{\Delta} = 0.95$, the TBE increases significantly at $H_{\mathrm{T}}$ decreases. \Cref{fig:TBE_HT_pdeltaconts} again highlights just how strongly spin polarization can enhance the TBE at fixed $p_{\Delta}$.

\begin{figure}[!tb]
    \centering
    \begin{subfigure}[t]{0.97\textwidth}
    \includegraphics[width=1.0\textwidth]{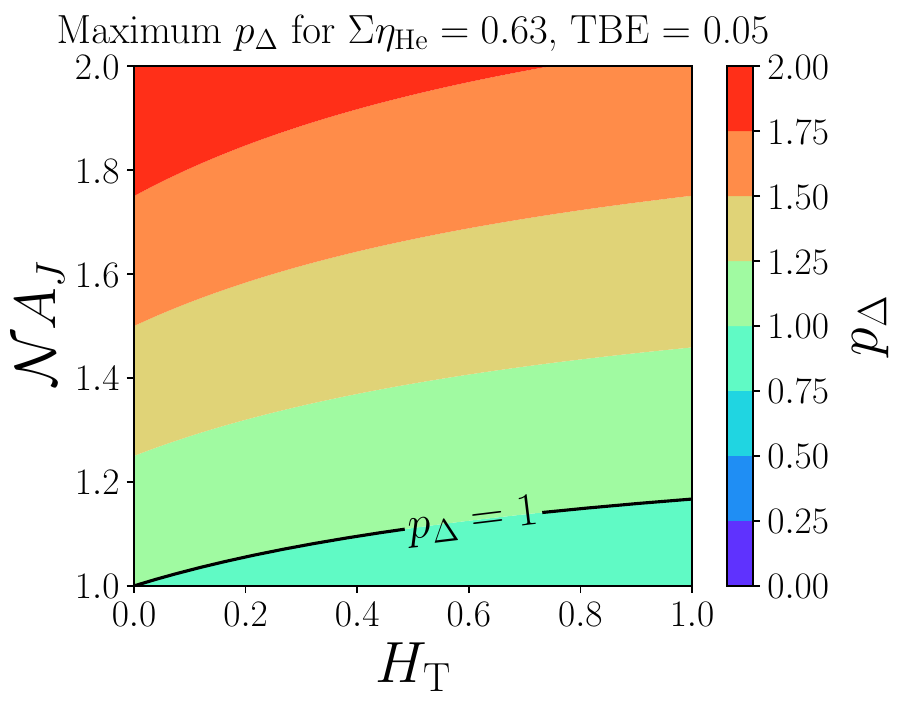}
    \caption{TBE = 0.40.}
    \end{subfigure}
     ~
    \centering
    \begin{subfigure}[t]{0.97\textwidth}
    \includegraphics[width=1.0\textwidth]{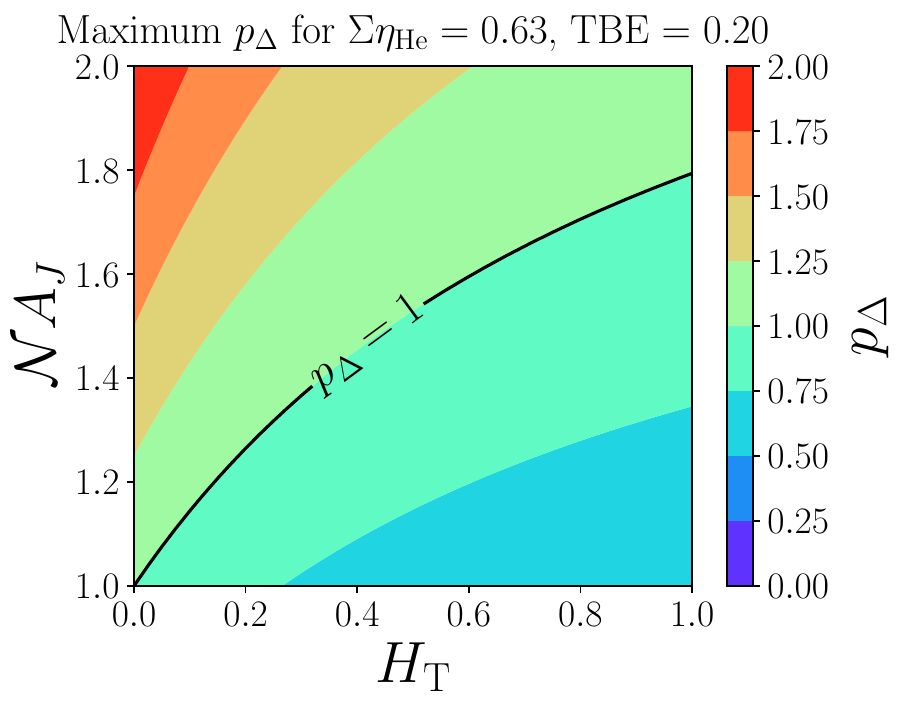}
    \caption{TBE = 0.20.}
    \end{subfigure}
    \centering
    \begin{subfigure}[t]{0.97\textwidth}
    \includegraphics[width=1.0\textwidth]{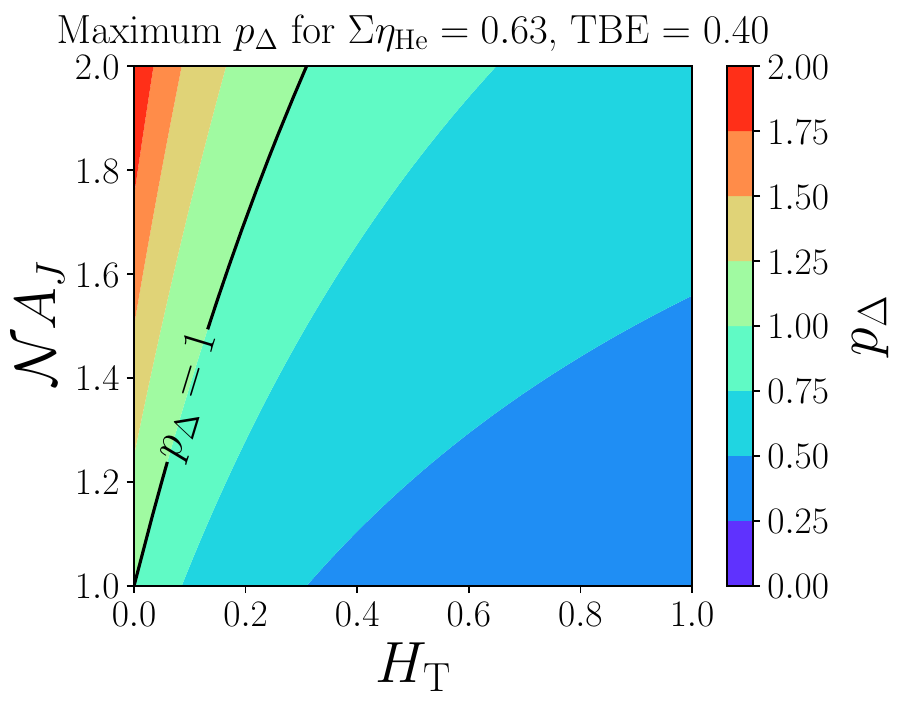}
    \caption{TBE = 0.40.}
    \end{subfigure}
    \caption{Contour plots of $p_{\Delta}$ versus $\mathcal{N} A_J$ and $H_{\mathrm{T} }$ for three TBE values: 0.05 (a), 0.20 (b), and 0.40 (c).}
    \label{fig:pDelta_ft_NAJ_enrichment}
\end{figure}

The fourth notable is that even small increases in polarization can increase the TBE significantly at fixed power multiplier. We plot solutions for the TBE in \Cref{fig:TBEfull}(a) versus $f_{\mathrm{T}}^{\mathrm{co}}$ for $\Sigma \eta_{\mathrm{He} } = 0.63$, $H_{\mathrm{T}} = 0.95$, and a small power degradation $p_{\Delta} = 0.95$. A 25\% increase in the polarization multiplier $\mathcal{N} A_J$ from 1.00 to 1.25 gives a 462\% increase in TBE from TBE =0.016 to TBE = 0.090. \Cref{fig:TBE_HT_pdeltaconts} also shows how factors of tens to hundreds of the TBE can be achieved by polarizing the fuel. This shows that unpolarized fuel gives a very poor TBE when only a minor power degradation is tolerated.

The result that pushing a plasma close to its maximum achievable power gives poor TBE was also explained in \cite{Whyte2023}. Here, we extend this idea to spin-polarized fuel with the following analogy. Suppose you are a driver in an endurance car race. Because of the tradeoff between speed and efficiency, the winning strategy involves a careful balance between speed and fuel economy. Suppose that a new combustion fuel is discovered with a 50\% higher reactivity. You could maintain the same top speed at much lower RPM, leading to big fuel savings, fewer pitstops, and less engine wear. Spin-polarized fuel shares some of these properties: at fixed fusion power (car speed) the tritium burn efficiency (fuel economy) increases substantially, leading to a lower tritium startup inventory (initial fuel) and operating at a fusion power that is comfortably within the device's capabilities (manageable RPM).

By choosing to operate a power plant with unpolarized fuel close to $p_{\Delta} = 1.0$, we are forcing an unacceptably low TBE. However, it might also economically very undesirable to operate at $p_{\Delta} \lesssim 0.8$ where the TBE is higher. Spin-polarized fuels offer a way out of this conundrum: the higher cross section allows one to operate at lower tritium fraction and enhance the TBE, while maintaining the power density as unpolarized fuels with a 50:50 D-T mix. This is summarized in \Cref{fig:TBE_HT_pdeltaconts}, where contours of constant $f_{\mathrm{T}}^{\mathrm{co}}$ correspond to contours of constant $\mathcal{N} A_J$ at fixed $p_{\Delta}$. This is the major new insight of this paper.

The very strong sensitivity of the TBE on small increases in polarization for $p_{\Delta}$ close to 1 results from a resonant denominator in the TBE of the form $\sqrt{4 f_{\mathrm{T}}^{\mathrm{co}} (1-f_{\mathrm{T}}^{\mathrm{co}}) \mathcal{N} A_J } - \sqrt{p_{\Delta}} $. In \Cref{sec:intuition}, the effect of $\mathcal{N} A_J$ on TBE is discussed in more detail. For now, it suffices to know that at fixed power and for $\Sigma \eta_{\mathrm{He} } \sim 1$, the TBE scales as
\begin{equation}
\mathrm{TBE} \sim \frac{1}{f_{\mathrm{T}}^{\mathrm{co}}}  \left( 1 + \frac{1}{ \sqrt{\frac{4 f_{\mathrm{T}}^{\mathrm{co}} (1-f_{\mathrm{T}}^{\mathrm{co}}) \mathcal{N} A_J }{p_{\Delta}}}  -1}  \right) ^{-1},
\label{eq:TBEscaling}
\end{equation}
where the exact form is given in \Cref{eq:TBEfull}. Because the terms in \Cref{eq:TBEscaling} are typically of order unity, further simplification is challenging. This phenomenon can be seen by following lines of constant $p_{\Delta}$ in \Cref{fig:TBEfull}(b). For example, for $p_{\Delta} = 1.0$, the TBE increases from TBE = 0.00 at $\mathcal{N} A_J = 1.0$ to TBE $\simeq 0.075$ at $\mathcal{N} A_J = 1.25$ and to TBE $\simeq 0.15$ for $\mathcal{N} A_J = 1.50$. The increase in TBE at $p_{\Delta} \approx 1$ (and in-fact, most $p_{\Delta}$ values) with spin-polarized fuel is extremely large.

However, even spin-polarizing the fuel can only increase the TBE so much. The tritium enrichment $H_{\mathrm{T} }$ also plays a big role, particularly at higher TBE. Other important parameters, $\Sigma$ and $\eta_{\mathrm{He}}$ have been identified in \cite{Whyte2023}, so we will not focus on them here. Shown in \Cref{fig:pDelta_ft_NAJ_enrichment}, we plot the maximum achievable $p_{\Delta}$ value (that is, finding the $f_{\mathrm{T} }^{\mathrm{co}}$ that maximizes $p_{\Delta}$) across $\mathcal{N} A_J$ and $H_{\mathrm{T} }$ values. Each subfigure corresponds to a different TBE value. In \Cref{fig:pDelta_ft_NAJ_enrichment}(a), which has TBE = 0.05, the $\mathcal{N} A_J$ value is far more important for increasing $p_{\Delta}$ than $H_{\mathrm{T}}$. However, at higher TBE values--- TBE = 0.20, 0.40 in \Cref{fig:pDelta_ft_NAJ_enrichment}(b) and (c) -- decreasing $H_{\mathrm{T} }$ can be much more important than increasing the spin polarization.

\begin{figure}[bt!]
    \centering
    \includegraphics[width=0.9\textwidth]{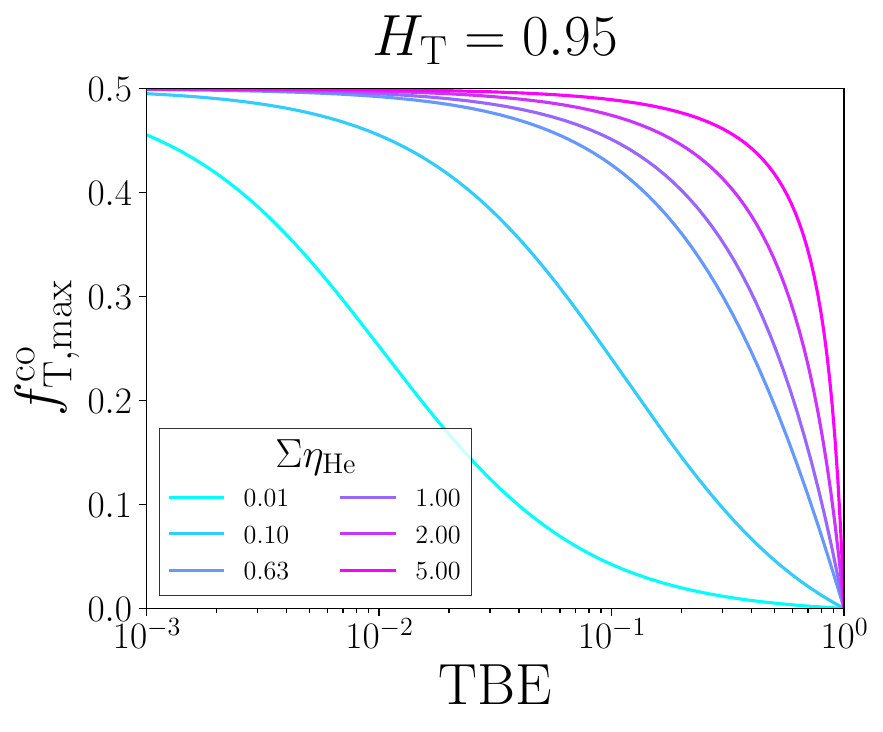}
    \caption{Tritium fraction for which the fusion power density is maximized $f_{\mathrm{T},\mathrm{max}}^{\mathrm{co}}$ versus TBE (see \Cref{eq:pDeltamaxfT}). We assumed $H_{\mathrm{T} } = 0.95$.}
    \label{fig:ftcomax}
\end{figure}

To gain more intuition for the effect of tritium fraction, we differentiate \Cref{eq:pDeltaform2_HT} to find $\partial p_{\Delta} / \partial f_{\mathrm{T}}^{\mathrm{co}} = 0$ at fixed $\Sigma, \eta_{\mathrm{He}}, \mathrm{TBE}, \mathcal{N} A_J, H_{\mathrm{T}},$ (but varying $f_{\mathrm{He},\mathrm{div}}$) in order to find the $f_{\mathrm{T}}^{\mathrm{co}}$ for which $p_{\Delta}$ is maximized,
\begin{equation}
f_{\mathrm{T},\mathrm{max}}^{\mathrm{co}} =\frac{1}{2} \frac{1}{1  + \frac{2 H_{\mathrm{T} }}{ 1 + H_{\mathrm{T} }}   \frac{ \mathrm{TBE}}{\Sigma \eta_{\mathrm{He}}(1-\mathrm{TBE})} }.
\label{eq:pDeltamaxfT}
\end{equation}
In \Cref{fig:ftcomax}, we plot $f_{\mathrm{T},\mathrm{max}}^{\mathrm{co}}$ versus the TBE for a range of $\Sigma \eta_{\mathrm{He}}$ values. Only for $\Sigma \eta_{\mathrm{He}} > 1$ is $f_{\mathrm{T},\mathrm{max}}^{\mathrm{co}} \simeq 1/2$ for a wide range of TBE values. Therefore, regardless of fuel polarization, if $\Sigma \eta_{\mathrm{He}}$ is not sufficiently large, the maximum fusion power for moderate-high TBE will be maximized for $f_{\mathrm{T}}^{\mathrm{co}}$ significantly less than 1/2. 

We again emphasize that the curves in \Cref{fig:ftcomax} are obtained by holding $\Sigma, \eta_{\mathrm{He}}, \mathcal{N} A_J, H_{\mathrm{T}}$ fixed, and allowing TBE and $f_{\mathrm{He},\mathrm{div}}$ to vary -- the effect on $f_{\mathrm{He},\mathrm{div}}$ in shown in \Cref{fig:pDelta_fT}(d), where $f_{\mathrm{He},\mathrm{div}}$ increases monotonically with $f_{\mathrm{T}}^{\mathrm{co}}$, although $p_{\Delta}$, shown in \Cref{fig:pDelta_fT}(c), does not necessarily increase monotonically with $f_{\mathrm{T}}^{\mathrm{co}}$.

The physical interpretation of $f_{\mathrm{T},\mathrm{max}}^{\mathrm{co}} < 1/2$ is that while a tritium fraction close to $1/2$ is beneficial for maximizing the power density neglecting helium dilution effects, its benefits can be outweighed once helium dilution effects that decrease the power density are included, particularly at high TBE when helium dilution tends to be higher. As the TBE increases, so do $f_{\mathrm{He},\mathrm{div}}$ and $f_{\mathrm{dil}}$, meaning that helium ash dilutes the plasma (see \Cref{eq:fdil,eq:fHediv}). According to \Cref{eq:pfform2}, $p_{\Delta} \propto f_{\mathrm{T}}^{\mathrm{co}} (1-f_{\mathrm{T}}^{\mathrm{co}}) (1-2 f_{\mathrm{dil}})^2$ (neglecting spin polarization). Therefore, the prefactor $f_{\mathrm{T}}^{\mathrm{co}} (1-f_{\mathrm{T}}^{\mathrm{co}})(1-2 f_{\mathrm{dil}})^2$, might not have a maximum at $f_{\mathrm{T}}^{\mathrm{co}} = 1/2$ at moderate to high TBE because helium dilution effects can make the $(1-2 f_{\mathrm{dil}})^2$ term very small for $f_{\mathrm{T}}^{\mathrm{co}} = 1/2$. Formally, only for TBE = 0 is $f_{\mathrm{T},\mathrm{max}}^{\mathrm{co}} = 1/2$.

We comment briefly on the relation between the total fusion power $P_f$ and the power density $p_f$. If one assumes that $f_{\mathrm{T}}^{\mathrm{co}}$ and $\mathcal{N} A_J$ are radially constant, the enhancement (or degradation) in the power density $p_{\Delta}$ is equal to the total fusion power enhancement (or degradation): $P_{\Delta} \equiv \frac{P_f}{P_{f,\mathrm{nom}}}$, where $P_{f,\mathrm{nom}}$ is $P_f$ with $f_{\mathrm{T}}^{\mathrm{co}} =1/2$ and $\mathcal{N} A_J = 1$. One can assume constant $f_{\mathrm{T}}^{\mathrm{co}}$ if $H_{\mathrm{T} }=1$. Shown in \Cref{fig:TBE1}(c), $H_{\mathrm{T} }=1$ is a decent approximation as long as $F_{\mathrm{T}}^{\mathrm{in}}$ is not too close to 0 or 1, and $f_{\mathrm{He},\mathrm{div}} \Sigma$ is not too large. Making the additional assumption that $\mathcal{N} A_J$ is radially constant means that the total power enhancement is equal to the power density enhancement in the core $P_{\Delta} = p_{\Delta}$.

\begin{figure}[!tb]
    \centering
    \begin{subfigure}[t]{0.97\textwidth}
    \includegraphics[width=1.0\textwidth]{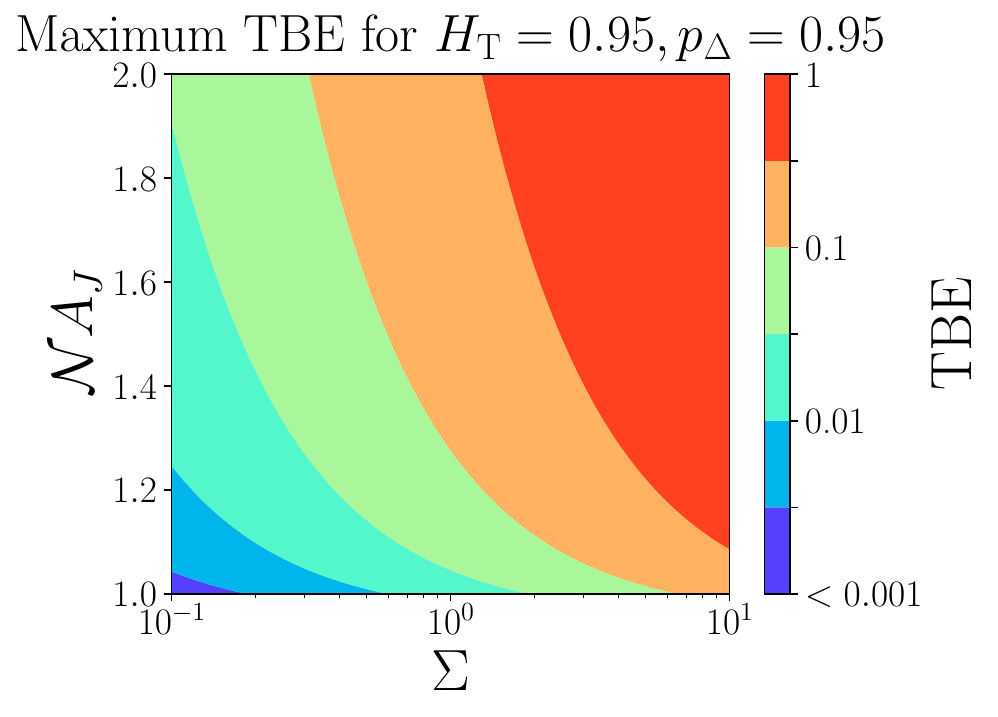}
    \caption{$H_{\mathrm{T}}=0.95$.}
    \end{subfigure}
     ~
    \centering
    \begin{subfigure}[t]{0.97\textwidth}
    \includegraphics[width=1.0\textwidth]{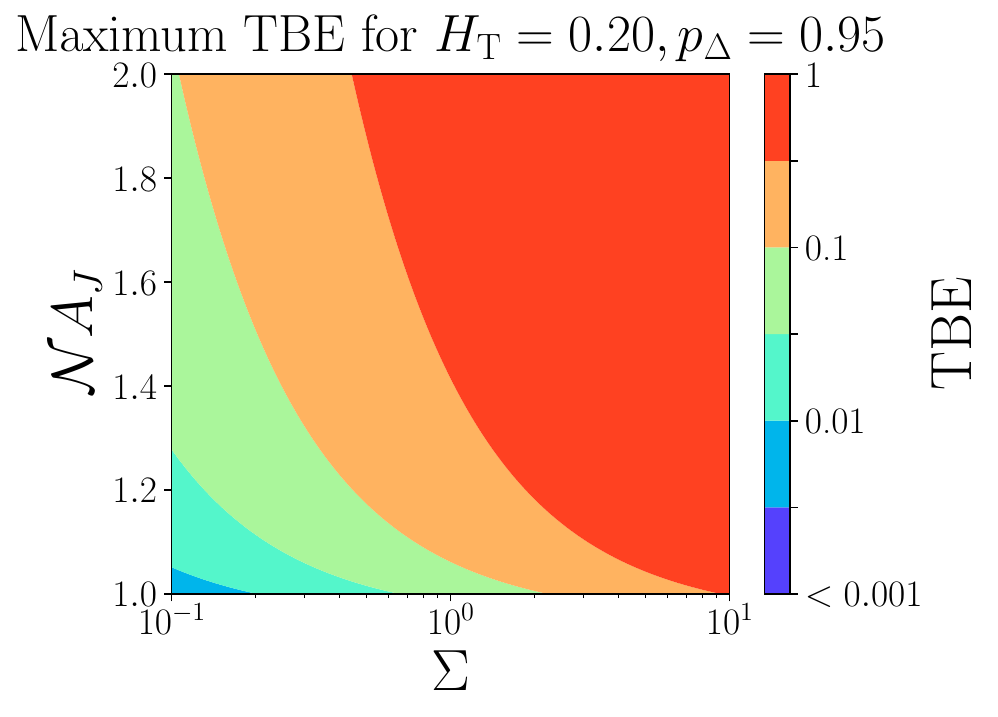}
    \caption{$H_{\mathrm{T}}=0.2$.}
    \end{subfigure}
    \caption{Maximum tritium burn efficiency (TBE) for a minimum power multiplier $p_{\Delta} > 0.95$ versus spin-polarization multiplier $\mathcal{N} A_J$ and relative helium pumping speed $\Sigma$. Each subfigure corresponds to a different tritium enrichment value, $H_{\mathcal{T}} = 0.95, 0.20$. The helium enrichment is $\eta_{\mathrm{He}} = 1.00$.}
    \label{fig:maxTBE_NAJ_Sigma}
\end{figure}

\begin{figure*}[t]
    \centering
    \begin{subfigure}[t]{0.42\textwidth}
    \centering
    \includegraphics[width=1.0\textwidth]{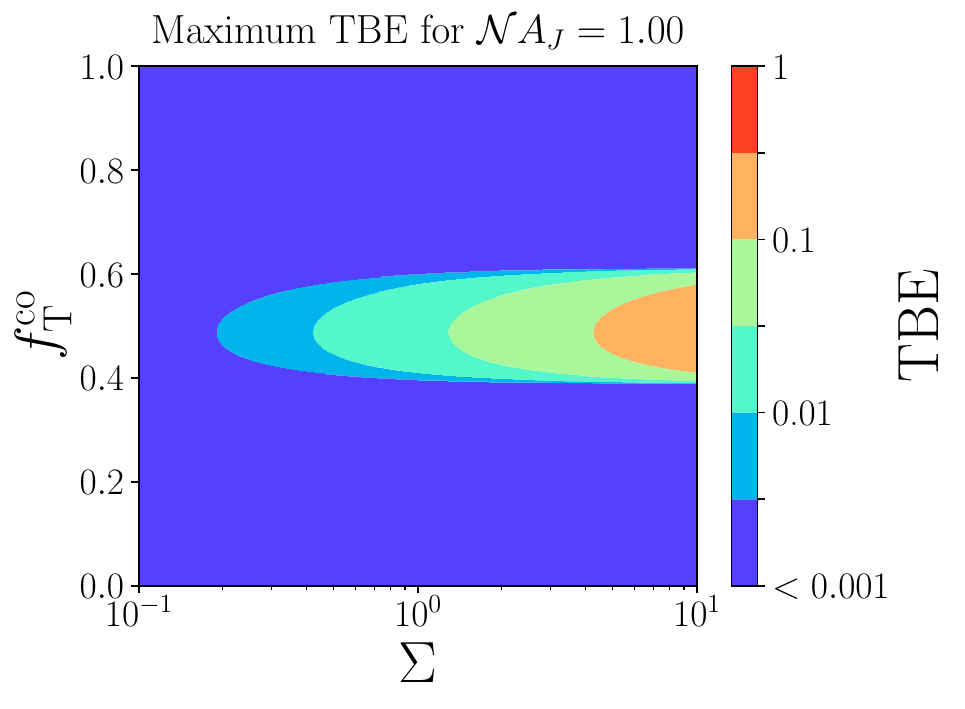}
    \caption{$\mathcal{N} A_J = 1.00$.}
    \end{subfigure}
     ~
    \begin{subfigure}[t]{0.42\textwidth}
    \centering
    \includegraphics[width=1.0\textwidth]{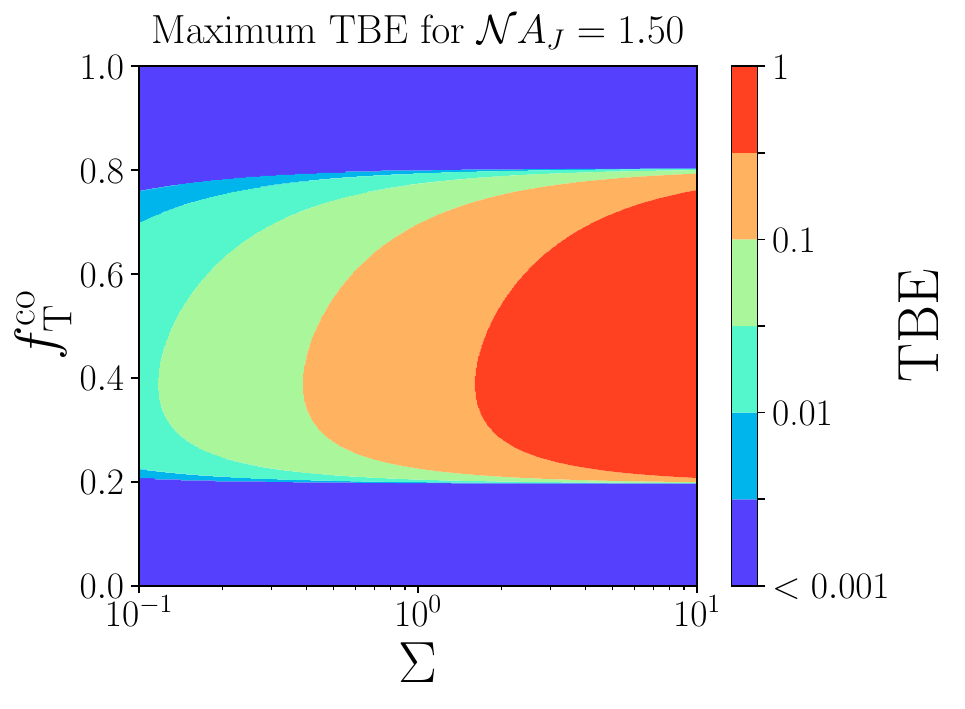}
    \caption{$\mathcal{N} A_J = 1.25$.}
    \end{subfigure}
     ~
    \begin{subfigure}[t]{0.42\textwidth}
    \centering
    \includegraphics[width=1.0\textwidth]{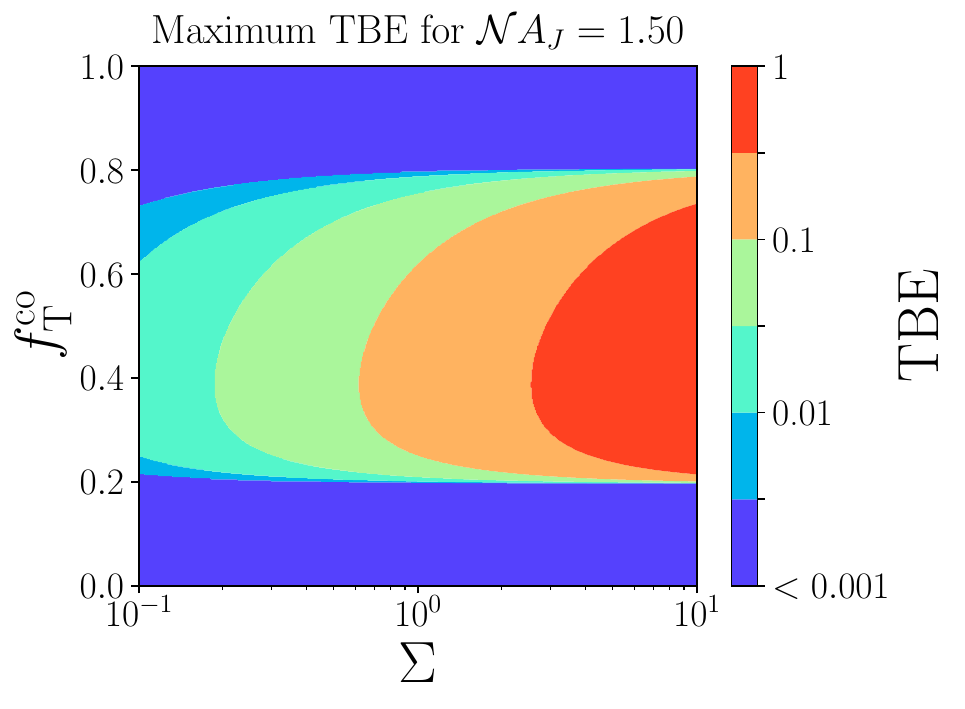}
    \caption{$\mathcal{N} A_J = 1.50$.}
    \end{subfigure}
     ~
    \centering
    \begin{subfigure}[t]{0.42\textwidth}
    \centering
    \includegraphics[width=1.0\textwidth]{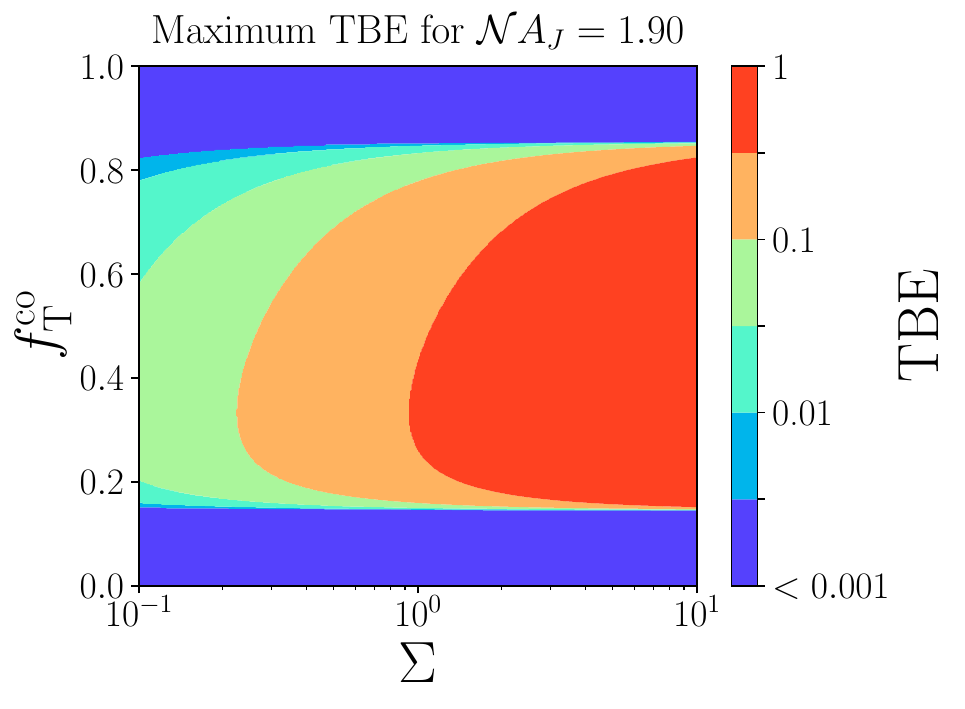}
    \caption{$\mathcal{N} A_J = 1.90$.}
    \end{subfigure}
    \caption{Maximum tritium burn efficiency (TBE) for a minimum power multiplier $p_{\Delta} > 0.95$ versus core tritium fraction $f_{\mathrm{T} }^{\mathrm{co} }$ and relative helium pumping speed $\Sigma$. Each subfigure corresponds to a different D-T polarization cross-section multiplier value, $\mathcal{N} A_J = 1.00, 1.25, 1.50, 1.90$. The helium enrichment is $\eta_{\mathrm{He}} = 1.00$. Because $f_{\mathrm{T} }^{co} \in [0.0, 1.0]$, it is assumed that $H_{\mathrm{T} } = 1.0$.}
    \label{fig:TBE3p6}
\end{figure*}

\begin{figure}[bt!]
    \centering
    \includegraphics[width=0.9\textwidth]{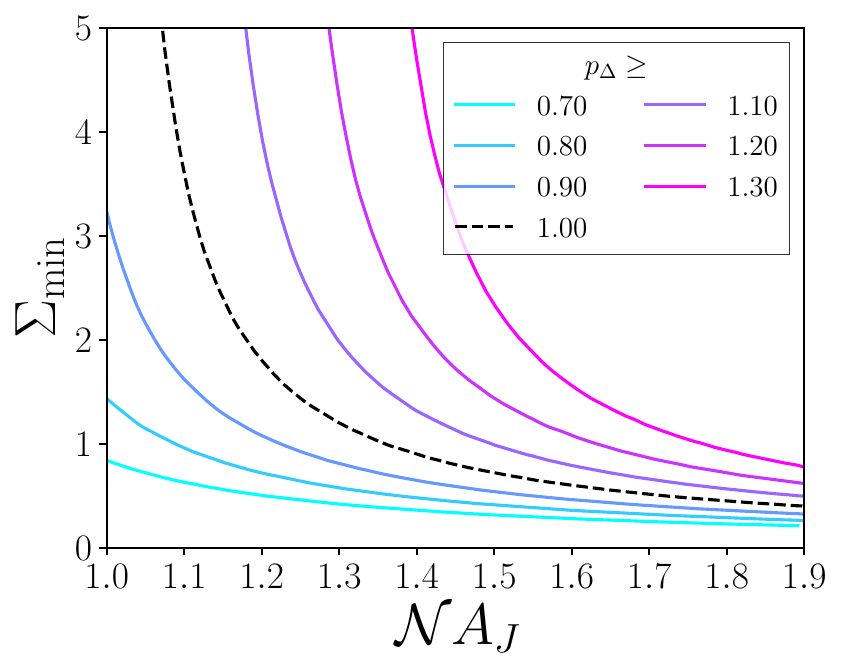}
    \caption{The minimum $\Sigma$ value required to achieve TBE = 0.10 for a range of $\mathcal{N} A_J$ values. Each curve represents a different minimum power degradation value $p_{\Delta}$. We assume $H_{\mathrm{T}} = \eta_{\mathrm{He}} = 1.0$}
    \label{fig:Sigma_Min}
\end{figure}

$\Sigma$, the ratio of helium to unburned hydrogen pumping speeds (see \Cref{eq:ashpumpspeed}) has been identified an important area for technological development, given that higher $\Sigma$ values can simultaneously enhance the TBE and fusion power \cite{Whyte2023}. We plot the maximum achievable TBE at a given $p_{\Delta}$ for two different tritium enrichments in \Cref{fig:maxTBE_NAJ_Sigma}: lower tritium enrichment and higher spin polarization $\mathcal{N} A_J$ increases the space of large TBE values significantly. The effect of spin polarization is also illustrated by comparing \Cref{fig:TBE3p6}(a)-(d) where the maximum achievable TBE for $\mathcal{N} A_J = 1.00, 1.25, 1.50,$ and $1.90$ is plotted for $p_{\Delta} \geq 0.95$. Optimistically, lack of progress in improving $\Sigma$ could be compensated for by SP fuels and by lower tritium enrichment.

Comparing \Cref{fig:TBE3p6}(a)-(d) shows that SP fuels have a maximum TBE at lower $f_{\mathrm{T}}^{\mathrm{co}}$ and lower $\Sigma$ than unpolarized fuels at fixed fusion power density, $p_{\Delta} = 0.95$. This demonstrates how the increased cross section given by SP fuels can be used to trade fusion power for tritium burn efficiency. For example, for unpolarized fuel in \Cref{fig:TBE3p6}(a), TBE = 0.10 can be achieved with $\Sigma \simeq 4-5$ and $f_{\mathrm{T}}^{\mathrm{co}} \simeq 0.49$. However, if the polarization is increased to $\mathcal{N} A_J = 1.5$, as in \Cref{fig:TBE3p6}(c), TBE = 0.10 can be achieved with $\Sigma \simeq 0.5-0.6$ and $f_{\mathrm{T}}^{\mathrm{co}} \simeq 0.37$. Because achieving $\mathcal{N} A_J = 1.5$ will likely be challenging -- this requires all of the D-T fuel undergoing fusion reactions to be spin polarized, but neglects benefits the nonlinear power enhancement -- we also plot the TBE for an intermediate polarization, $\mathcal{N} A_J = 1.25$ in \Cref{fig:TBE3p6}(d). This value of $A_J = 1.25$ is comparable to the upcoming D-He3 spin-polarization experiments in DIII-D \cite{Baylor2023,Garcia2023,Heidbrink2024}, but it is not known what $\mathcal{N}$ would correspond to in an equivalent D-T experiment. For $\mathcal{N} A_J = 1.25$, TBE = 0.10 can be achieved with $\Sigma \simeq 0.40$ and $f_{\mathrm{T}}^{\mathrm{co}} \simeq 0.4$. Finally, for the optimistic spin-polarization case, $\mathcal{N} A_J = 1.9$, a value of TBE = 0.1 can be achieved with $\Sigma \simeq 0.12$. Therefore, the required $\Sigma$ for a given TBE (assuming $f_{\mathrm{T}}^{\mathrm{co}}$ can be varied) is highly non-linear in $\mathcal{N} A_J$.

In \Cref{fig:Sigma_Min} we plot the minimum required $\Sigma$ for TBE = 0.1, $\Sigma_{\mathrm{min} }$, across different polarizations $\mathcal{N} A_J$ for seven minimum $p_{\Delta}$ values and $\eta_{\mathrm{He}} = 0.63$. \Cref{fig:Sigma_Min} demonstrates just how beneficial very high values of $\mathcal{N} A_J$ are, given how closely spaced the seven curves for different $p_{\Delta}$ values are at $\mathcal{N} A_J = 1.9$. Conversely, for lower $p_{\Delta}$ values, $\mathcal{N} A_J$ is not as useful for reducing $\Sigma_{\mathrm{min} }$ at fixed TBE. For example, for $p_{\Delta} = 0.7$, the $\Sigma_{\mathrm{min} }$ value is roughly only double for $\mathcal{N} A_J = 1.0$ than for $\mathcal{N} A_J = 1.5$. If one is willing to accept a relatively severe power degradation of $p_{\Delta} = 0.7$ but can achieve polarized fuel with $\mathcal{N} A_J = 1.5$, the increase in $\Sigma_{\mathrm{min} }$ to increase the TBE would be relatively small, but would result in a large payoff in tritium burn efficiency. Fully exploring the parameter space of TBE, $\Sigma$, $\eta_{\mathrm{He}}$, $\mathcal{N} A_J$, $f_{\mathrm{T}}^{\mathrm{co}}$, $p_{\Delta}$, and $H_{\mathrm{T} }$, and  is beyond the scope of this initial work, but our results suggest that increasing $\mathcal{N} A_J$ using spin polarization combined with variable tritium fraction has surprisingly large and nonlinear benefits for the TBE,  $p_{\Delta}$, and $\Sigma_{\mathrm{min} }$.

\subsection{Fusion Gain} \label{sec:fusion_gain}

\begin{figure}[t]
    \centering
    \begin{subfigure}[t]{0.88\textwidth}
    \centering
    \includegraphics[width=1.0\textwidth]{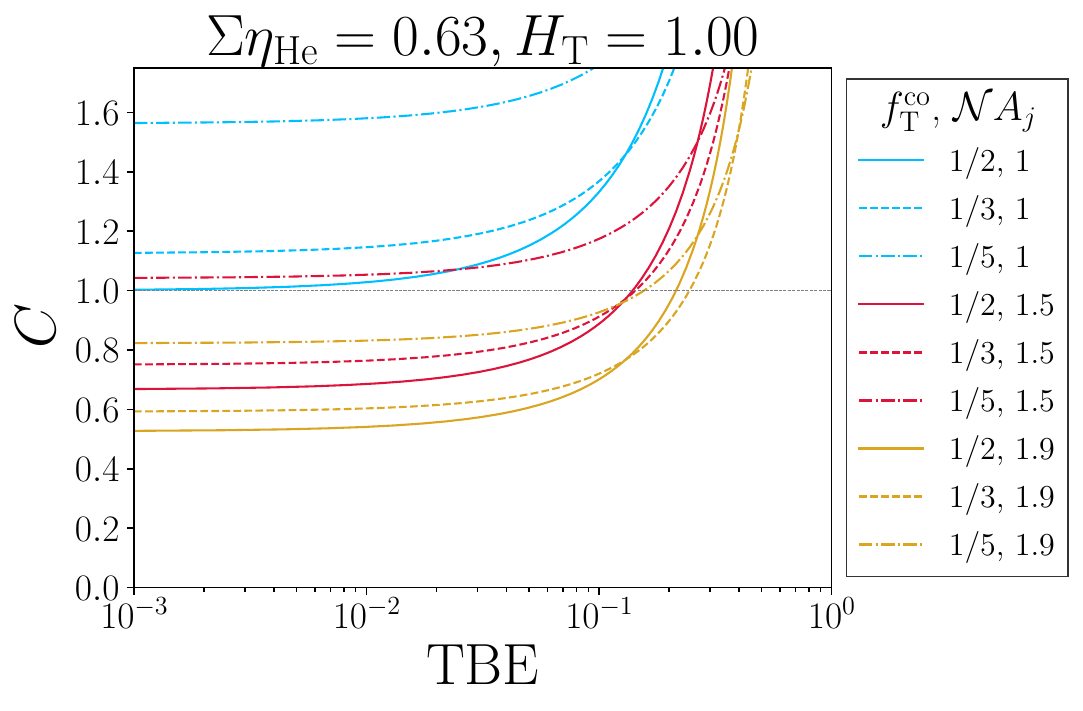}
    \caption{}
    \end{subfigure}
     ~
    \centering
    \begin{subfigure}[t]{0.88\textwidth}
    \centering
    \includegraphics[width=1.0\textwidth]{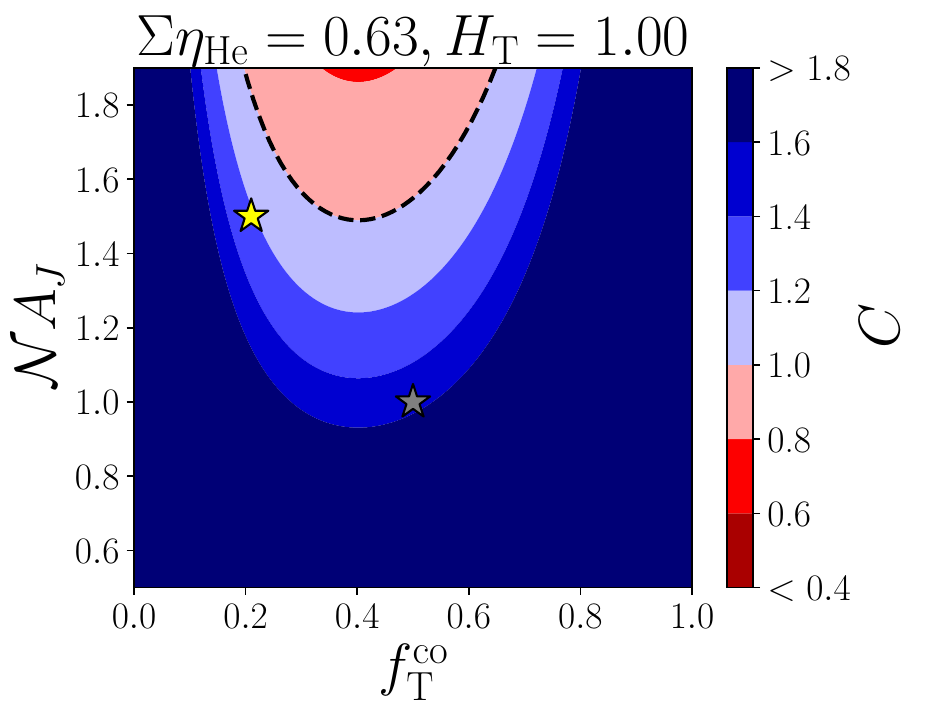}
    \caption{TBE $=0.15$.}
    \end{subfigure}
     ~
    \begin{subfigure}[t]{0.88\textwidth}
    \centering
    \includegraphics[width=1.0\textwidth]{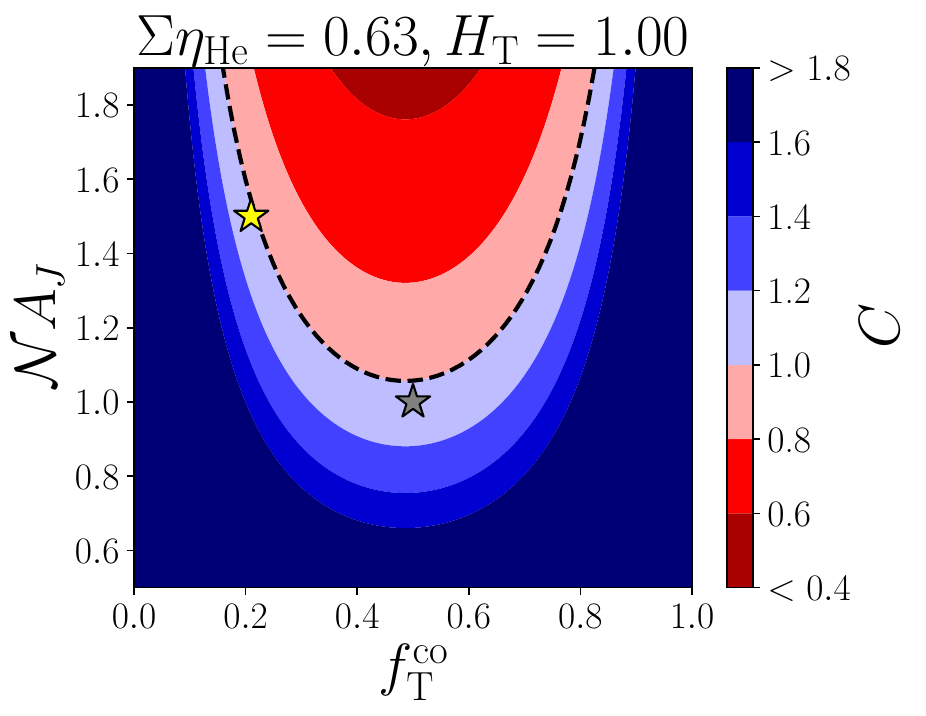}
    \caption{TBE $=0.02$.}
    \end{subfigure}
    \caption{Required $n_e \tau_E$ multiplication factor $C$ (\Cref{eq:C}) versus TBE (a), and $\mathcal{N} A_J$ and $f_{\mathrm{T}}^{\mathrm{co}}$ ((b) and (c)) to maintain a fixed plasma gain. Smaller values indicate higher performance. Dashed contour in (b) and (c) is $C = 1$ and we use $H_{\mathrm{T}} = \Sigma \eta_{\mathrm{He}} = 1.0$ for all figures.}
    \label{fig:TBE4}
\end{figure}

In this section, we demonstrate how for a fixed plasma gain $Q$, spin polarization can give a TBE tens or even hundreds of times larger than that of unpolarized fuel. For simplicity, throughout this section we assume a tritium enrichment of $H_{\mathrm{T} } = 1.0$. According to results in previous section, using $H_{\mathrm{T} } = 1.0$ is also the most conservative value, which will tend to underestimate the fusion power and TBE.

\begin{figure}[!tb]
    \centering
    \begin{subfigure}[t]{0.93\textwidth}
    \centering
    \includegraphics[width=1.0\textwidth]{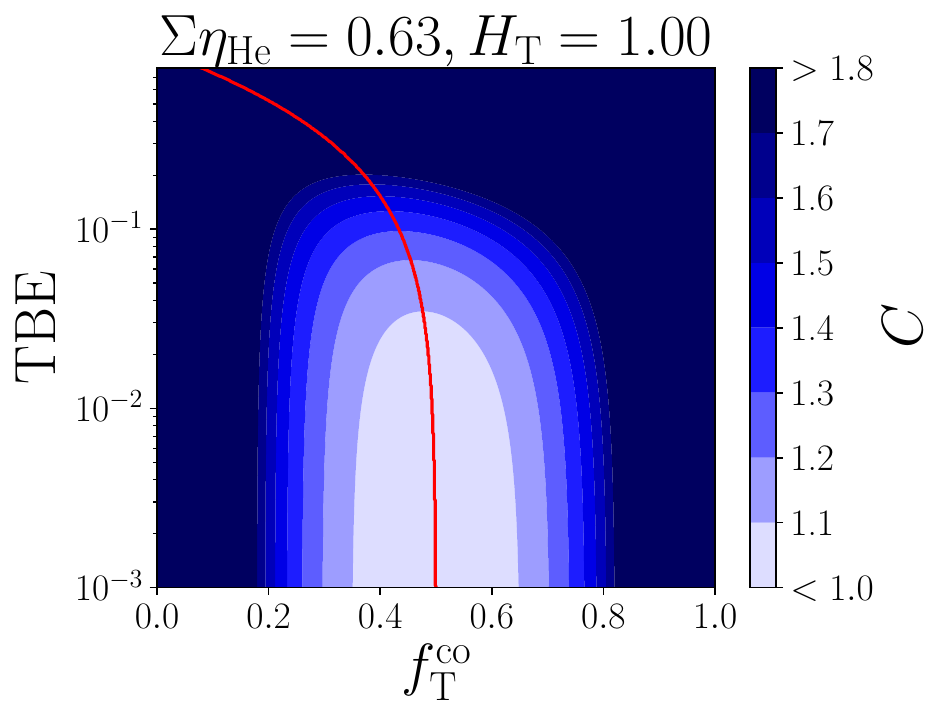}
    \caption{$\mathcal{N} A_J = 1.0$.}
    \end{subfigure}
     ~
    \centering
    \begin{subfigure}[t]{0.93\textwidth}
    \centering
    \includegraphics[width=1.0\textwidth]{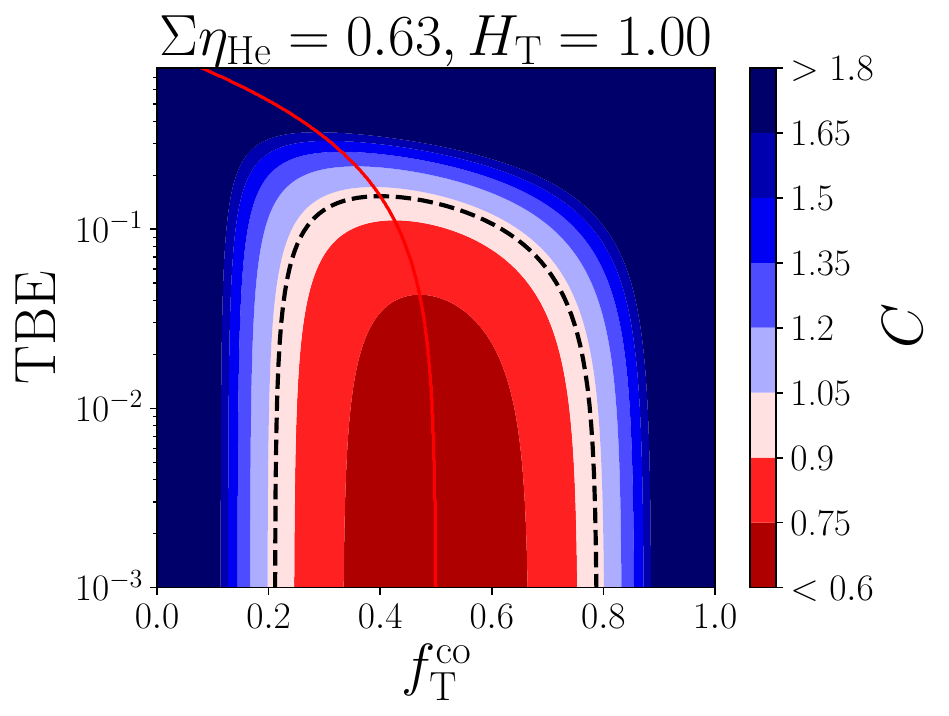}
    \caption{$\mathcal{N} A_J = 1.5$}
    \end{subfigure}
    \caption{Required $n_e \tau_E$ multiplication factor $C$ (\Cref{eq:C}) versus TBE and $f_{\mathrm{T}}^{\mathrm{co}}$to maintain a fixed plasma gain, for unpolarized fuel (a) and polarized fuel with $\mathcal{N} A_J = 1.5$ (b). The maximum fusion gain occurs for $f_{\mathrm{T}}^{\mathrm{co}} < 0.5$ as the TBE increases, shown by the red curve that gives the maximum $C$ value at any given TBE. The dashed blue curve (absent in (a), but appearing in (b)) is a contour of $C = 1.0$. We use $\Sigma \eta_{\mathrm{He}} = 0.63$, $H_{\mathrm{T}} = 1.0$ in both figures.}
    \label{fig:TBE5}
\end{figure}

Power balance in the core approximately requires
\begin{equation}
p_{\alpha} \left( 1 + 5 / Q \right)  = \frac{w_{\mathrm{th} }}{\tau_E},
\label{eq:corepower}
\end{equation}
where $p_{\alpha}$ is the alpha heating power density, $Q = p_f / p_{\mathrm{heat} }$ is the local plasma fusion gain (on a flux surface), $p_{\mathrm{heat} }$ is the heating power density absorbed by the plasma, $\tau_E$ is the energy confinement time and $w_{\mathrm{th} }$ is the stored thermal energy density. Also expressing the core thermal energy density as
\begin{equation}
w_{\mathrm{th} } = \frac{3}{2} \left(n_e + n_{Q,\mathrm{co}} \right) k_B T + \frac{3}{2} n_{\alpha} k_B \langle T_{\alpha} \rangle,
\end{equation}
where the alpha particle pressure is
\begin{equation}
w_{\mathrm{th} , \alpha} = \frac{3}{2} n_{\alpha} k_B \langle T_{\alpha} \rangle,
\label{eq:w1}
\end{equation}
and the average temperature $\langle T_{\alpha} \rangle$ accounts for the range of alpha particle energies. Here, we assume that the electron and ion temperatures have an equal value $T$. Using quasineutrality, $n_{Q, \mathrm{co} } = n_e - 2n_{\alpha}$, \Cref{eq:w1} is
\begin{equation}
w_{\mathrm{th} } = 3 k_B T D,
\end{equation}
where the dilution factor is
\begin{equation}
D \equiv \left( 1 + f_{\mathrm{dil}} \left( \frac{1}{2} \frac{\langle T_{\alpha} \rangle}{T} -1 \right)  \right),
\end{equation}
which is equal to 1 when there are no helium dilution effects. We obtain
\begin{equation}
\begin{aligned}
& n_e \tau_E (1 + 5/Q) \\ & = \frac{15 k_B T}{\langle v \overline{\sigma} \rangle E /4} \frac{\left( 1 + f_{\mathrm{dil}} \left( \frac{1}{2} \frac{\langle T_{\alpha} \rangle}{T} -1 \right)  \right)}{4 f_{\mathrm{T}}^{\mathrm{co}} (1-f_{\mathrm{T}}^{\mathrm{co}}) (1-2f_{\mathrm{dil}})^2 \mathcal{N} A_J}.
\label{eq:fusiongainform1}
\end{aligned}
\end{equation}
The second fraction on the RHS of \Cref{eq:fusiongainform1},
\begin{equation}
C \equiv \frac{ 1 + f_{\mathrm{dil}} \left( \frac{1}{2} \frac{\langle T_{\alpha} \rangle}{T} -1 \right)  }{4 f_{\mathrm{T}}^{\mathrm{co}} (1-f_{\mathrm{T}}^{\mathrm{co}}) (1-2f_{\mathrm{dil}})^2 \mathcal{N} A_J},
\label{eq:C}
\end{equation}
is the required multiplier to keep $Q$ constant at fixed $T$. For simplicity, in this work we use $\langle T_{\alpha} \rangle = T$ as in \cite{Whyte2023}. In \Cref{fig:TBE4}(a), we plot $C$ versus TBE for different $f_{\mathrm{T}}^{\mathrm{co}}$ and $\mathcal{N} A_J$ values. The results here are perhaps even more revealing than earlier comparisons of how polarized fuel affects the TBE at fixed $p_{\Delta}$. If we wish to maintain $Q$ at fixed T, this requires $C=1$. Shown by the blue dash-dotted line in \Cref{fig:TBE4}(a), this formally give TBE = 0.00 for unpolarized fuel -- it is impossible to achieve $C =1$ with unpolarized fuel. However, for polarized fuel with $\mathcal{N} A_J =1.5$, a reasonable TBE is achieved of TBE $\approx 0.14$. Ideally, one would operate a plant at a TBE value where TBE is as large as possible and $C$ is as small as possible. SP fuel makes the tradeoff between TBE and $C$ much more favorable.

In \Cref{fig:TBE4}(b) and (c), we plot $C$ versus $\mathcal{N} A_J$ and $f_{\mathrm{T}}^{\mathrm{co}}$ for TBE = 0.15 (b) and TBE = 0.02 (c). At the given value of $\Sigma \eta_{\mathrm{He}} = 0.63$, for a TBE = 0.15 (b), only a very narrow range of $f_{\mathrm{T}}^{\mathrm{co}}$ and $\mathcal{N} A_J$ values can achieve $C \leq 1$. Decreasing the TBE to 0.02 (c), only plasmas with $\mathcal{N} A_J \gtrsim 1.05$ can achieve $C \leq 1$.

In \Cref{fig:TBE5} we plot $C$ versus the TBE and $f_{\mathrm{T}}^{\mathrm{co}}$ for an unpolarized fuel (\Cref{fig:TBE5}(a)) and a polarized fuel (\Cref{fig:TBE5}(b)). The maximum fusion gain occurs for $f_{\mathrm{T}}^{\mathrm{co}} < 0.5$ as the TBE increases, shown by the red curve that gives the maximum $C$ value at any given TBE. With the value of $\Sigma \eta_{\mathrm{He}} = 0.63$ in \Cref{fig:TBE5}(a), for $\mathcal{N} A_J=1.0$, there is no value of $C \leq 1$, indicating that regardless of the TBE value (in the range plotted), the plasma gain will always be lower than the ideal undiluted case. However, for polarized fuel in \Cref{fig:TBE5}(b), $C = 1$ can be achieved with a TBE $\simeq 0.14$ whereas for unpolarized fuel, $\mathcal{N} A_J = 1.0$, $C=1$ is obtainable only in the limit that TBE$\to 0$.

\section{Minimum Startup Inventory} \label{sec:min_startup_inventory}

\begin{figure}[t]
    \centering
    \begin{subfigure}[t]{0.82\textwidth}
    \centering
    \includegraphics[width=1.0\textwidth]{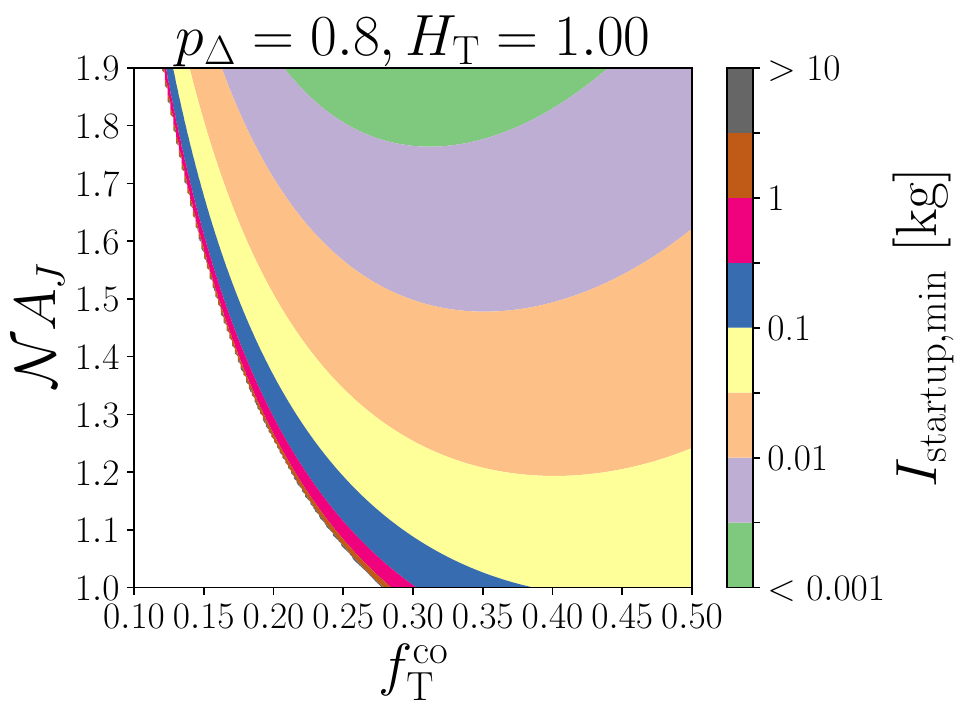}
    \caption{$p_{\Delta} = 0.8$.}
    \end{subfigure}
     ~
    \centering
    \begin{subfigure}[t]{0.82\textwidth}
    \centering
    \includegraphics[width=1.0\textwidth]{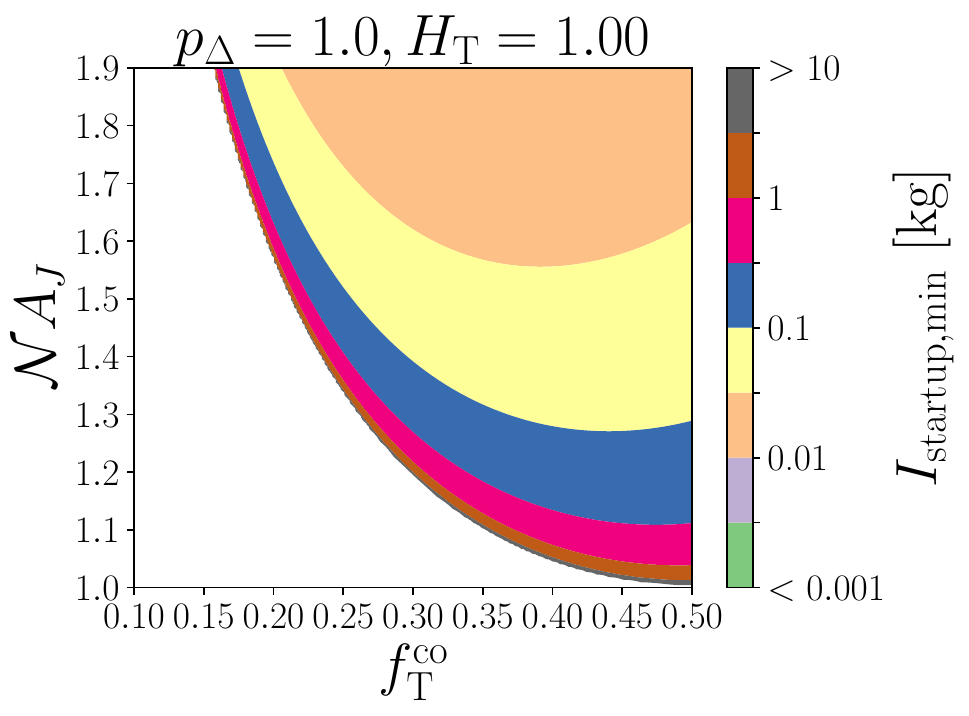}
    \caption{$p_{\Delta} = 1.0$.}
    \end{subfigure}
     ~
    \begin{subfigure}[t]{0.82\textwidth}
    \centering
    \includegraphics[width=1.0\textwidth]{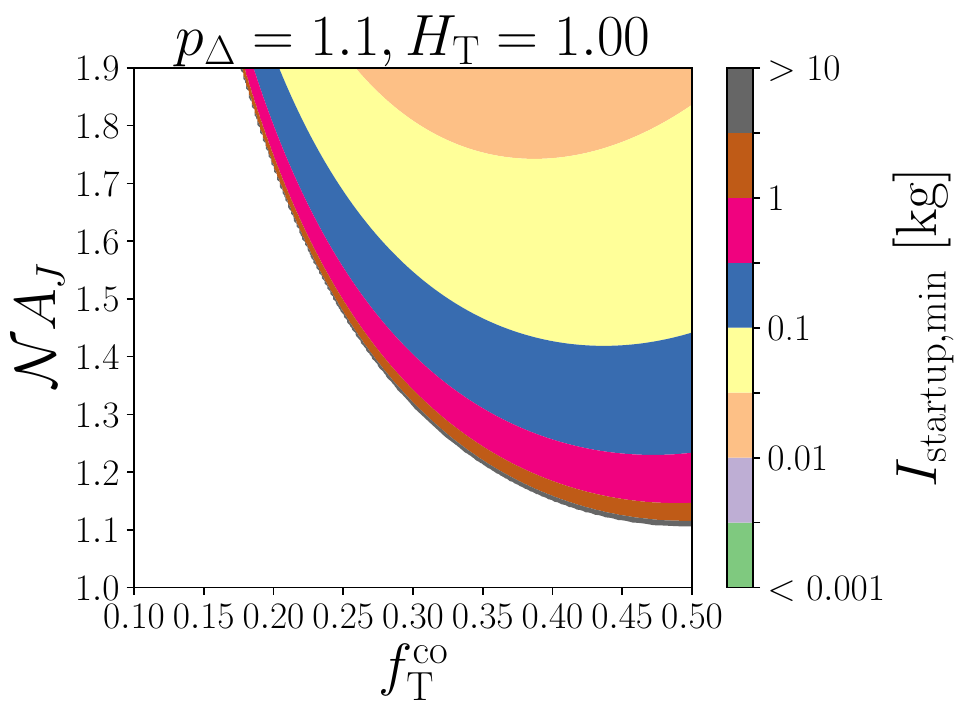}
    \caption{$p_{\Delta} = 1.1$.}
    \end{subfigure}
    \caption{Minimum tritium startup inventory $I_\mathrm{startup,min}$ (see \Cref{eq:tritium_storage} and surrounding discussion) as a function of tritium fraction $f_{\mathrm{T}}^{\mathrm{co}}$ and spin-polarization multiplier $\mathcal{N} A_J$. (a)-(c) show three values of power degradation/enhancement $p_{\Delta}$ (\Cref{eq:pDeltaform2_HT}). White color indicates that the tritium burn efficiency is less than 0 for a given $f_{\mathrm{T}}^{\mathrm{co}}$,  $\mathcal{N} A_J$, $p_{\Delta}$, and unphysical. We use $\Sigma \eta_{\mathrm{He}} = 0.63$ and $H_{\mathrm{T}} = 1.0$.}
    \label{fig:minstartup_inventory}
\end{figure}

As a short illustrative example, we calculate the effect of tritium fraction and spin-polarization multiplier on the minimum tritium startup inventory. As discussed in the introduction, when a D-T-fueled power plant starts up, it takes some time before tritium is bred in sufficient quantities to fully fuel the plant. During this initial phase, the power plant draws from a tritium startup inventory with mass $I_\mathrm{startup}$. As shown in Equation (33) of \cite{Meschini2023}, the total time-dependent tritium inventory $I_\mathrm{storage}$ can be expressed as
\begin{equation}
\begin{aligned}
& I_\mathrm{storage} = I_\mathrm{startup} + \dot{N}_{\mathrm{T},\mathrm{burn}} (\mathrm{TBR} - 1) t \\
& + \dot{N}_{\mathrm{T},\mathrm{burn}} \mathrm{TBR} \frac{\tau^2_{\mathrm{IFC} } }{\tau_{\mathrm{OFC}} - \tau_{\mathrm{IFC}}} (1 - \rm{e}^{-t/\tau_{\mathrm{IFC}} } ) \\
& + \dot{N}_{\mathrm{T},\mathrm{burn}} \mathrm{TBR} \frac{\tau^2_{\mathrm{OFC} } }{\tau_{\mathrm{OFC}} - \tau_{\mathrm{IFC}}} (1 - \rm{e}^{-t/\tau_{\mathrm{OFC}} } ) \\
& + \dot{N}_{\mathrm{T},\mathrm{burn}} \tau_{\mathrm{IFC} } \frac{1 - \mathrm{TBE} }{\mathrm{TBE} }  (1 - \rm{e}^{-t/\tau_{\mathrm{IFC}} } ).
\end{aligned}
\label{eq:tritium_storage}
\end{equation}
Here, TBR is the tritium breeding ratio, $t$ is the time in seconds, and $\tau_{\mathrm{IFC}}$ and $\tau_{\mathrm{OFC}}$ are the tritium residence times for the inner and outer fuel cycle. The goal of this short exercise is to calculate the minimum startup inventory $I_\mathrm{startup,min}$ so that $I_\mathrm{storage} (t) $ always satisfies $I_\mathrm{storage} (t) \geq 0$. For three values of power degradation $p_{\Delta}$ (\Cref{eq:pDeltaform2_HT}), we plot $I_\mathrm{startup,min}$ in \Cref{fig:minstartup_inventory}
versus tritium fraction $f_{\mathrm{T}}^{\mathrm{co}}$ and polarization multiplier $\mathcal{N} A_J$. We assume a set of parameters used in \cite{Meschini2023}: $\tau_{\mathrm{IFC}} = 4$h, $\tau_{\mathrm{OFC}} = 24$h, TBR = 1.08, and $9 \times 10^{-7}$ kg s$^{-1}$ of tritium are burned for the zero power degradation case (corresponding to 507 MW of fusion power for an ARC-class device \cite{Sorbom2015}). We allow the TBE to vary based on the combination of $f_{\mathrm{T}}^{\mathrm{co}}$, $\mathcal{N} A_J$, and $p_{\Delta}$, and also assume that $\Sigma \eta_{\mathrm{He}} = 0.63$. For each value of $f_{\mathrm{T}}^{\mathrm{co}}$, $\mathcal{N} A_J$, and $p_{\Delta}$, the TBE is self-consistently calculated from \Cref{eq:pDeltaform2_HT}.

The results in \Cref{fig:minstartup_inventory} are striking: with zero power degradation $p_{\Delta} = 1.0$ in \Cref{fig:minstartup_inventory}(b), for $\mathcal{N} A_J = 1.0$ there is no tritium fraction where the minimum tritium startup inventory is less than 10 kg. For $\mathcal{N} A_J = 1.5$ and a suitable $f_{\mathrm{T}}^{\mathrm{co}}$ value, $I_\mathrm{startup,min}$ is less than 100 grams. Additionally, for $p_{\Delta} \gtrsim 0.9$, $I_\mathrm{startup,min}$ is extremely sensitive to $\mathcal{N} A_J$ around $\mathcal{N} A_J \simeq 1.0$ and $f_{\mathrm{T}}^{\mathrm{co}} \simeq 0.5$. This suggests that even a 5\% increase in $\mathcal{N} A_J$ from $\mathcal{N} A_J = 1.00$ to $\mathcal{N} A_J = 1.05$ could decrease the required minimum tritium startup inventory by a factor of ten.

If the power is enhanced by 10\% ($p_{\Delta} = 1.1$), shown in \Cref{fig:minstartup_inventory}(c), at high polarization multiplier values, $\mathcal{N} A_J > 1.4$, $I_\mathrm{startup,min}$ is less than 100 grams. And if the power is degraded by 20\% ($p_{\Delta} = 0.8$), shown in \Cref{fig:minstartup_inventory}(a), for $\mathcal{N} A_J \simeq 1.5$, $I_\mathrm{startup,min}$ is predicted to be roughly 10 grams at $f_{\mathrm{T}}^{\mathrm{co}} \simeq 0.35$. At such small tritium startup inventory values, there are likely other effects, not captured in \Cref{eq:tritium_storage}, which set the minimum inventory.

\begin{table}
\caption{Key fusion power and tritium self-sufficiency parameters for different operating scenarios in an ARC-like device. In addition to listed all parameters, all cases have $H_{\mathrm{T}} = 1.0$, $\eta_{\mathrm{He}} = 1.00$, $\Sigma = 0.63$, $\tau_{\mathrm{IFC}} = 4$h, $\tau_{\mathrm{OFC}} = 24$h, and TBR = 1.08. Note that $I_{\mathrm{startup,min}}$ values below a certain value are likely not well-described by the model in \Cref{eq:tritium_storage}.}
\begin{ruledtabular}
  \begin{tabular}{| c || c|c|c|c|c|c |c | }
   Case & $ \mathcal{N} A_J$ & \shortstack{$P_f$ \\ (MW)} & Q & TBE & \shortstack{$I_{\mathrm{startup,min}}$ \\ (kg)} & $f_{\mathrm{T}}^{\mathrm{co}}$  & $f_{\mathrm{He},\mathrm{div}}$ \\
    \hline
    base & 1.00 & 482 & 20.0  & 0.016 & 0.677 & 0.49 & 0.013 \\
    A & 1.25 & 482 & 23.9  & 0.090 & 0.076 & 0.43 & 0.068 \\
    B & 1.50 & 482 & 26.4 & 0.154  & 0.028 & 0.39  & 0.112 \\
    C & 1.25 & 602 & $\infty$ & 0.016 & 0.846 & 0.49  & 0.013 \\
    D & 1.50 & 722 & $\infty$ & 0.016 & 1.015 & 0.49  & 0.013 \\
    E & 1.50 & 253 & 4.27 & 0.387 & $<$0.001 & 0.25  & 0.251 \\
    F & 1.50 & 101 & 1.17 & 0.672 & $<$0.001 & 0.12  & 0.384 \\
    G & 1.90 & 482 & 29.6 & 0.240 & 0.007 & 0.33  & 0.167 \\
    H & 1.90 & 915 & $\infty$ & 0.016 & 1.286 & 0.49  & 0.013 \\
    I & 1.50 & 562 & 150 & 0.100 & 0.075 & 0.43  & 0.075 \\
    J & 1.90 & 712 & $\infty$ & 0.100 & 0.096 & 0.43  & 0.075 \\
  \end{tabular}
\end{ruledtabular}
\label{tab:tab1}
\end{table}

The benefits of smaller tritium startup inventory with spin-polarized fuel are substantial. Not only could lower startup inventory requirements prevent possible tritium availability shortages, but combined with higher tritium burn efficiency, could reduce the total amount of tritium retained in the plant. ITER, for example, has an administrative limit on the total tritium retainment of 700 grams \cite{Roth2008}, which limits the number of discharges, constraining other parts of the design, notably the wall and divertor materials \cite{Shimada2009}.

\section{ARC-like Power Plant} \label{sec:casestudy}

\begin{figure*}[!tb]
    \centering
    \begin{subfigure}[t]{0.75\textwidth}
    \centering
    \includegraphics[width=1.0\textwidth]{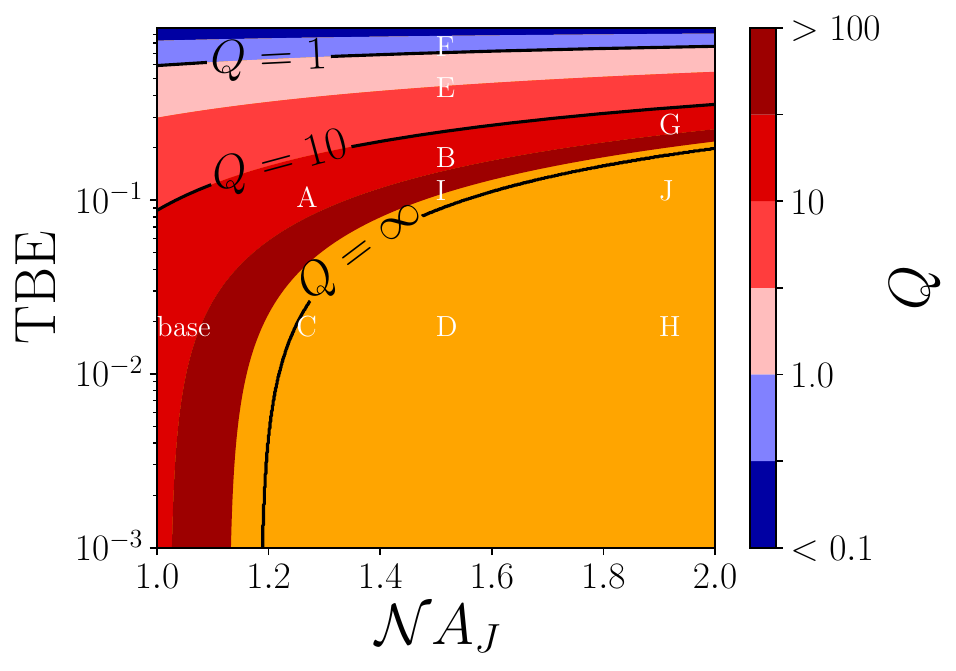}
    \caption{Plasma gain $Q$.}
    \end{subfigure}
     ~
    \centering
    \begin{subfigure}[t]{0.75\textwidth}
    \centering
    \includegraphics[width=1.0\textwidth]{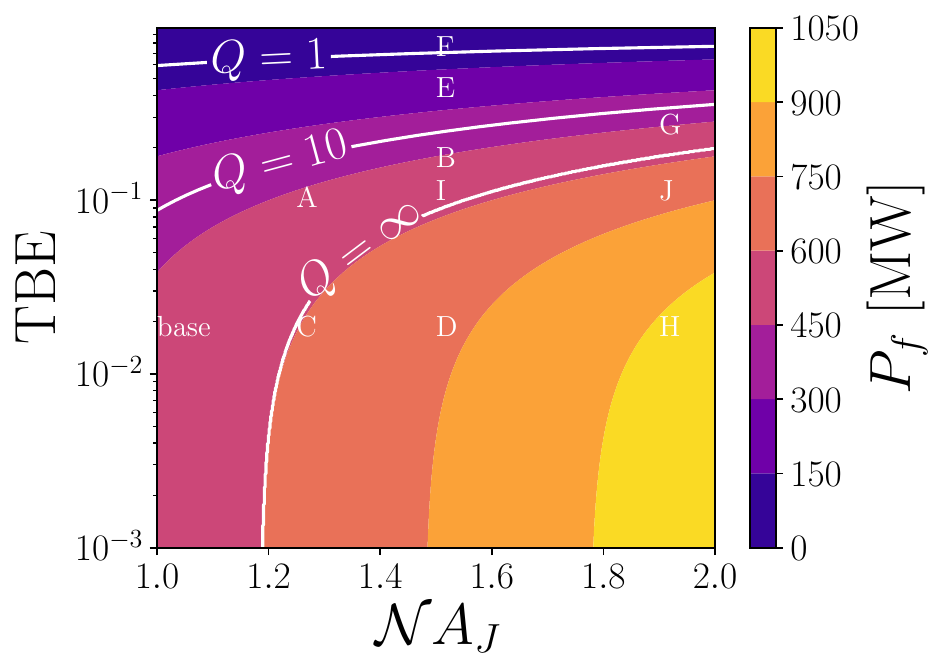}
    \caption{Fusion power $P_f$.}
    \end{subfigure}
    \caption{Plasma gain $Q$ (a) and fusion power $P_f$ (b) versus tritium burn efficiency (TBE) and spin-polarization multiplier $\mathcal{N} A_J$ for an ARC-like fusion power plant. Plasmas below the $Q=\infty$ curve are predicted to ignite. The location of operating cases corresponding to \Cref{tab:tab1} are labeled with white text.}
    \label{fig:Qplasma}
\end{figure*}

In this section, we briefly summarize the main effects of spin polarization on the primary metrics for fusion power and tritium self-sufficiency by analyzing an ARC-like device \cite{Sorbom2015}. Using similar parameters to those in \cite{Meschini2023} (also used in the previous section), we calculate key parameters for seven operating scenarios with varying levels of spin polarization. These are summarized in \Cref{tab:tab1}. All scenarios in \Cref{tab:tab1} have $\tau_{\mathrm{IFC}} = 4$h, $\tau_{\mathrm{OFC}} = 24$h, TBR = 1.08, $\Sigma = 0.63$, $\eta_{\mathrm{He}} = 1.0$, and $H_{\mathrm{T}} = 1.0$. More details for this section's workflow are provided in \Cref{sec:workflow}. For a quick summary, compare the base scenario and cases I and J in \Cref{fig:arcclass_spider}. The relative location of all the cases is also plotted in the plasma gain graph \Cref{fig:Qplasma}.

Cases A and B have the same power as Base of 482 MW, and use spin polarization to significantly increase the TBE and decrease the minimum startup inventory $I_\mathrm{startup,min}$. Increasing $\mathcal{N} A_J$ from 1.0 to 1.5 decreases $I_\mathrm{startup,min}$ from 0.68 kg to 0.03 kg. In Base, A, and B, the tritium fraction $f_{\mathrm{T}}^{\mathrm{co}}$ is chosen to maximize the TBE. As expected by previous discussions (e.g. surrounding \Cref{fig:TBEfull}), the improvement of the TBE with $\mathcal{N} A_J$ is nonlinear. Notably, at higher $\mathcal{N} A_J$ values, the divertor helium fraction $f_{\mathrm{He},\mathrm{div}}$ increases significantly. The most promising result between Base, A, and B is that $I_\mathrm{startup,min}$ already decreases by 88\% when increasing $\mathcal{N} A_J$ from 1.0 to 1.25 (50\% fuel polarization). This suggests that achieving 100\% fuel polarization is not necessary to significantly increase the TBE and therefore decrease $I_\mathrm{startup,min}$. 

In C and D, spin polarization is used to maximize the fusion power constrained to the same TBE value as the Base scenario, TBE = 0.016. In our model, the fusion power scales with $\mathcal{N} A_J$ at fixed TBE. Case C with $\mathcal{N} A_J = 1.25$ has $P_f = 602$ MW and case D with $\mathcal{N} A_J = 1.50$ has $P_f = 722$ MW. In C and D, the tritium fraction $f_{\mathrm{T}}^{\mathrm{co}}$ is chosen to maximize the fusion power. Notably, in the tritium model in \Cref{eq:tritium_storage}, tritium startup inventory increases linearly with $\mathcal{N} A_J$ in C and D compared with the Base scenario. This is because we have assumed the TBR is constant, and therefore because $P_f \propto \dot{N}_{\mathrm{T},\mathrm{burn}}$ in \Cref{eq:tritium_storage}, $I_{\mathrm{startup,min}} \propto A_J$. This may not be realistic.

\begin{figure}[tb]
    \centering
    \includegraphics[width=0.98\textwidth]{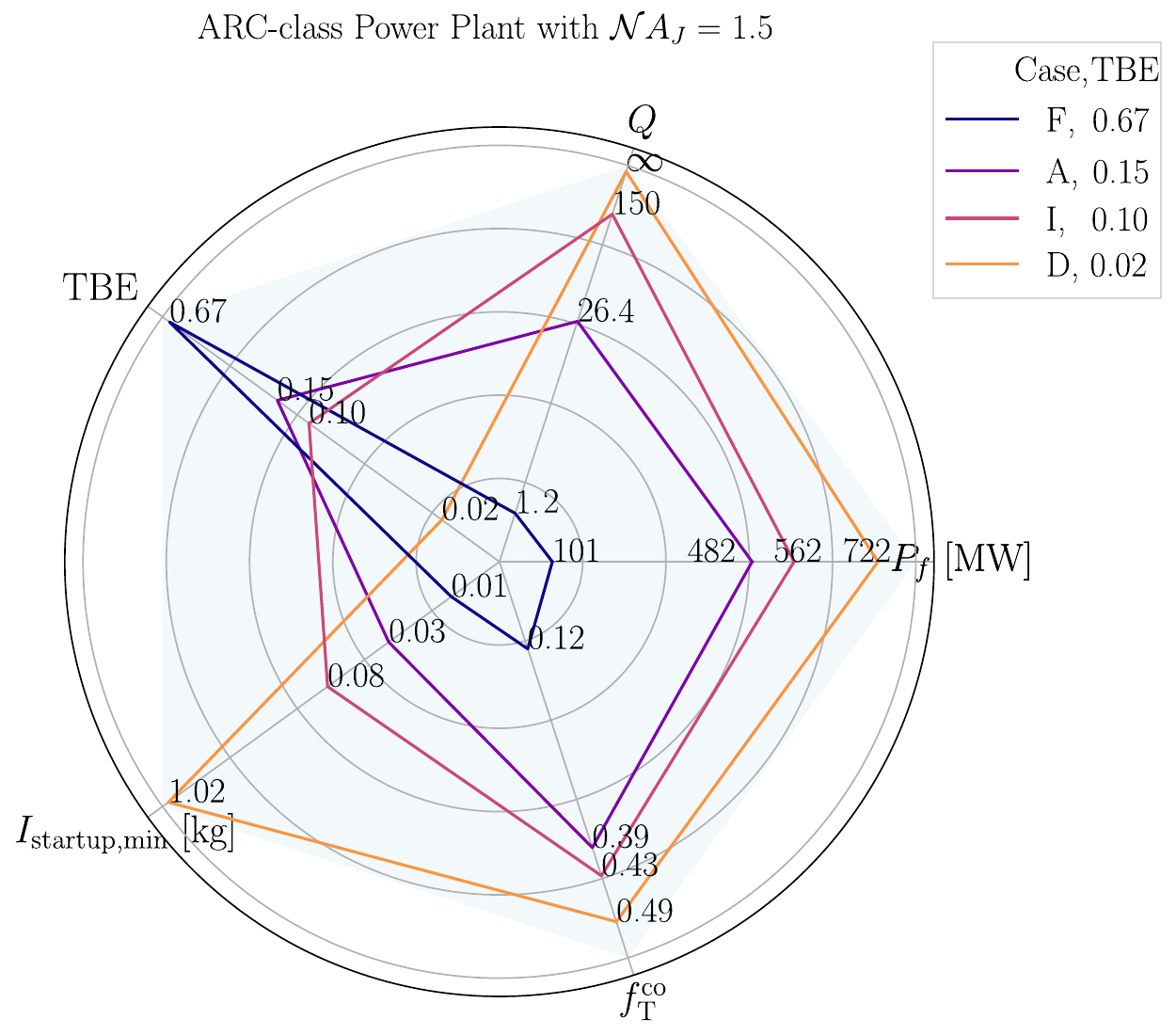}
    \caption{Tritium self-sufficiency and fusion power parameters of an ARC-like device for four cases with four TBE values and fixed spin polarization $\mathcal{N} A_J = 1.5$. The fusion power is $P_f$, the plasma gain is $Q$, the tritium burn efficiency is TBE, the tritium startup inventory is $I_{\mathrm{startup,min}}$, and the tritium fraction is $f_{\mathrm{T}}^{\mathrm{co}}$. The $f_{\mathrm{T}}^{\mathrm{co}}$ and $P_f$ axes have a value of zero at the origin and are linearly scaled; TBE, $Q$, and $I_{\mathrm{startup,min}}$ are log-scaled.}
    \label{fig:TBE_scan_spider}
\end{figure}

Cases E and F in \Cref{tab:tab1} are included to illustrate extremely high tritium burn up regimes. Case F operates with a very low tritium fraction, $f_{\mathrm{T}}^{\mathrm{co}} = 0.12$, has a very low fusion power, $P_f = 101$ MW, but achieves TBE = 0.672: for every three tritium particles injected in the reaction chamber, over two of them undergo fusion events.

As a demonstration of just how far spin polarization positively affects tritium self-sufficiency and fusion power, we consider cases G and H with very high spin polarization and power enhancement, $\mathcal{N} A_J = 1.9$. Case G is designed to maximize the TBE and minimize $I_{\mathrm{startup,min}}$, and case H is designed the maximize the fusion power. Case G achieves TBE = 0.240 and $I_{\mathrm{startup,min}} = 0.007$ kg, and case H achieves $P_f = 915$ MW but requires $I_{\mathrm{startup,min}} = 1.29$ kg according to the tritium inventory model in \Cref{eq:tritium_storage}.

Finally, we consider two compromise cases, I and J, where both the tritium self-sufficiency and fusion power are improved, rather than the cases in cases A-H where either the tritium self-sufficiency or fusion power is maximized. We did not perform an optimized search in the TBE but chose TBE = 0.10.

Case I has $\mathcal{N} A_J = 1.50$ and TBE = 0.10. This gives a fusion power of 562 MW, $I_{\mathrm{startup,min}} = 0.075$ kg and $f_{\mathrm{T}}^{\mathrm{co}} = 0.43$. Case J has the same inputs as case I except its spin-polarization multiplier is higher, $\mathcal{N} A_J = 1.9$. This gives a fusion power of 712 MW, $I_{\mathrm{startup,min}} = 0.096$ kg and $f_{\mathrm{T}}^{\mathrm{co}} = 0.43$. In both of these cases, $f_{\mathrm{He},\mathrm{div}} = 0.075$. Cases I and J are compared with the base case in \Cref{fig:arcclass_spider}.

In \Cref{fig:TBE_scan_spider}, we use a spider plot to summarize the effect of TBE at fixed $\mathcal{N} A_J = 1.5$ for cases A, D, F, and I. Increasing the TBE decreases $f_{\mathrm{T}}^{\mathrm{co}}$, $Q$, $P_f$, and $I_{\mathrm{startup,min}}$. Cases A and I are likely good candidate cases that compromise between increased TBE and increased $P_f$, $Q$.

The plasma gain values in \Cref{tab:tab1} are particularly striking. For a detailed description of how these $Q$ values were calculated, see \Cref{sec:workflow}. Briefly, the $Q$ values were calculated at fixed $n_e T \tau_E$, and account for helium dilution, tritium fraction, and spin-polarization effects (see \Cref{eq:C}). In order to better compare the different cases, we plot their $Q$ values against TBE and $\mathcal{N} A_J$ in \Cref{fig:Qplasma}(a). As shown by cases C, D, and H in \Cref{fig:Qplasma}, the power plant ignites for a wide range of $\mathcal{N} A_J$ values. Inspection of \Cref{fig:Qplasma}(a) at very low TBE values reveals that for the parameters chosen here, this power plant could ignite at $\mathcal{N} A_J \simeq 1.18$. Optimistically, this value of spin polarization, $\mathcal{N} A_J \simeq 1.18$, is very low compared with the spin-polarization power boost from recent simulations $\mathcal{N} A_J = 1.9$ \cite{Heidbrink2024}. Notably, as shown by the fusion power in \Cref{fig:Qplasma}(b), $P_f$ still increases significantly in ignited plasmas, from a minimum of $P_f \approx 600$ MW to a maximum of $P_f \approx 960$ MW for the TBE and $\mathcal{N} A_J$ values shown.

Additionally, these results suggest that the upcoming SPARC experiment \cite{Creely2020} could achieve ignition with moderately high values of $\mathcal{N} A_J \gtrsim 1.42$. In \Cref{sec:Q10_scan}, we perform the same exercise but with a nominal plasma gain of $Q=10$ and $Q=40$ rather than the $Q=20$ we used here. At very low TBE, for the nominal $Q=10$ case, ignition is predicted for $\mathcal{N} A_J \gtrsim 1.42$ and for the nominal $Q=40$ case, ignition is predicted for $\mathcal{N} A_J \gtrsim 1.06$.

This short case study has demonstrated that spin polarization can simultaneously improve a fusion power plant's tritium burn efficiency, increase the fusion power, and decrease the tritium startup inventory. For case J in \Cref{tab:tab1}, operating with spin-polarized fuel increases the fusion power by 52\%, increases the TBE by 525\%, and decreases the startup tritium inventory by 84\%. A future question to answer is whether the increased $f_{\mathrm{He},\mathrm{div}}$ at higher TBE is feasible. Although operating at higher $f_{\mathrm{He,div}}$ could be concerning, in this exercise $f_{\mathrm{He,div}}$ is increasing because the hydrogen divertor density is falling, not because the helium divertor density is increased; the helium divertor removal rate $\dot{N}_{\mathrm{He,div} }$ and density are fixed because the fusion power is constant. Operating at high spin polarization and high TBE might be challenging because the plasma is very efficient with both tritium and deuterium, resulting in low hydrogen divertor density. If spin-polarized plasmas were to be operated at high TBE, mitigating the potentially negative impact on the power exhaust would be crucial \cite{Wischmeier2015,Asakura2017,Kuang2020}.

See \Cref{sec:intuition} for further discussion of the TBE and $f_{\mathrm{He,div}}$ and see \Cref{sec:constantTBR} for a discussion of the limitations of the constant TBR assumption used in this study. In \cite{Whyte2023}, it was suggested that a high TBE could give an unreasonably long helium particle confinement time. This is of particular concern for spin-polarized plasmas with very high TBE. We study this question in \Cref{sec:heliumparticleconfinement} and conclude that while spin-polarized plasmas with high TBE do have longer helium particle confinement times, they do not exceed dimensionless thresholds measured in current experiments. Additionally, in this study we assumed the most pessimistic case for enrichment, $H_{\mathrm{T}} = 1.00$. With lower enrichment the TBE and plasma gain improve significantly; see \Cref{sec:tritium_enrichment} for more details.

\section{Discussion} \label{sec:discussion}

Drawing from recent work \cite{Whyte2023,Meschini2023}, we have demonstrated that burning plasmas using D-T polarized fuel can increase the tritium burn efficiency by at least an order of magnitude over unpolarized fuels and with no degradation in power density. If the power density is allowed to increase significantly with spin-polarized fuel, the tritium burn efficiency can still increase by a factor of five to ten. This could significantly improve the economic viability of D-T power plants with high tritium self-sufficiency.

In one example based on the power increase using spin-polarized fuels \cite{Heidbrink2024}, an ARC-like fusion power plant with spin-polarized fuel increases the fusion power by 52\%, increases the TBE by 525\%, and decreases the startup tritium inventory by 84\% relative to the plant operating with unpolarized fuel (see case J in \Cref{tab:tab1} and \Cref{fig:arcclass_spider}). In another example where a power plant using spin-polarized fuel maintains the same fusion power as unpolarized fuel and the increased cross section is used solely for improved tritium burn efficiency (see case G in \Cref{tab:tab1}), the TBE increases by 1400\% to TBE = 0.24 and the predicted tritium startup inventory falls 99\% to 7 grams. This demonstrates that the increased cross section conferred by spin-polarized fuels gives a significant advantage in tritium burn efficiency, in addition to the widely recognized increase in power density. Operating with spin-polarized fuel can also drastically reduce improvements in the helium pumping efficiency required to obtain a given fusion power and tritium burn efficiency (see \Cref{fig:Sigma_Min}), sometimes by an order of magnitude. The unexpectedly large increase in the TBE for spin-polarized D-T is due to a double benefit of operating at lower tritium fraction and lower overall hydrogen fueling (see discussion surrounding \Cref{fig:TBEfull,sec:intuition}), as well as the fusion power scaling linearly with spin polarization.

We also showed that in some points of parameter space, very small increases in the fuel's spin polarization could decrease the tritium startup inventory by an order of magnitude. For an ARC-class device, the tritium startup inventory could be reduced to less than one hundred grams, with a moderate power increase (see \Cref{fig:minstartup_inventory,tab:tab1,fig:arcclass_spider}). If spin-polarized fusion is demonstrated with high polarization survivability \cite{Heidbrink2024}, the results in this work suggest that any impending tritium shortages for deuterium-tritium power plants would be solved. This indicates that spin-polarized fusion fuels, although undemonstrated and speculative, could significantly lower the technological requirements and costs for a power plant with high tritium self-sufficiency. %

Spin-polarized low-tritium-fraction fuels have other useful properties. For example, spin-polarized fuels increase the fusion power without necessarily increasing the plasma density or $\beta$, allowing plasmas to have higher fusion power output without getting closer to Greenwald density or $\beta$ limits \cite{Greenwald1988,Sauter1997,Loizu2017}. Additionally, as described in \Cref{sec:ignitionstab}, spin-polarized fuels can decrease the triple product required for ignition by a factor of two \cite{Mitarai1992}, and lower-tritium-fraction fuels might provide passive stabilization against thermal ignition runaway events. A particularly useful property of spin-polarized fueling systems is that they are unlikely to add significant complexity and cost to a fusion power plant \cite{Garcia2023,Baylor2023,Heidbrink2024} and are complementary to other innovations designed to improve tritium self-sufficiency \cite{Ubeyli2007,Chen2016,Pearson2018,Igitkhanov2018,Federici2019,Forsberg2020,Abdou2021,Vladimirov2021,Ferry2023,Zhou2023,Bohm2023,Whyte2023,Cohen2024}.

There are important outstanding questions directly related to this work. One is the effect of spin polarization and tritium fraction on the alpha heating. Studies have found that when alpha heating effects are included, the total fusion power increased 80-90\% when operating with spin-polarized fuels with a 50\% cross-section enhancement \cite{Smith2018IAEA,Heidbrink2024}. However, it is not yet known how the total fusion power scales for a wide range of polarization fractions and tritium fractions. For example, a simple estimate in \Cref{eq:pf_pola_runpolar} showed that a spin-polarized fuel with a 50\% cross-section enhancement and a 79:21 D-T mix will have a comparable power density to an unpolarized fuel with a 50:50 D-T mix. However, this estimate neglected nonlinear heating effects, which is further complicated by the anisotropic fusion-borne alpha particle distribution \cite{Bittoni1983}. Answering such questions in the context of tritium burn efficiency is important because, in this work, we primarily studied the effect of spin polarization and tritium fraction on the fusion power density, not on the total fusion power. A second closely related and well-known challenge is the feasibility of keeping a high core polarization fraction. Wave resonances with deuterium and tritium precession frequencies and metallic wall recycling are particularly concerning \cite{Kulsrud1986,Heidbrink2024}. Recent ideas for keeping the core tritium extremely well-confined \cite{Boozer2024stellarators} and for achieving low particle recycling with lithium wall coatings \cite{Krasheninnikov2003,Boyle2011} might be fruitfully applied to the wall depolarization problem as well as for increasing $\Sigma$. A third challenge from this work is effect of relatively high ratio of the neutral helium ash to hydrogenic fuel in the divertor $f_{\mathrm{He},\mathrm{div}}$ (see \Cref{eq:fHediv,tab:tab1}) that results from high TBE operation. While we have self-consistently included the effects of $f_{\mathrm{He},\mathrm{div}}$ on plasma performance (e.g. \Cref{eq:pDeltaform2_HT,eq:C}), there may be additional challenges when operating with sufficiently large $f_{\mathrm{He},\mathrm{div}}$ values such as the power exhaust due to low neutral hydrogen divertor density. A fourth outstanding question is the effect of radial profiles. For example, in this work, we have not carefully treated the radial dependence of the tritium fraction or spin polarization -- higher fidelity integrated modeling is needed to better understand these potentially important effects.

This work primarily focuses on magnetic confinement fusion, but it may also have applications in other fusion approaches, such as inertial confinement fusion. For instance, a recent experiment at the National Ignition Facility achieved ignition with a tritium burnup fraction of 1.9\% \cite{AbuShawareb2022}. By spin-polarizing the fuel \cite{More1983,Temporal2012} and optimizing the tritium fraction, it may be possible to enhance both the burnup fraction and the overall fusion yield.

Finally, the arguments presented here might reasonably be made with improvements in plasma confinement resulting from schemes other than spin polarization. In this work, the fusion power increases due to spin polarization, but the form of the equations would be similar (albeit with some important differences) if the spin-polarization multiplier $\mathcal{N} A_J$ were substituted appropriately with a parameter that measures improvement in plasma confinement from other mechanisms \cite{Yushmanov1990,Mukhovatov2003}. This could motivate using high-confinement regimes \cite{Wagner1982,Goldston1994,Galambos1995,Taylor1997,Stambaugh1998,Oyama2009,Snyder2019,Creely2020,Kaye2021,Lunsford2021,Nelson2023,Ding2024} more generally to increase tritium burn efficiency by decreasing the tritium fraction.

\section{Data Availability Statement}

The data that support the findings of this study are openly available in the Princeton Data Commons at \cite{parisi_data4_2024}.

\section{Acknowledgements}

We are grateful to A. H. Boozer, J. E. Menard, and M. C. Zarnstorff for insightful conversations, and to T. M. Qian for reading the manuscript. The US Government retains a non-exclusive, paid-up, irrevocable, world-wide license to publish or reproduce the published form of this manuscript, or allow others to do so, for US Government purposes.

\appendix

\section{Core Tritium Density and Flow Fraction} \label{sec:core_trit_frac}

In this section, we show the relation between the core tritium density fraction $f_{\mathrm{T}  }^{\mathrm{co} }$ and the core tritium flow fraction $F_{\mathrm{T}  }^{\mathrm{co} }$. The tritium flow rate through a flux surface is
\begin{equation}
\dot{N}_{\mathrm{T}  }^{\mathrm{co} } = V' \langle \nabla r \rangle \Gamma_{\mathrm{T}},
\end{equation}
where $V' = dV / dr$ for the flux-surface volume $V$ and minor radial flux-surface coordinate $r$, $\langle \nabla r \rangle$ is the flux-surface average of $\nabla r$, and $\Gamma_{\mathrm{T}}$ is the particle flux density. Using a diffusive model for the flux density
\begin{equation}
\Gamma_{\mathrm{T}} = - D_{\mathrm{T}} \nabla n_{\mathrm{T}},
\label{eq:diffusivetransp}
\end{equation}
where $D_{\mathrm{T}}$ is a diffusion coefficient, the tritium flow rate becomes
\begin{equation}
\dot{N}_{\mathrm{T}  }^{\mathrm{co} } = - V' \langle \nabla r \rangle D_{\mathrm{T}} \nabla n_{\mathrm{T}}.
\end{equation}
The total hydrogen flow rate is
\begin{equation}
\dot{N}_{\mathrm{Q}  }^{\mathrm{co} } = - V' \langle \nabla r \rangle \left( D_{\mathrm{T}} \nabla n_{\mathrm{T}} + D_{\mathrm{D}} \nabla n_{\mathrm{D}} \right).
\end{equation}
We now make several simplifying assumptions. First, we assume that the deuterium and tritium densities satisfy
\begin{equation}
\nabla n_{\mathrm{D}} \simeq \frac{1 - f_{\mathrm{T}  }^{\mathrm{co} }}{f_{\mathrm{T}  }^{\mathrm{co} }} \nabla n_{\mathrm{T}},
\label{eq:assum1}
\end{equation}
where we assumed that
\begin{equation}
\nabla \left( \frac{1 - f_{\mathrm{T}  }^{\mathrm{co} }}{f_{\mathrm{T}  }^{\mathrm{co} }} n_{\mathrm{T}} \right) \simeq  \frac{1 - f_{\mathrm{T}  }^{\mathrm{co} }}{f_{\mathrm{T}  }^{\mathrm{co} }} \nabla n_{\mathrm{T}},
\label{eq:nablaD}
\end{equation}
and used
\begin{equation}
n_{\mathrm{D}} = \frac{1 - f_{\mathrm{T}  }^{\mathrm{co} }}{f_{\mathrm{T}  }^{\mathrm{co} }} n_{\mathrm{T}}.
\end{equation}
We also assume that deuterium and tritium diffusion coefficients are equal $D = D_D = D_{\mathrm{T}}$, which while not disproven for previous D-T experiments \cite{Balet1993,Tala2023}, may not hold for burning plasmas. This gives
\begin{equation}
\dot{N}_{\mathrm{Q}  }^{\mathrm{co} } = - V' \langle \nabla r \rangle D \frac{\nabla n_{\mathrm{T}}}{f_{\mathrm{T}  }^{\mathrm{co} }}.
\end{equation}
Using these assumptions to write the tritium flow rate,
\begin{equation}
\dot{N}_{\mathrm{T}  }^{\mathrm{co} } = - V' \langle \nabla r \rangle D \nabla n_{\mathrm{T}},
\end{equation}
results in the tritium flow rate fraction being equal to the tritium density fraction,
\begin{equation}
F_{\mathrm{T}  }^{\mathrm{co} } = \frac{\dot{N}_{\mathrm{T}  }^{\mathrm{co} }}{\dot{N}_{\mathrm{Q}  }^{\mathrm{co} }} = f_{\mathrm{T}  }^{\mathrm{co} }.
\label{eq:FTco_ftco}
\end{equation}
In summary, to derive \Cref{eq:FTco_ftco}, we used three main simplifications: diffusive particle transport in \Cref{eq:diffusivetransp}, neglecting the spatial gradient of $F_{\mathrm{T}  }^{\mathrm{co} }$ in \Cref{eq:assum1} and  $D = D_{\mathrm{D}} = D_{\mathrm{T}}$. Interesting future extensions of this model might study the effect of $D_{\mathrm{D}} \neq D_{\mathrm{T}}$, keeping terms proportional to $\nabla (1 / f_{\mathrm{T}  }^{\mathrm{co} })$ in \Cref{eq:nablaD}, and adding a particle pinch term.

\section{Tritium Fraction and Exhaust} \label{sec:tritfueling_exhaust}

We now calculate the relative tritium removal rate in the divertor $F_{\mathrm{T}}^{\mathrm{div}}$ compared to the relative injection rate $F_{\mathrm{T}}^{\mathrm{in}}$. When the D-T fuel mixture is no longer 50:50, new constraints arise.

\begin{figure}[tb]
    \centering
    \includegraphics[width=1.0\textwidth]{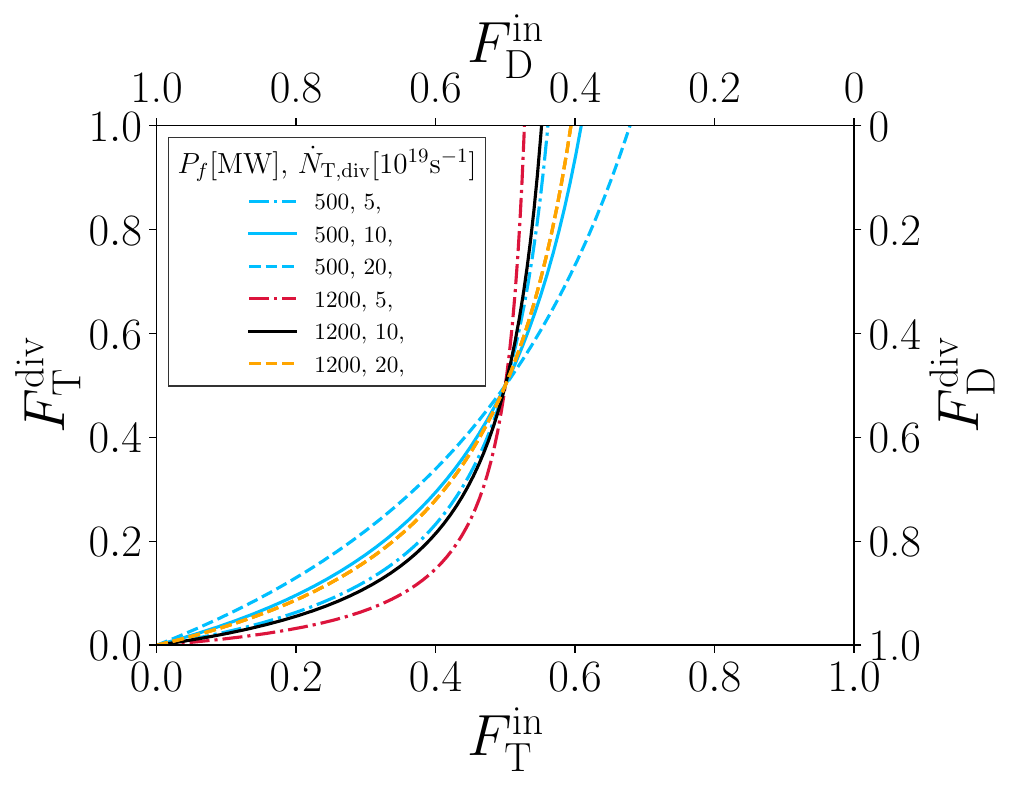}
    \caption{Tritium and deuterium divertor fractions $F_{\mathrm{T}}^{\mathrm{div}}$ and $F_{\mathrm{D}}^{\mathrm{div}}$ versus injection fractions $F_{\mathrm{T}}^{\mathrm{in}}$ and $F_{\mathrm{D}}^{\mathrm{in}}$ (see \Cref{eq:FTdiv1,eq:FDdiva}) for different fusion power $P_f$ and divertor removal rates $\dot{N}_{\mathrm{T},\mathrm{div}}$. Only for $F_{\mathrm{T}}^{\mathrm{in}} = 1/2$ are all of these quantities equal.}
    \label{fig:trtitiumdeuterium_inj_div}
\end{figure}

Using the definitions in \Cref{eq:fueling_ratio,eq:divertor_ratio} and particle conservation in \Cref{eq:particle_conservation,eq:individualtritium}, and that $\dot{N}_{\mathrm{He},\mathrm{div}} = P_f / E$, the tritium divertor fraction is
\begin{equation}
F_{\mathrm{T}}^{\mathrm{div}} = F_{\mathrm{T}}^{\mathrm{in}} \frac{1}{1 + \frac{P_f}{E} \frac{1}{\dot{N}_{\mathrm{T},\mathrm{div}}} \left( 1 - 2 F_{\mathrm{T}}^{\mathrm{in}} \right)},
\label{eq:FTdiv1}
\end{equation}
and the deuterium divertor fraction is
\begin{equation}
F_{\mathrm{D}}^{\mathrm{div}} = 1 - F_{\mathrm{T}}^{\mathrm{in}} \frac{1}{1 + \frac{P_f}{E} \frac{1}{\dot{N}_{\mathrm{T},\mathrm{div}}} \left( 1 - 2 F_{\mathrm{T}}^{\mathrm{in}} \right)}.
\label{eq:FDdivinitial}
\end{equation}
\Cref{eq:FTdiv1,eq:FDdivinitial} reduce to the intuitive result that if the tritium and deuterium input rates are equal, $F_{\mathrm{T}}^{\mathrm{in}} = 1/2$, then the injection and divertor fractions are equal,
\begin{equation}
F_{\mathrm{T}}^{\mathrm{div}} = F_{\mathrm{T}}^{\mathrm{in}} = 1/2.
\end{equation}
However, we wish to study the consequences of $F_{\mathrm{T}}^{\mathrm{in}} \neq 1/2$. In \Cref{fig:trtitiumdeuterium_inj_div}, we plot solutions for $F_{\mathrm{T}}^{\mathrm{div}}$, $F_{\mathrm{D}}^{\mathrm{div}}$, $F_{\mathrm{T}}^{\mathrm{in}}$, and $F_{\mathrm{D}}^{\mathrm{in}}$ for different fusion power levels tritium divertor removal levels. The quantities $F_{\mathrm{T}}^{\mathrm{div}}$, $F_{\mathrm{D}}^{\mathrm{div}}$, $F_{\mathrm{T}}^{\mathrm{in}}$, and $F_{\mathrm{D}}^{\mathrm{in}}$ are equal only when $F_{\mathrm{T}}^{\mathrm{in}} = 1/2$, but small deviations from $F_{\mathrm{T}}^{\mathrm{in}} = 1/2$ can sometimes lead to relatively large differences in $F_{\mathrm{T}}^{\mathrm{div}}$ and $F_{\mathrm{D}}^{\mathrm{div}}$. For example, at very high power, $P_f = 1200$ MW, and divertor removal rate $\dot{N}_{\mathrm{T},\mathrm{div}} = 5 \times 10^{19}$ s${}^{-1}$, reducing the tritium input fraction from $F_{\mathrm{T}}^{\mathrm{in}} =0.5$ to $F_{\mathrm{T}}^{\mathrm{in}} =0.45$ decreases the tritium divertor fraction from $F_{\mathrm{T}}^{\mathrm{in}} =0.5$ to $F_{\mathrm{T}}^{\mathrm{in}} =0.24$. This is shown by the red curve in \Cref{fig:trtitiumdeuterium_inj_div}.

We now rewrite $P_f /E\dot{N}_{\mathrm{T},\mathrm{div}} = \dot{N}_{\mathrm{He},\mathrm{div}}/\dot{N}_{\mathrm{T},\mathrm{div}}$ in terms of dimensionless variables. Using \Cref{eq:neutralpump,eq:ashpumpspeed,eq:ashtofuel} we find
\begin{equation}
\frac{P_f}{E} \frac{1}{\dot{N}_{\mathrm{T},\mathrm{div}}} = \frac{\dot{N}_{\mathrm{He},\mathrm{div}}}{\dot{N}_{\mathrm{T},\mathrm{div}}} = \frac{ f_{\mathrm{He},\mathrm{div}} \Sigma}{F_{\mathrm{T}}^{\mathrm{in}}},
\label{eq:helium_exhaust}
\end{equation}
which substituted into \Cref{eq:FTdiv1} gives
\begin{equation}
F_{\mathrm{T}}^{\mathrm{div}} = F_{\mathrm{T}}^{\mathrm{in}} \frac{1}{1 +   f_{\mathrm{He},\mathrm{div}} \Sigma \left( 1 /F_{\mathrm{T}}^{\mathrm{in}} - 2 \right)  },
\label{eq:FTdiv0}
\end{equation}
and the deuterium divertor fraction is
\begin{equation}
F_{\mathrm{D}}^{\mathrm{div}} = 1 - \frac{\left(F_{\mathrm{T}}^{\mathrm{in}}\right)^2}{F_{\mathrm{T}}^{\mathrm{in}} + f_{\mathrm{He},\mathrm{div}} \Sigma \left( 1 - 2 F_{\mathrm{T}}^{\mathrm{in}} \right) }.
\label{eq:FDdiva}
\end{equation}

\subsection{Divertor Constraints}

Physically, the divertor tritium and deuterium fractions must satisfy
\begin{equation}
F_{\mathrm{T}}^{\mathrm{div}} \leq 1, \;\;\; F_{\mathrm{D}}^{\mathrm{div}} \leq 1,
\label{eq:divertor_constraint}
\end{equation}
When $F_{\mathrm{T}}^{\mathrm{div}} = 1$, the plasma has burned all of the deuterium, and when $F_{\mathrm{D}}^{\mathrm{div}} = 1$, the plasma has burned all of the tritium. \Cref{eq:divertor_constraint} constrains $F_{\mathrm{T}}^{\mathrm{in}}$ and $f_{\mathrm{He},\mathrm{div}} \Sigma$.

When $F_{\mathrm{T}}^{\mathrm{in}} > 1/2$, requiring that $F_{\mathrm{T}}^{\mathrm{div}} \leq 1$ gives
\begin{equation}
f_{\mathrm{He},\mathrm{div}} \Sigma \leq \frac{F_{\mathrm{T}}^{\mathrm{in}} - 1}{1/F_{\mathrm{T}}^{\mathrm{in}} -2}. 
\label{eq:FTdivlessthanone}
\end{equation}
When $F_{\mathrm{T}}^{\mathrm{in}} < 1/2$, inspection of \Cref{eq:FTdiv0} reveals that $F_{\mathrm{T}}^{\mathrm{div}}$ always satisfies $0 \leq F_{\mathrm{T}}^{\mathrm{div}}\leq F_{\mathrm{T}}^{\mathrm{in}}$ for $ f_{\mathrm{He},\mathrm{div}} \Sigma \geq 0$. Hence, there are no constraints in this region on $ f_{\mathrm{He},\mathrm{div}} \Sigma$.
For $F_{\mathrm{T}}^{\mathrm{in}} = 1/2$, there is no analogous constraint on $f_{\mathrm{He},\mathrm{div}} \Sigma$.

A useful quantity is the tritium enrichment $H_{\mathrm{T}}$ defined in \Cref{eq:HT_expression},
\begin{equation}
H_{\mathrm{T}} = \frac{F_{\mathrm{T}}^{\mathrm{div}}}{F_{\mathrm{T}}^{\mathrm{in}}} = \frac{1}{1 + \frac{P_f}{E} \frac{1}{\dot{N}_{\mathrm{T},\mathrm{div}}} \left( 1 - 2 F_{\mathrm{T}}^{\mathrm{in}} \right)}.
\label{eq:tritium_enrichment}
\end{equation}
Values of $H_{\mathrm{T}} > 1$ are tritium inefficient, indicating a relatively higher deuterium than tritium burn fraction, and $H_{\mathrm{T}} < 1$ is tritium efficient. If $H_{\mathrm{T}} = 0$, this indicates that all of the tritium has been burned. Higher $P_f$ and lower $\dot{N}_{\mathrm{T},\mathrm{div}}$ decrease the enrichment for $F_{\mathrm{T}}^{\mathrm{in}} < 1/2$ and increase the enrichment for $F_{\mathrm{T}}^{\mathrm{in}} > 1/2$. For $F_{\mathrm{T}}^{\mathrm{in}} = 1/2$, the enrichment is always equal to 1. It will also be helpful to write the enrichment as
\begin{equation}
H_{\mathrm{T}} = \frac{1}{1 +   f_{\mathrm{He},\mathrm{div}} \Sigma \left( 1 /F_{\mathrm{T}}^{\mathrm{in}} - 2 \right)  }.
\label{eq:FTdiv_new}
\end{equation}

\begin{figure}[!tb]
    \centering
    \begin{subfigure}[t]{\textwidth}
    \centering
    \includegraphics[width=1.0\textwidth]{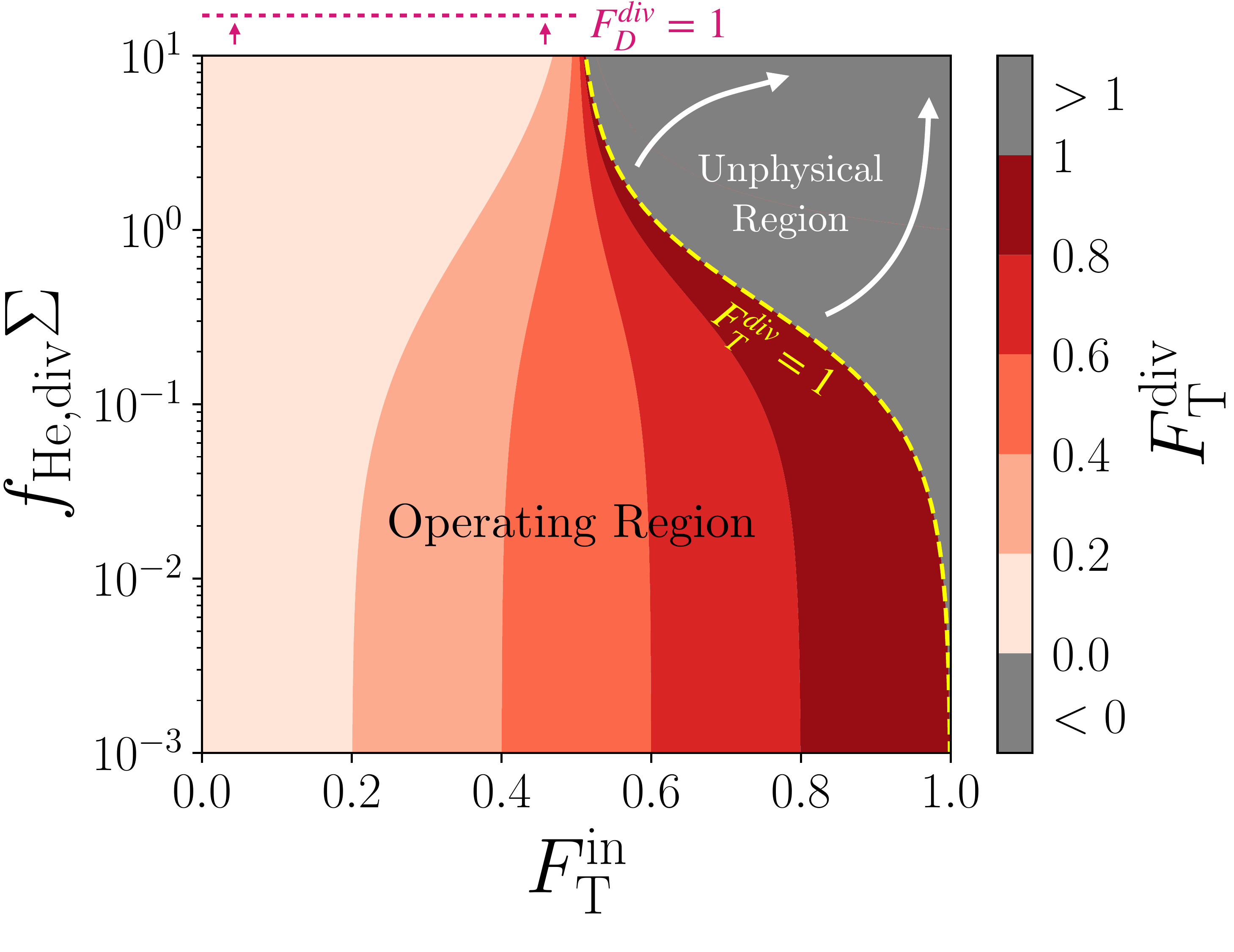}
    \caption{$F_{\mathrm{T}}^{\mathrm{div}}$ (\Cref{eq:FTdiv0}) versus $f_{\mathrm{He},\mathrm{div}} \Sigma$ and $F_{\mathrm{T}}^{\mathrm{in}}$.}
    \end{subfigure}
     ~
    \begin{subfigure}[t]{\textwidth}
    \centering
    \includegraphics[width=1.0\textwidth]{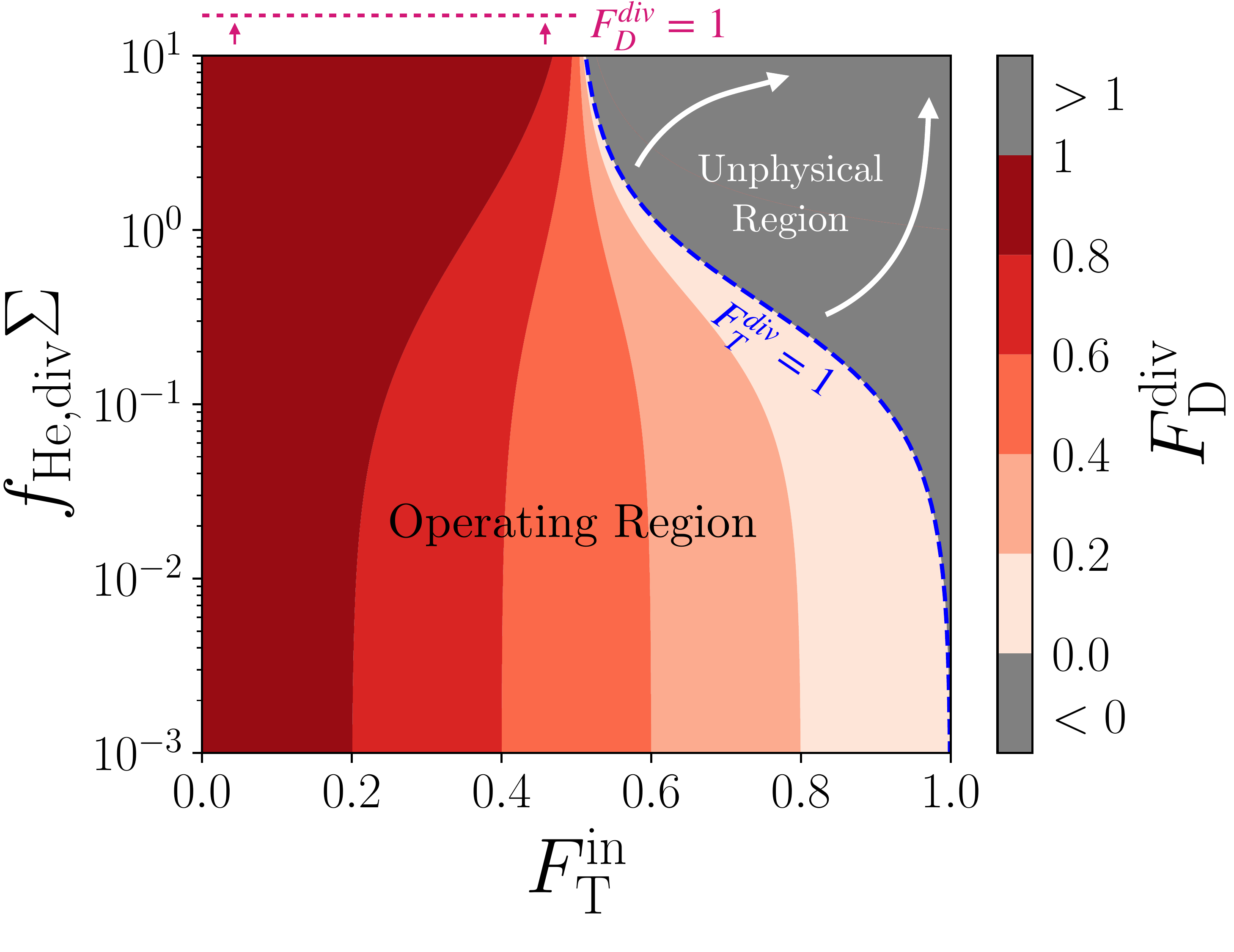}
    \caption{$F_{\mathrm{D}}^{\mathrm{div}}$ (\Cref{eq:FDdiva}) versus $f_{\mathrm{He},\mathrm{div}} \Sigma$ and $F_{\mathrm{T}}^{\mathrm{in}}$.}
    \end{subfigure}
    \caption{Tritium and deuterium divertor fraction ($F_{\mathrm{T}}^{\mathrm{div}}$ (a) and $F_{\mathrm{D}}^{\mathrm{div}}$ (b)) versus $f_{\mathrm{He},\mathrm{div}} \Sigma$ and $F_{\mathrm{T}}^{\mathrm{in}}$.}
    \label{fig:FTD_div}
\end{figure}

\begin{figure}[!tb]
    \centering
    \begin{subfigure}[t]{1.0\textwidth}
    \centering
    \includegraphics[width=1.0\textwidth]{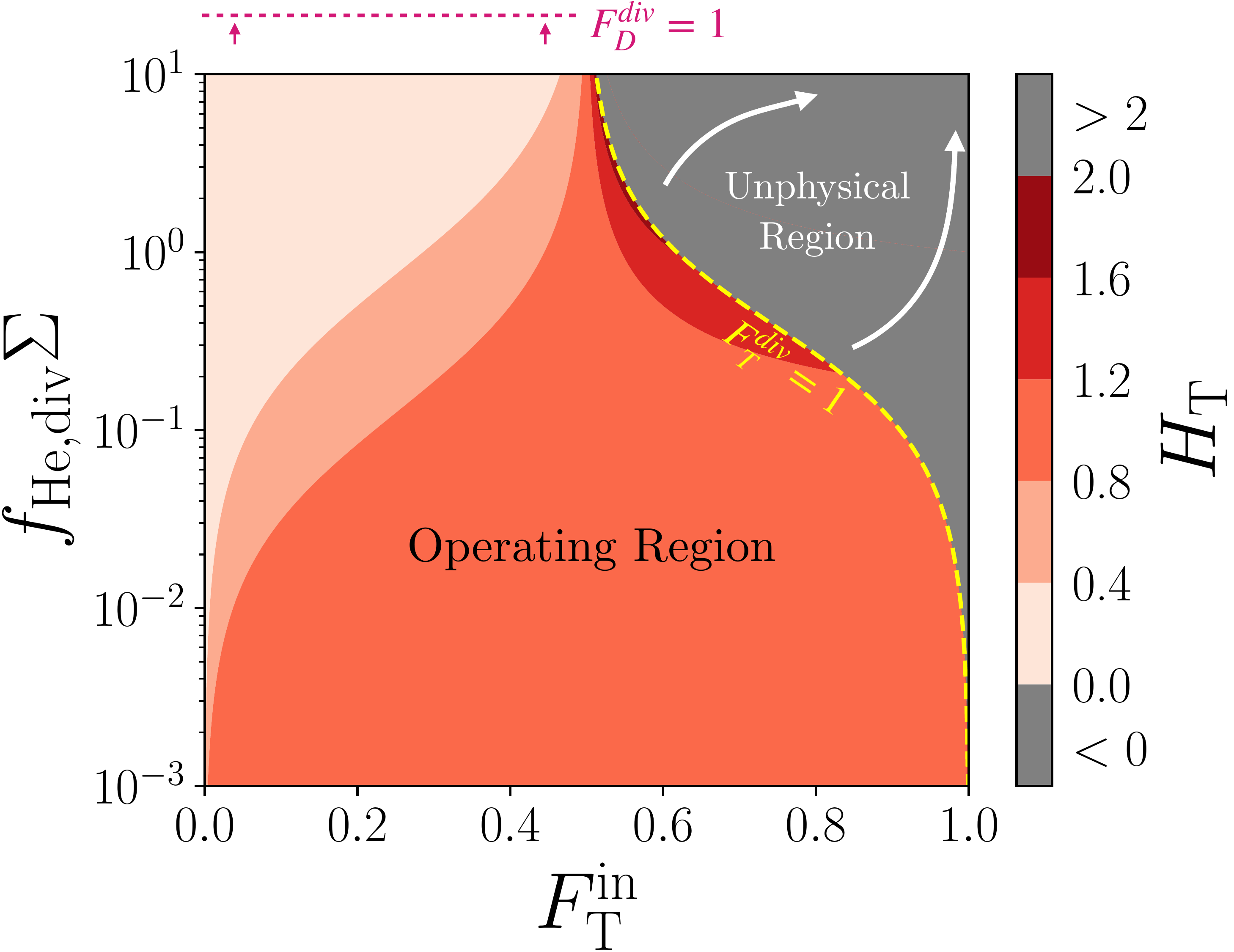}
    \caption{$H_{\mathrm{T}}$ (\Cref{eq:FTdiv_new}) versus $f_{\mathrm{He},\mathrm{div}} \Sigma$ and $F_{\mathrm{T}}^{\mathrm{in}}$.}
    \end{subfigure}
     ~
    \centering
    \begin{subfigure}[t]{1.0\textwidth}
    \centering
    \includegraphics[width=1.0\textwidth]{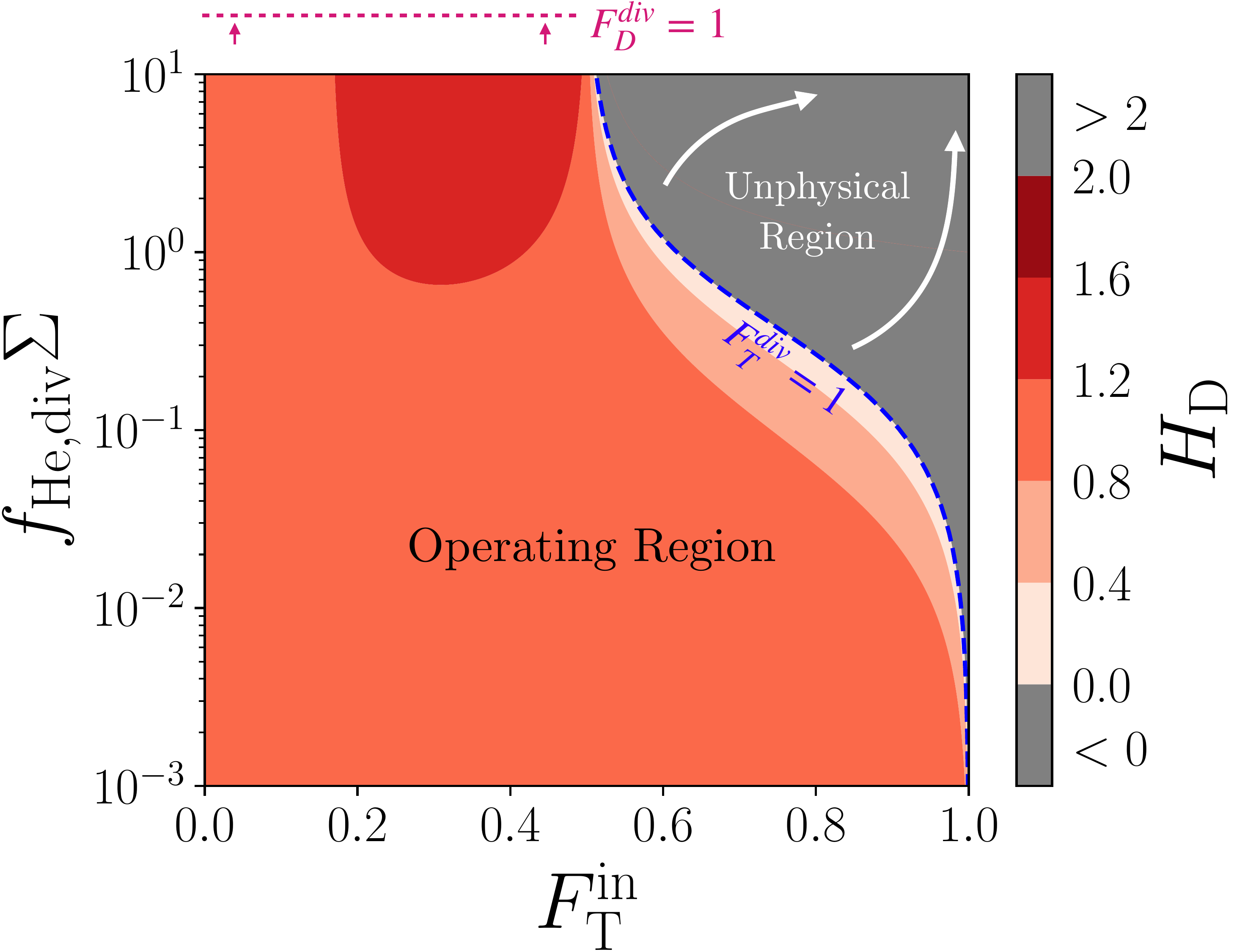}
    \caption{$H_D$ (\Cref{eq:HD}) versus $f_{\mathrm{He},\mathrm{div}} \Sigma$ and $F_{\mathrm{T}}^{\mathrm{in}}$.}
    \end{subfigure}
    \caption{Tritium and deuterium enrichment ($H_{\mathrm{T}}$, (a) and $H_D$, (b)) versus $f_{\mathrm{He},\mathrm{div}} \Sigma$ and $F_{\mathrm{T}}^{\mathrm{in}}$.}
    \label{fig:trtitiumenrichment}
\end{figure}

\subsection{Regions in $f_{\mathrm{He},\mathrm{div}} \Sigma$} \label{sec:op_reg_fHeSigma}

There are distinct regions in $F_{\mathrm{T}}^{\mathrm{in}}$, $f_{\mathrm{He},\mathrm{div}} \Sigma$ space where the divertor pumping rate is consistent with the tritium and deuterium core and divertor. These regions are shown in \Cref{fig:FTD_div,fig:trtitiumenrichment}, where we plot $F_{\mathrm{T}}^{\mathrm{div}}$ and $F_{\mathrm{D}}^{\mathrm{div}}$, and $H_{\mathrm{T}}$ and $H_D$.

\textit{Operating Region.} In this region, both divertor fractions satisfy $0 \leq F_{\mathrm{T}}^{\mathrm{div}} \leq 1$, $0 \leq F_{\mathrm{D}}^{\mathrm{div}} \leq 1$. In this region, both the tritium and deuterium have minimum values of $H_{\mathrm{T}}, H_D = 0.0$ and maximum values of $H_{\mathrm{T}}, H_D = 2.0$.

\textit{Unphysical Region.} -- In this inaccessible region  $F_{\mathrm{T}}^{\mathrm{in}} > 1/2$, the divertor fractions are out of bounds. This corresponds to plasmas that have run out of deuterium or tritium fuel.

There are several things to note about \Cref{fig:trtitiumenrichment}. First, the upper bound on $f_{\mathrm{He},\mathrm{div}} \Sigma$ shown in \Cref{fig:trtitiumenrichment} has a value typically beyond what current technologies can achieve \cite{Whyte2023}, indicating that most values of $F_{\mathrm{T}}^{\mathrm{in}}$ would not be limited by $f_{\mathrm{He},\mathrm{div}} \Sigma$.  \Cref{fig:trtitiumenrichment}(a) and (b) also reveal that the lowest possible enrichment is $H_{\mathrm{T}} = 0.0$, which we prove in the following section. However, obtaining such a low enrichment is generally hard and requires helium pumping beyond current technological capabilities.

\section{Divertor and Fusion Power} \label{sec:divertor_fusionpower}

In this section we constrain the fusion power using the deuterium and tritium divertor pumping. Using $\dot{N}_{\alpha} = \dot{N}_{\mathrm{He},\mathrm{div}}$ we find
\begin{equation}
\frac{\dot{N}_{\mathrm{He},\mathrm{div}}}{\dot{N}_{\mathrm{T},\mathrm{div}}} = \frac{ f_{\mathrm{He},\mathrm{div}} \Sigma}{F_{\mathrm{T}}^{\mathrm{in}}} = \frac{P_f}{E} \frac{1}{\dot{N}_{\mathrm{T},\mathrm{div}}}, 
\label{eq:justsomerelations}
\end{equation}
This gives the required tritium divertor removal rate for a given fusion power, helium pumping, and tritium input fraction,
\begin{equation}
\dot{N}_{\mathrm{T},\mathrm{div}} = F_{\mathrm{T}}^{\mathrm{in}} \frac{P_f}{E} \frac{1}{f_{\mathrm{He},\mathrm{div}} \Sigma}. 
\label{eq:NTdivertor}
\end{equation}
\Cref{eq:NTdivertor} shows that if $P_f$ improves due to plasma conditions, the divertor pumping must also increase. Physically, this corresponds to a divertor pump that must remove sufficient helium ash as more tritium is burnt in the core (to see this, consider that $f_{\mathrm{He},\mathrm{div}} \Sigma$ held fixed).

We can also substitute $P_f / E$ in \Cref{eq:justsomerelations} into \Cref{eq:FTdiv1} to bound the fusion power given the tritium divertor fraction in \Cref{eq:FTdiv0}. When $F_{\mathrm{T}}^{\mathrm{in}} \neq 1/2$, the difference $|F_{\mathrm{T}}^{\mathrm{in}} - F_{\mathrm{T}}^{\mathrm{div}}|$ is increased by the fusion power $P_f$ due to a higher conversion rate of tritium and deuterium into alpha particles. 

To constrain the fusion power with divertor pumping, we again require $F_{\mathrm{T}}^{\mathrm{in}} \leq 1$, $F_{\mathrm{T}}^{\mathrm{div}} \leq 1$. We first consider $1/2 < F_{\mathrm{T}}^{\mathrm{in}} \leq 1$. By requiring $F_{\mathrm{T}}^{\mathrm{div}}\leq 1$, the power is bounded by
\begin{equation}
P_f \leq P_{f,max,1} \equiv (1 - F_{\mathrm{T}}^{\mathrm{in}}) \frac{E \dot{N}_{\mathrm{T},\mathrm{div}}}{ 2 F_{\mathrm{T}}^{\mathrm{in}} - 1}.
\label{eq:Pfmin1}
\end{equation}
Physically, \Cref{eq:Pfmin1} states that the fusion power cannot exceed a maximum bound for a fueling rate $\dot{N}_{\mathrm{T},\mathrm{in}}$ and tritium input fraction $F_{\mathrm{T}}^{\mathrm{in}}$. The closer the tritium injection fraction is to 1, the lower the maximum power can be because the plasma will run out of deuterium at a lower power. Shown in \Cref{fig:trtitiumenrichment}, because the plasma approaches the $F_{\mathrm{D}}^{\mathrm{div}}= 1$ limit (occurring when $H_{\mathrm{T} } \to 0$) when $f_{\mathrm{He,div} }\Sigma \to \infty$ for $F_{\mathrm{T}}^{\mathrm{in}} < 1/2$.

\begin{figure}[]
    \centering
    \includegraphics[width=0.99\textwidth]{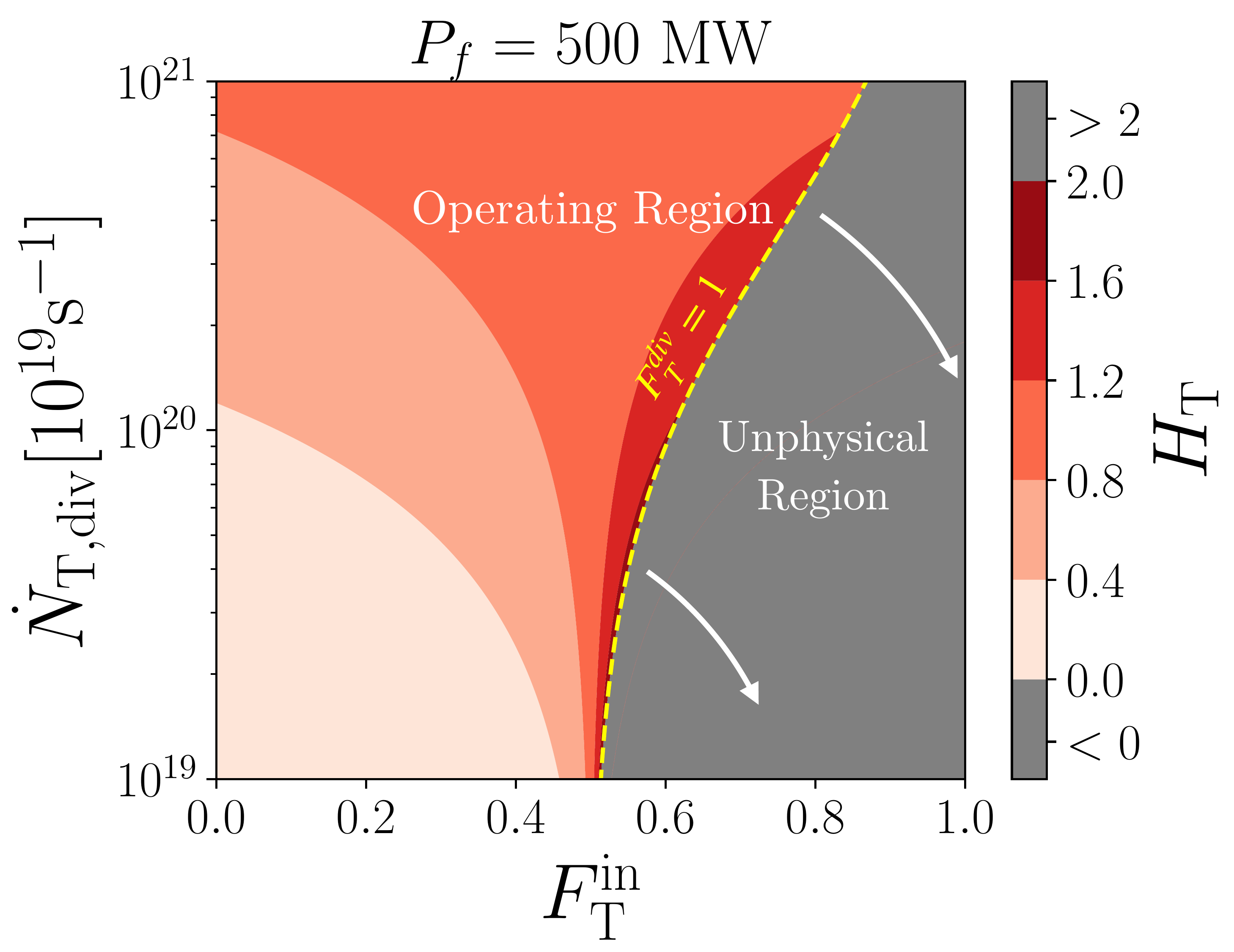}
    \caption{Tritium enrichment $H_{\mathrm{T}}$ versus tritium input fraction $F_{\mathrm{T}}^{\mathrm{in}}$ and divertor removal rate $\dot{N}_{\mathrm{T},\mathrm{div}}$. The five regions A-E are equivalent to those in \Cref{fig:trtitiumenrichment}, but appear different due to the different coordinates used.}
    \label{fig:enrichment_NDot_FTin}
\end{figure}

\subsection{Enrichment Bounds} \label{sec:enrichmentbounds}

In this section, we briefly study the minimum and maximum achievable deuterium and tritium enrichment.

The deuterium enrichment and tritium enrichment are related by,
\begin{equation}
H_D = \frac{1 - F_{\mathrm{T}}^{\mathrm{div}}}{1-F_{\mathrm{T}}^{\mathrm{in}}} = \frac{1/F_{\mathrm{T}}^{\mathrm{in}} - H_{\mathrm{T}}}{1/F_{\mathrm{T}}^{\mathrm{in}} - 1}.
\label{eq:HD}
\end{equation}
The maximum enrichment for $F_{\mathrm{T}}^{\mathrm{in}} < 1/2$ is
\begin{equation}
H_{\mathrm{T},max,1} = 1 - \epsilon,
\end{equation}
occurring when $F_{\mathrm{T}}^{\mathrm{in}} = 1/2 - \epsilon$, where $\epsilon \ll 1$ is a small real number. The minimum enrichment for $F_{\mathrm{T}}^{\mathrm{in}} < 1/2$ is $H_{\mathrm{T}} = 0$; it is theoretically possible for all of the tritium fuel to be burned.

\begin{figure}[tb]
    \centering
    \includegraphics[width=0.99\textwidth]{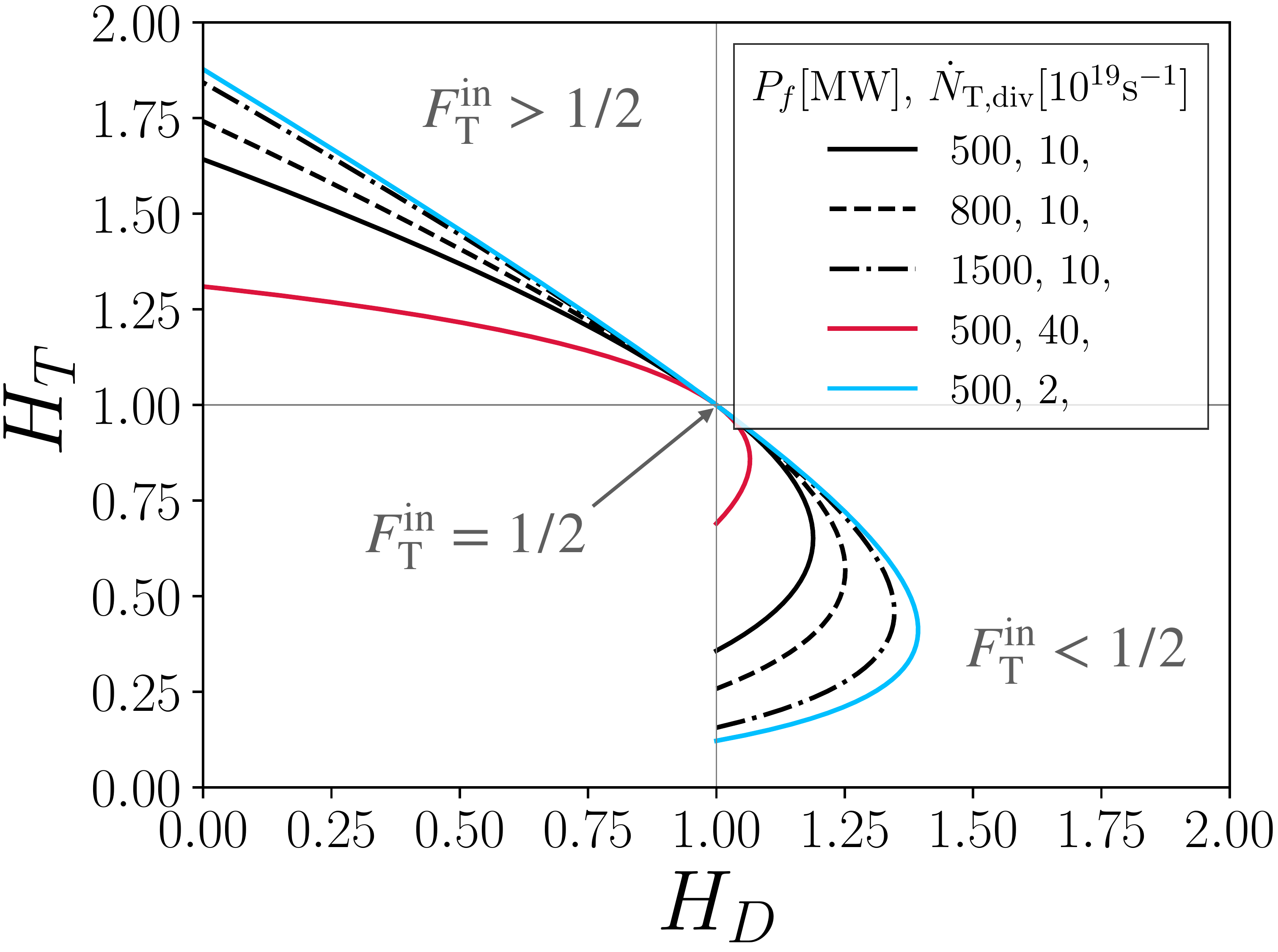}
    \caption{Tritium enrichment ($H_{\mathrm{T}}$) versus deuterium enrichment ($H_D$) for different fusion power and tritium divertor pumping rates.}
    \label{fig:trtitiumenrichment2}
\end{figure}

In \Cref{fig:trtitiumenrichment2}, we plot $H_{\mathrm{T}}$ versus $H_D$ for different $P_f$ and $\dot{N}_{\mathrm{T},\mathrm{div}}$ values. Only for $F_{\mathrm{T}}^{\mathrm{in}}=1/2$ does $H_{\mathrm{T}} = H_D = 1$. The upper left quadrant of \Cref{fig:trtitiumenrichment2} corresponds to $F_{\mathrm{T}}^{\mathrm{in}}>1/2$ and the lower right quadrant corresponds to $F_{\mathrm{T}}^{\mathrm{in}}<1/2$.

For $F_{\mathrm{T}}^{\mathrm{in}} > 1/2$, substituting the condition for $F_{\mathrm{T}}^{\mathrm{div}} = 1$ into the enrichment one finds
\begin{equation}
H_{\mathrm{T},max,2} = \frac{1}{F_{\mathrm{T}}^{\mathrm{in}}},
\label{eq:HTmax2}
\end{equation}
indicating that the maximum enrichment for $F_{\mathrm{T}}^{\mathrm{in}} > 1/2$ occurs for $F_{\mathrm{T}}^{\mathrm{in}} = 1/2 + \epsilon$, $H_{\mathrm{T},max,2} = 2 - \epsilon$ where $\epsilon$ is a small positive number. The minimum enrichment for $F_{\mathrm{T}}^{\mathrm{in}} > 1/2$ is
\begin{equation}
H_{\mathrm{T},min,2} = \frac{1}{1 - \epsilon},
\end{equation}
occurring when $F_{\mathrm{T}}^{\mathrm{in}} = 1/2 + \epsilon$, where $\epsilon \ll 1$ is a small real number.

\section{Deuterium and Tritium Pumping Efficiency} \label{sec:trit_deut_pump_efficiency}

In this section, we briefly describe the effects of different deuterium and tritium removal efficiencies. The total unburned fuel divertor removal rate is
\begin{equation}
\dot{N}_{Q,\mathrm{div}} = n_{\mathrm{T},\mathrm{div}} \left[ S_T + \frac{1 - F_{\mathrm{T}}^{\mathrm{div}}}{F_{\mathrm{T}}^{\mathrm{div}}} S_D \right],
\end{equation}
and the tritium removal rate is $\dot{N}_{\mathrm{T},\mathrm{div}} = n_{\mathrm{T},\mathrm{div}} S_T$. Therefore, $F_{\mathrm{T}}^{\mathrm{div}}$ in \Cref{eq:divertor_ratio} is
\begin{equation}
F_{\mathrm{T}}^{\mathrm{div}} = \left(1 + \frac{1 - F_{\mathrm{T}}^{\mathrm{div}}}{F_{\mathrm{T}}^{\mathrm{div}}} \frac{S_D}{S_T} \right)^{-1},
\end{equation}
where
\begin{equation}
F_{\mathrm{T}}^{\mathrm{div}} \equiv \frac{n_{\mathrm{T},\mathrm{div}}}{n_{Q,\mathrm{div}}} = \left( 1 + \frac{S_T}{S_D} \left( 1 - \frac{1}{F_{\mathrm{T}}^{\mathrm{div}}} \right) \right)^{-1}.
\label{eq:fTdiv}
\end{equation}
Defining the ratio of the helium to tritium pumping efficiency as
\begin{equation}
\Sigma_{\mathrm{ He}, \mathrm{T}} \equiv \frac{S_{\mathrm{He}}}{S_T}, 
\end{equation}
one finds
\begin{equation}
\frac{\dot{N}_{\mathrm{He},\mathrm{div}}}{\dot{N}_{\mathrm{T},\mathrm{div}}} = \frac{f_{\mathrm{He},\mathrm{div}}}{f_{\mathrm{T},\mathrm{div}}} \Sigma_{\mathrm{ He}, \mathrm{T}}.
\label{eq:HeTdivratio}
\end{equation}
Finally, substituting \Cref{eq:HeTdivratio,eq:fTdiv} into \Cref{eq:FTdiv1} and solving for $F_{\mathrm{T}}^{\mathrm{div}}$ gives
\begin{equation}
F_{\mathrm{T}}^{\mathrm{div}} = \frac{F_{\mathrm{T}}^{\mathrm{in}} - f_{\mathrm{He},\mathrm{div}} \left( 1 - 2 F_{\mathrm{T}}^{\mathrm{in}} \right) \Sigma_{\mathrm{ He}, \mathrm{T}} S_T/S_D }{1 + f_{\mathrm{He},\mathrm{div}} \left( 1 - 2 F_{\mathrm{T}}^{\mathrm{in}} \right) (S_D - S_T) \Sigma_{\mathrm{ He}, \mathrm{T}}}.
\label{eq:diffFtdiv_pumping}
\end{equation}
When $F_{\mathrm{T}}^{\mathrm{in}} = 1/2$, this reduces to $F_{\mathrm{T}}^{\mathrm{div}} = 1/2$.

\section{ARC-like Device Workflow} \label{sec:workflow}

In this section, we outline the steps to calculate the parameters in \Cref{tab:tab1}. The model inputs are $\mathcal{N} A_J$, $\Sigma = 1.0$, $\eta_{\mathrm{He}} = 0.63,$, $\tau_{\mathrm{IFC}} = 4$h, $\tau_{\mathrm{OFC}} = 24$h, and TBR = 1.08.

For the Base Case, we set the fusion power $P_f = 482$ MW, the nominal plasma gain $Q_0 = 10.0$, and the polarization multiplier $\mathcal{N} A_J = 1.0$. For these parameters, we aim to maximize the TBE by rearranging \Cref{eq:pDeltaform2_HT} for TBE and assuming that $p_{\Delta} = 0.95$. We then find the $f_{\mathrm{T}}^{\mathrm{co}}$ value that gives the highest TBE, which is equivalent to allowing $f_{\mathrm{He},\mathrm{div}}$ to vary (see \Cref{eq:fHediv}). Using the procedure outlined in \Cref{sec:min_startup_inventory}, the minimum tritium startup $I_{\mathrm{startup,min}}$ is calculated. $f_{\mathrm{He},\mathrm{div}}$ is calculated using \Cref{eq:fHediv}.

Cases A and B in \Cref{tab:tab1} follow a similar procedure to the Base Case. The objective is to maximize the TBE at fixed power. We fix the total fusion power $P_f = 482$ MW but choose higher values of $\mathcal{N} A_J$. Again, using \Cref{eq:pDeltaform2_HT}, we find the $f_{\mathrm{T}}^{\mathrm{co}}$ value that gives the highest TBE. Because $\mathcal{N} A_J$ is higher, a higher $f_{\mathrm{T}}^{\mathrm{co}}$ value can be achieved at fixed $P_f$. To find the new plasma gain $Q$, we assume that $n_e \tau_E$, and $T$ in \Cref{eq:C} are constant. Thus, one can write $1 + Q_0 / 5 = k C_0$, where $C_0$ is the nominal multiplier to keep $Q$ constant in \Cref{eq:C} and $k$ is a constant of value
\begin{equation}
k = \frac{1 + 5/Q_0}{C_0}. 
\end{equation}
This allows one to write an equation for $Q$,
\begin{equation}
Q = \frac{5}{ k C - 1}.
\end{equation}
The plasma ignition criterion is
\begin{equation}
k C < 1,
\end{equation}
which some of the ARC-like cases are shown to satisfy in \Cref{fig:Qplasma}.

For cases C and D, the objective is to find the maximum fusion power for a given $\mathcal{N} A_J$ using the TBE value for the base case. For these cases with two different $\mathcal{N} A_J$ values, \Cref{eq:pDeltaform2_HT} is used to find the $f_{\mathrm{T}}^{\mathrm{co}}$ value that gives the highest fusion power. Following this, $I_{\mathrm{startup,min}}$, $Q$, and $f_{\mathrm{He},\mathrm{div}}$ are found.

For cases E and F, the objective is to find the maximum TBE for a given power degradation $p_{\Delta} = 0.50, 0.25$ respectively. Again, using \Cref{eq:pDeltaform2_HT} the $f_{\mathrm{T}}^{\mathrm{co}}$ value that maximizes the TBE is found and $I_{\mathrm{startup,min}}$, $Q$, and $f_{\mathrm{He},\mathrm{div}}$ are found.

Cases G and H are illustrative examples with very high polarization multiplier $\mathcal{N} A_J = 1.9$. Case G maximizes the TBE at fixed power in the same way as Cases A and B. Case H maximizes the fusion power for a given $\mathcal{N} A_J$ using the TBE value for the base case. This is the same workflow as Cases C and D.

For cases I and J that simultaneously increase the fusion power and the TBE, we follow the same workflow as C and D, only in this case we increase the TBE to TBE = 0.10.

\section{Limitations} \label{sec:limitations}

\begin{figure*}[!tb]
    \centering
    \begin{subfigure}[t]{0.48\textwidth}
    \centering
    \includegraphics[width=1.0\textwidth]{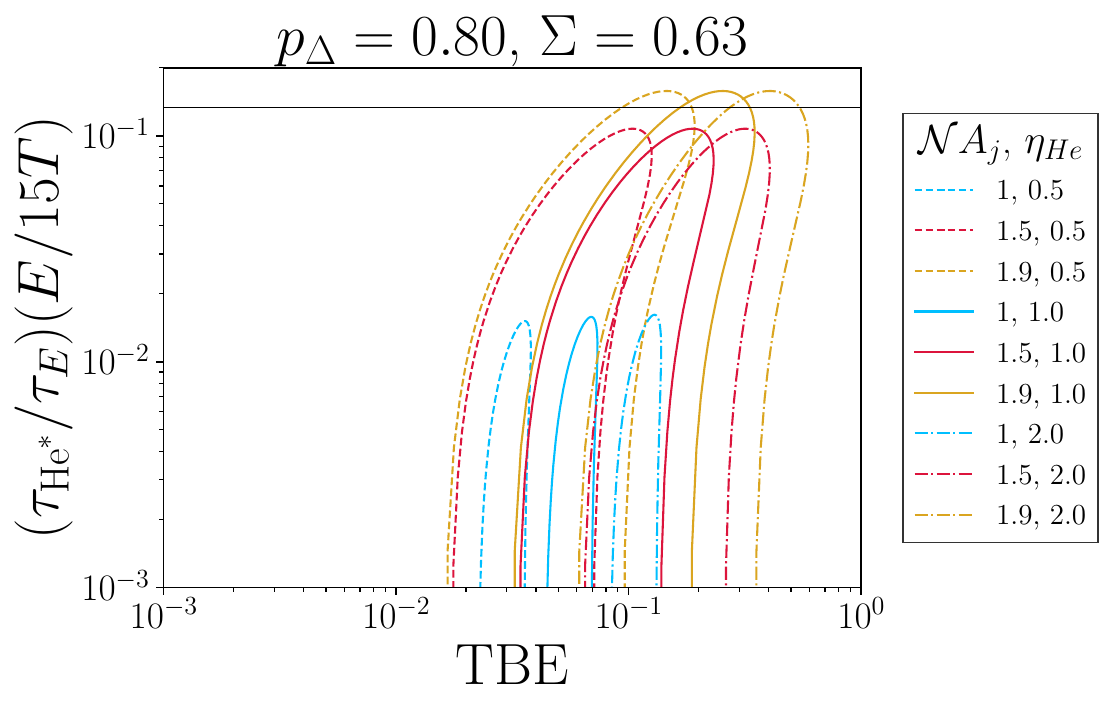}
    \caption{$p_{\Delta} = 0.80$.}
    \end{subfigure}
     ~
    \centering
    \begin{subfigure}[t]{0.48\textwidth}
    \centering
    \includegraphics[width=1.0\textwidth]{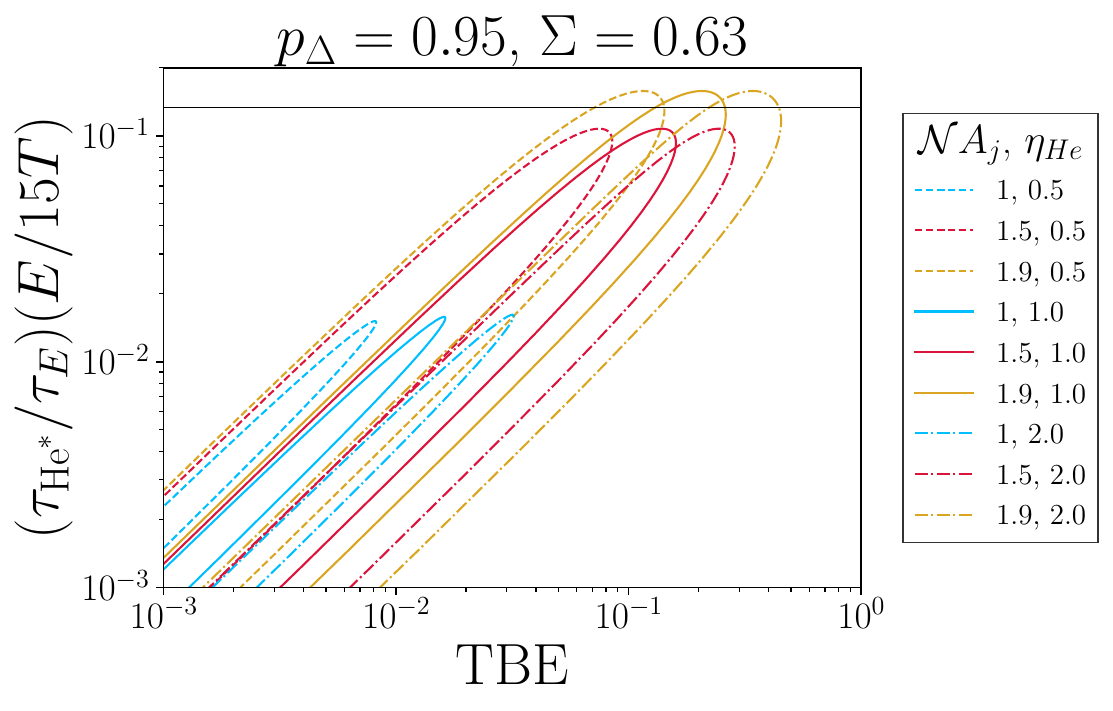}
    \caption{$p_{\Delta} = 0.95$.}
    \end{subfigure}
     ~
    \centering
    \begin{subfigure}[t]{0.48\textwidth}
    \centering
    \includegraphics[width=1.0\textwidth]{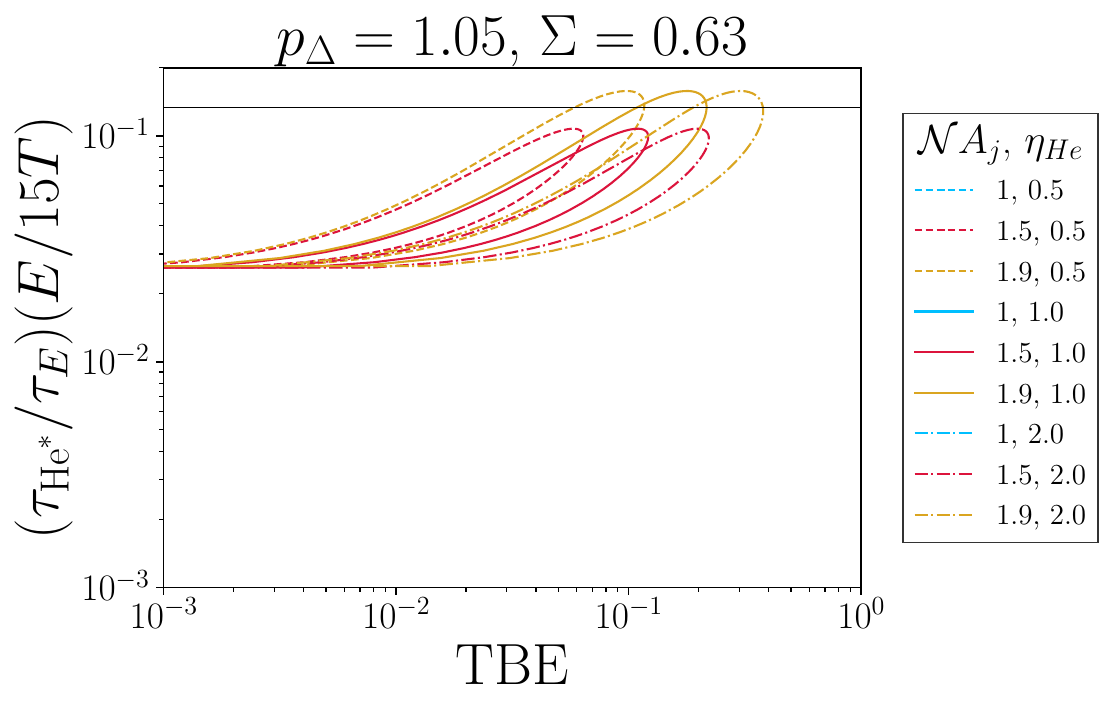}
    \caption{$p_{\Delta} = 1.05$.}
    \end{subfigure}
     ~
    \centering
    \begin{subfigure}[t]{0.48\textwidth}
    \centering
    \includegraphics[width=1.0\textwidth]{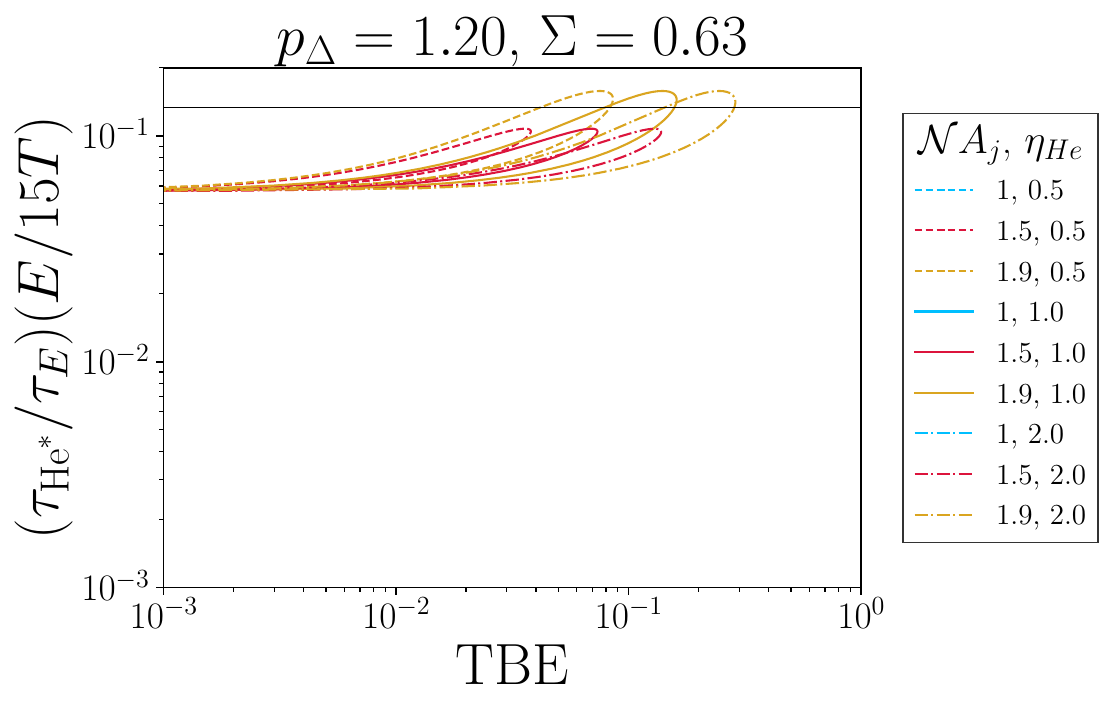}
    \caption{$p_{\Delta} = 1.2$.}
    \end{subfigure}
    \caption{$Y = (\tau_{\mathrm{He}}^*/\tau_E) (15 T/E)$ (see \Cref{eq:tauHe}) versus tritium burn efficiency (TBE) for four power degradation/enhancement factors $p_{\Delta}$. Each subplot has nine curves with three $\mathcal{N} A_J$ values and three $\eta_{\mathrm{He}}$ values. Each subplot is at a constant fusion power. Horizontal black line at $Y = 10/75$ indicates a rough upper bound for $(\tau_{\mathrm{He}}^*/\tau_E) (15 T/E)$ assuming that $\tau_{\mathrm{He}}^*/\tau_E$ can have a maximum value of 10 and $(15 T/E) = 75$, corresponding to a plasma temperature $T = 15.6$keV. We use a tritium enrichment value of $H_{\mathrm{T}} = 1.0$. For each curve, arclength is parameterized non-uniformly by $f_\mathrm{T}^{\mathrm{co} }$.}
    \label{fig:Y_versusTBE_degradation}
\end{figure*}

In this section, we briefly describe some of the limitations of the analysis in this work.

\subsection{Constant T Assumption} \label{sec:constantT}

When calculating the fusion gain multiplication factor $C$ in \Cref{eq:fusiongainform1}, we have assumed that $T$ is constant. This will cause errors when  $\mathcal{N} \neq 1$ because \Cref{eq:C} is derived assuming constant $T$. Physically, $\mathcal{N} \neq 1$ describes the effects of alpha heating changing the temperature and hence the fusion reactivity. To incorporate effects of changing temperature one could use that for $T$ between 10 and 20 keV that $\langle v \overline{ \sigma} \rangle \sim T^2$, and hence use $T \sim \sqrt{\mathcal{N}}$. However, given that the largest value of $\mathcal{N}$ in this work is bounded by $\mathcal{N} < 1.9 / A_J$ (since we consider $\mathcal{N} A_J \leq 1.9$), if we set $A_J = 1.5$, the maximum value of $\mathcal{N}$ is $\mathcal{N} \simeq 1.27$, and thus $\sqrt{\mathcal{N}} \lesssim 1.13$. Such an effect is beyond the scope of this work but may be important. 

\subsection{Constant TBR Assumption} \label{sec:constantTBR}

When calculating the tritium startup inventory in \Cref{sec:min_startup_inventory,sec:casestudy}, we assume that the tritium breeding ratio (TBR) is independent of tritium burn efficiency (TBE). This is despite work showing that the required TBR, $\mathrm{TBR_r}$ is reduced most by the TBE and the tritium doubling time
\cite{Meschini2023}. As a quick sanity check, we calculate $I_{\mathrm{startup,min}}$ for case J of the ARC-like power plant in \Cref{sec:casestudy}. For the nominal TBR value we used, TBR = 1.08, $I_{\mathrm{startup,min}} = 0.096$ kg. Using TBR = 1.04 and keeping everything else constant, the startup requirement increases by 2\% to $I_{\mathrm{startup,min}} = 0.098$ kg. Increasing the TBR significantly to TBR = 1.15, the startup requirement decreases by 6\% to $I_{\mathrm{startup,min}} = 0.090$ kg. Within the model used in this work, keeping TBR constant has a relatively small effect on $I_{\mathrm{startup,min}}$ compared with other parameters, most notably the tritium fraction $f_{\mathrm{T}}^{\mathrm{co}}$ and the spin-polarization multiplier $\mathcal{N} A_J$.

\subsection{Other Limitations}

There are other limitations inherent in our approach here: we have no radial profiles, no impurities, no alpha particle deposition model, no fueling model. Higher fidelity modeling is needed to study these effects.

\section{Helium Particle Confinement} \label{sec:heliumparticleconfinement}

\begin{figure}[!tb]
    \centering
    \begin{subfigure}[t]{0.95\textwidth}
    \centering
    \includegraphics[width=1.0\textwidth]{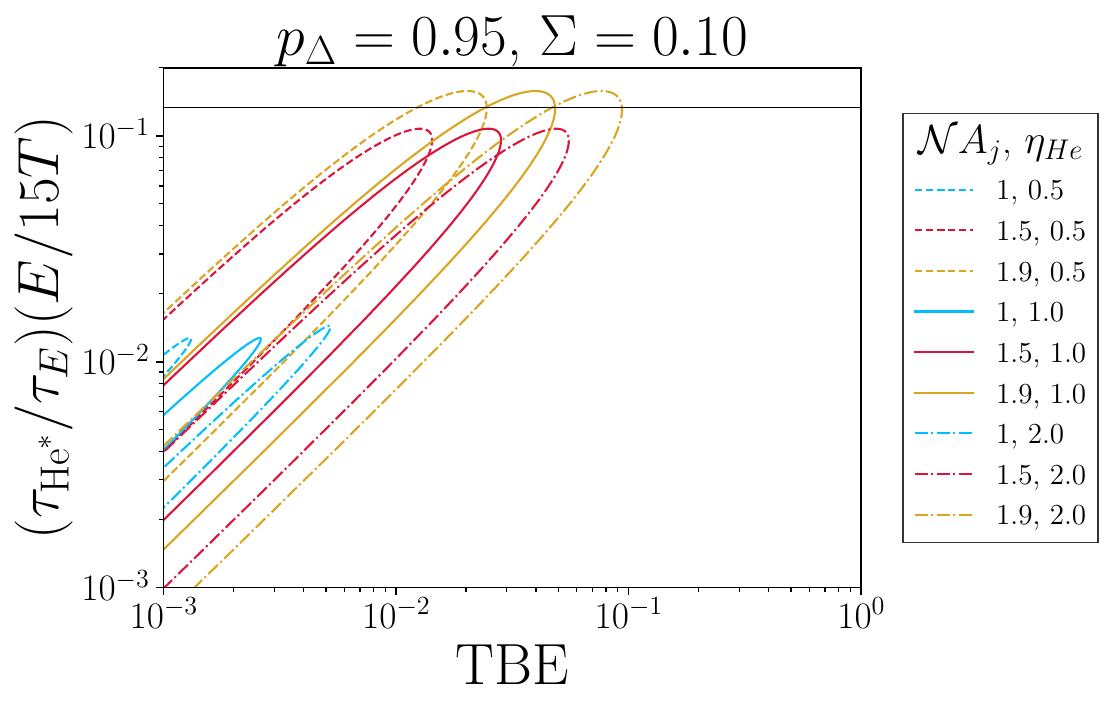}
    \caption{$\Sigma = 0.10$.}
    \end{subfigure}
     ~
    \centering
    \begin{subfigure}[t]{0.95\textwidth}
    \centering
    \includegraphics[width=1.0\textwidth]{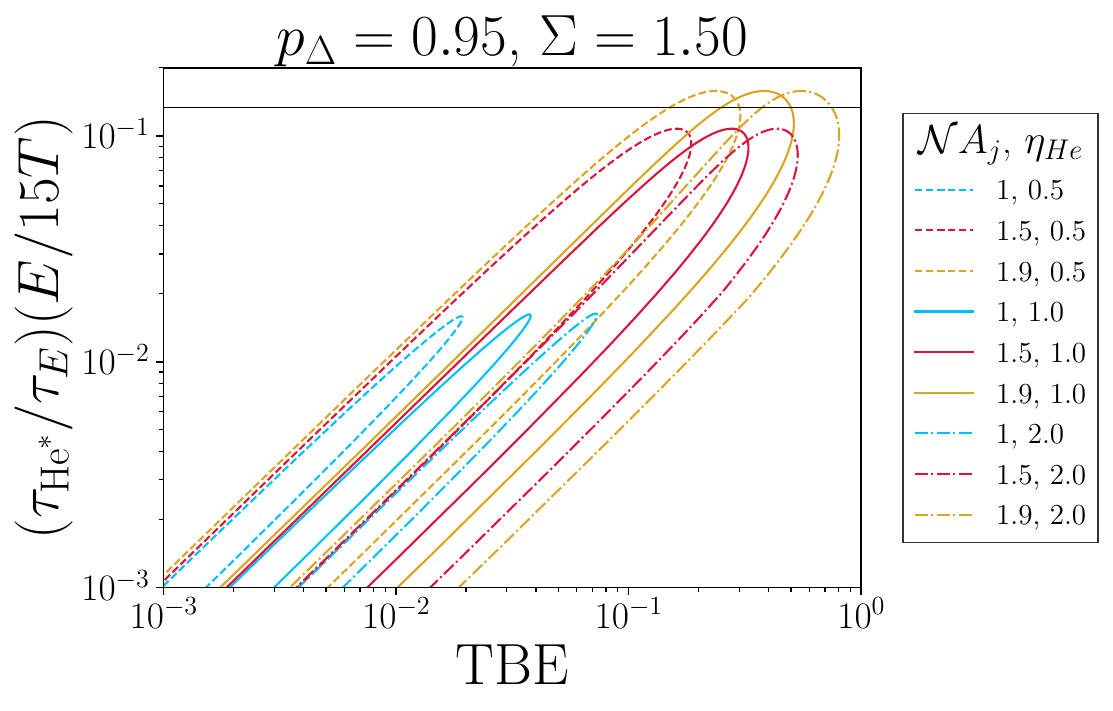}
    \caption{$\Sigma = 1.50$.}
    \end{subfigure}
    \caption{$Y = (\tau_{\mathrm{He}}^*/\tau_E) (15 T/E)$ (see \Cref{eq:tauHe}) versus tritium burn efficiency (TBE) for two helium pumping efficiency ratios $\Sigma$. Each subplot has nine curves with three $\mathcal{N} A_J$ values and three $\eta_{\mathrm{He}}$ values. Each subplot is at a constant fusion power. Horizontal black line at $Y = 10/75$ indicates a rough upper bound for $(\tau_{\mathrm{He}}^*/\tau_E) (15 T/E)$ assuming that $\tau_{\mathrm{He}}^*/\tau_E$ can have a maximum value of 10 and $(15 T/E) = 75$, corresponding to a plasma temperature $T = 15.6$keV. We use a tritium enrichment value of $H_{\mathrm{T}} = 1.0$. For each curve, arclength is parameterized non-uniformly by $f_\mathrm{T}^{\mathrm{co} }$.}
    \label{fig:Y_versusTBE_Sigma}
\end{figure}

In this section, we calculate the helium confinement time ratio $\tau_{\mathrm{He}}^*/\tau_E$, following the method of section 4 in \cite{Whyte2023}. A helium exhaust study of JT-60U found that $\tau_{\mathrm{He}}^*/\tau_E \lesssim 10$ \cite{Sakasai1999} -- a plasma operating regime that significantly exceeding this bound would require justification. We wish to test whether $\tau_{\mathrm{He}}^*/\tau_E$ exceeds  $\tau_{\mathrm{He}}^*/\tau_E \simeq 10$ for very high TBE values that are possible with spin polarization (see \Cref{fig:TBE3p6,fig:TBEfull}). Defining the helium particle confinement time,
\begin{equation}
\tau_{\mathrm{He} }^* \equiv \frac{n_{\alpha}}{\dot{n}_{\alpha}},
\label{eq:tauHestardef}
\end{equation}
where the alpha particle birth rate density is
\begin{equation}
\dot{n}_{\alpha} \equiv f_{\mathrm{T}}^{\mathrm{co}} (1-f_{\mathrm{T}}^{\mathrm{co}}) (1-2 f_{\mathrm{dil}})^2 \mathcal{N} A_J n_e^2 \langle v \overline{ \sigma} \rangle.
\label{eq:ndotalpha}
\end{equation}
Substituting $1-2 f_{\mathrm{dil}}$ from \Cref{eq:ndotalpha} into \Cref{eq:fusiongainform1} and using \Cref{eq:tauHestardef}, we find that $\tau_{\mathrm{He}}^*/\tau_E$ is
\begin{equation}
Y \equiv \frac{\tau_{\mathrm{He}}^*}{\tau_E} \frac{15 T}{E} = M \left(1 + 5/Q\right),
\label{eq:tauHe}
\end{equation}
where $M$ is defined as
\begin{equation}
M \equiv \frac{ f_{\mathrm{dil} }}{ 1 + f_{\mathrm{dil}} \left( \frac{1}{2} \frac{\langle T_{\alpha} \rangle}{T} -1 \right) }.
\label{eq:M}
\end{equation}
In \Cref{fig:Y_versusTBE_degradation}, we plot $Y$ versus TBE for a range of $\eta_{\mathrm{He} }$, $\mathcal{N} A_J$, and $p_{\Delta}$ values. While the power is fixed in each subfigure, the plasma gain is not; $Q$ is self-consistently calculated by assuming the nominal $Q=20$ plasma has $\Sigma = 0.63$, $\eta_{\mathrm{He} } = 1.0$, TBE = 0.016, and $f_{\mathrm{T} }^{\mathrm{co} } = 0.49$, just as assumed for the ARC-like device in \Cref{sec:casestudy}. We also assumed $\langle T_{\alpha} \rangle = T$ in \Cref{eq:M}. For each subfigure in \Cref{fig:Y_versusTBE_Sigma}, we plot $Y$ for a different $\Sigma$ value, demonstrating how $\Sigma$ improves the TBE at fixed $Y$.

In each \Cref{fig:Y_versusTBE_degradation,fig:Y_versusTBE_Sigma} subfigure, the horizontal black line at $Y = 10/75$ indicates a rough upper bound for $\tau_{\mathrm{He}}^*/\tau_E$ assuming that $\tau_{\mathrm{He}}^*/\tau_E$ can have a maximum value of 10 and $15 T/E = 75$, corresponding to a temperature of $T = 15.6$ keV. For very high polarizations, there are a few TBE values that correspond to $\tau_{\mathrm{He}}^*/\tau_E > 10$, but only by a very small factor of at most $\sim$ 20\% for $\mathcal{N} A_J = 1.9$. We therefore conclude that for spin-polarized plasmas with $\mathcal{N} A_J \lesssim 2$, there are some TBE values satisfying $\tau_{\mathrm{He}}^*/\tau_E \approx 10$. While these higher values of $\tau_{\mathrm{He}}^*/\tau_E$ may be challenging to achieve, they have been demonstrated previously \cite{Sakasai1999}, and this challenge appears relatively small compared with others in building and operating power plants that use spin-polarized fuels.

\section{Intuition for Burn Efficiency Across Polarization} \label{sec:intuition}

\begin{figure*}[!tb]
    \centering
    \begin{subfigure}[t]{0.8\textwidth}
    \centering
    \includegraphics[width=1.0\textwidth]{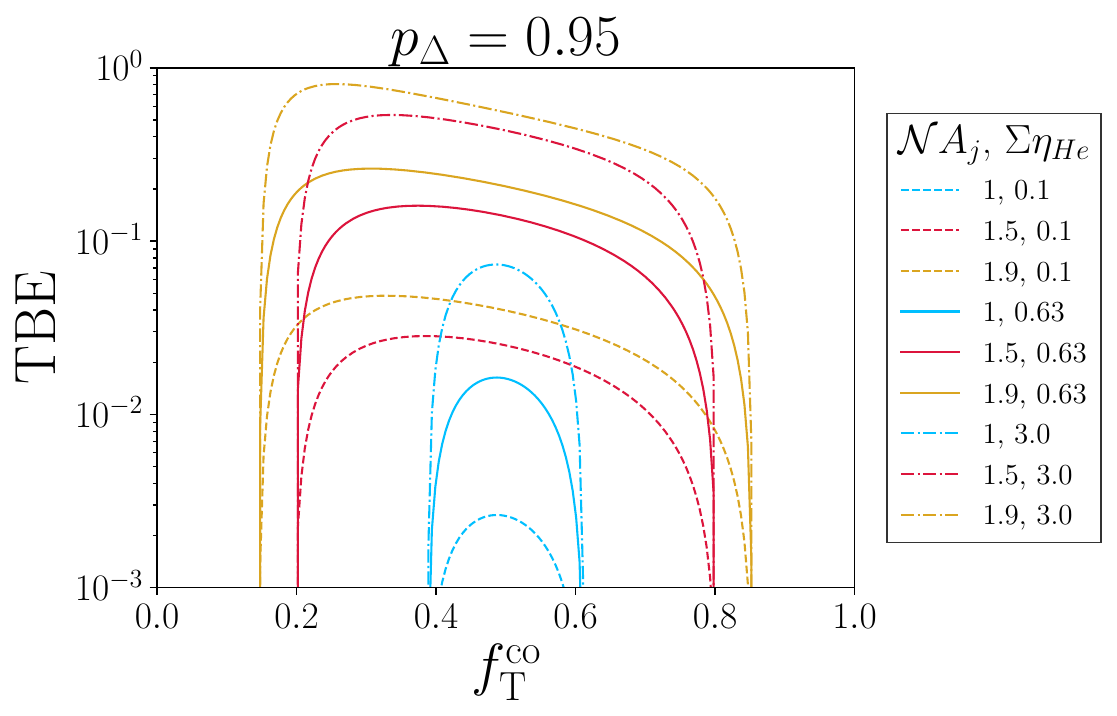}
    \caption{TBE.}
    \end{subfigure}
     ~
    \centering
    \begin{subfigure}[t]{0.8\textwidth}
    \centering
    \includegraphics[width=1.0\textwidth]{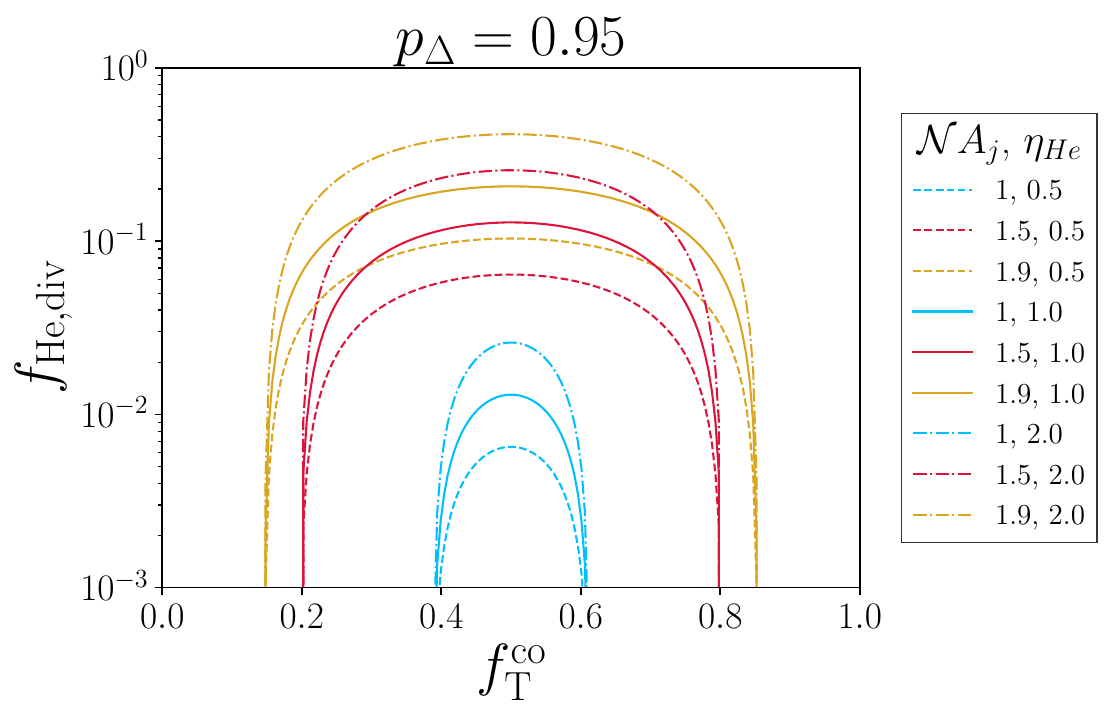}
    \caption{$f_{\mathrm{He,div}}$.}
    \end{subfigure}
    \caption{Tritium burn efficiency (TBE) (\Cref{eq:TBEfull}) (a) and helium divertor fraction $f_{\mathrm{He,div}}$ (\Cref{eq:fHediv_new}) (b) versus tritium core fraction $f_{\mathrm{T}}^{\mathrm{co}}$ for different spin-polarization multiplier $\mathcal{N} A_J$ and $\Sigma \eta_{\mathrm{He} }$ values. We use a tritium enrichment value of $H_{\mathrm{T}} = 1.0$.}
    \label{fig:TBE_fHediv_intuition}
\end{figure*}

In this section, we give some intuition for how the tritium burn efficiency (TBE) can be increased strongly using spin polarization without a corresponding power decrease.

The reason that the TBE increases by much more than suggested in \Cref{fig:TBE1} is because at lower tritium fraction and higher $\mathcal{N} A_J$, not only does $f_{\mathrm{T}}^{\mathrm{co}}$ decrease but the total D-T fuel injected decreases, even though the plasma electron density remains constant. To see this, consider the TBE in \Cref{eq:TBEfirst},
\begin{equation}
\mathrm{TBE} \equiv \frac{\dot{N}_{\mathrm{T},\mathrm{burn}}}{\dot{N}_{\mathrm{T},\mathrm{in}}} = \frac{\overline{P} }{F_{\mathrm{T}}^{\mathrm{in}} \dot{N}_{\mathrm{Q},\mathrm{in}}},
\label{eq:TBEintution}
\end{equation}
where we used $\overline{P} = \dot{N}_{\mathrm{T},\mathrm{burn}} =  P_f / E$ and $\dot{N}_{\mathrm{T},\mathrm{in}} = F_{\mathrm{T}}^{\mathrm{in}} \dot{N}_{\mathrm{Q},\mathrm{in}}$. Next, we obtain an expression for $\dot{N}_{\mathrm{Q},\mathrm{in}}$, first using particle conservation from \Cref{eq:particle_conservation},
\begin{equation}
\dot{N}_{\mathrm{Q},\mathrm{in}} = 2 \overline{ P} + \dot{N}_{\mathrm{Q},\mathrm{div}}.
\label{eq:NQdotin}
\end{equation}
Next, we find an expression for $\dot{N}_{\mathrm{Q},\mathrm{div}}$ using \Cref{eq:ashtofuel}, giving
\begin{equation}
\dot{N}_{\mathrm{Q},\mathrm{div}} = \frac{\overline{ P}}{f_{\mathrm{He,div}  } \Sigma}.
\label{eq:NQdiv}
\end{equation}
We find a new expression for $f_{\mathrm{He,div}}$ by solving \Cref{eq:pDeltaform1},
\begin{equation}
f_{\mathrm{He,div}} = \frac{\eta_{\mathrm{He} }}{2} \left( 2 \frac{\sqrt{f_{\mathrm{T}}^{\mathrm{co}} (1-f_{\mathrm{T}}^{\mathrm{co}}) \mathcal{N} A_J )} }{\sqrt{p_{\Delta}} } - 1 \right).
\label{eq:fHediv_new}
\end{equation}
Therefore, $\dot{N}_{\mathrm{Q},\mathrm{div}}$ is
\begin{equation}
\dot{N}_{\mathrm{Q},\mathrm{div}} = \frac{2\overline{ P}}{\Sigma \eta_{\mathrm{He} } } \left( 2 \frac{\sqrt{f_{\mathrm{T}}^{\mathrm{co}} (1-f_{\mathrm{T}}^{\mathrm{co}}) \mathcal{N} A_J )} }{\sqrt{p_{\Delta}} } - 1 \right)^{-1},
\end{equation}
which substituted into \Cref{eq:NQdotin} for $\dot{N}_{\mathrm{Q},\mathrm{in}}$ gives,
\begin{equation}
\begin{aligned}
& \dot{N}_{\mathrm{Q},\mathrm{in}} = \\
&2 \overline{ P} \left( 1 + \frac{1}{\Sigma \eta_{\mathrm{He} }} \left( 2 \frac{\sqrt{f_{\mathrm{T}}^{\mathrm{co}} (1-f_{\mathrm{T}}^{\mathrm{co}}) \mathcal{N} A_J )} }{\sqrt{p_{\Delta}} } - 1 \right)^{-1}  \right).
\end{aligned}
\end{equation}
Finally, using $F_{\mathrm{T} }^{\mathrm{in}} = 2 f_{\mathrm{T}}^{\mathrm{co}} / (1+H_{\mathrm{T}} )$ to the find the TBE,
\begin{equation}
\mathrm{TBE} = \frac{1}{4 f_{\mathrm{T}}^{\mathrm{co}}} \frac{1 + H_{\mathrm{T}}}{ 1 + \frac{1}{\Sigma \eta_{\mathrm{He} }} \left( 2 \frac{\sqrt{f_{\mathrm{T}}^{\mathrm{co}} (1-f_{\mathrm{T}}^{\mathrm{co}}) \mathcal{N} A_J )} }{\sqrt{p_{\Delta}} } - 1 \right)^{-1}  }.
\label{eq:TBEfull}
\end{equation}
We are particularly interested in how the TBE varies with $\mathcal{N} A_J$. We plot solutions to \Cref{eq:TBEfull} for $p_{\Delta} = 0.95$ for different values of $\mathcal{N} A_J$ and $\Sigma \eta_{\mathrm{He}}$ in \Cref{fig:TBE_fHediv_intuition}(a). The spin-polarization multiplier increases the TBE nonlinearly. Curiously, at very high $\Sigma \eta_{\mathrm{He}}$ values, here $\Sigma \eta_{\mathrm{He}} = 3.0$, and larger $\mathcal{N} A_J $ values, the derivative $d \mathrm{TBE}/ d f_{\mathrm{T}}^{\mathrm{co}}$ can become extremely large near the maximum TBE value. Under such conditions, operating a fusion plant near the maximum TBE value would require precise $f_{\mathrm{T}}^{\mathrm{co}}$ control such that the TBE does not decrease rapidly. Furthermore, because $f_{\mathrm{T}}^{\mathrm{co}}$ will likely have some radial dependence, the expected window of $f_{\mathrm{T}}^{\mathrm{co}}$ values across high power density regions would benefit from falling in high TBE regions.

In this exercise, it is important to recognize what is being held fixed and what is varying: in deriving the TBE in \Cref{eq:TBEfull}, we eliminated $f_{\mathrm{He,div}}$, allowing it to vary. Shown in \Cref{fig:TBE_fHediv_intuition}(b), plasmas with higher $f_{\mathrm{He,div}}$ have much higher TBE. While operating at higher $f_{\mathrm{He,div}}$ may be concerning, it is important to note that the helium divertor removal rate $\dot{N}_{\mathrm{He,div} }$ is fixed because the fusion power is constant. Therefore, the increase in $f_{\mathrm{He,div}}$ comes from the divertor hydrogenic divertor density $n_{\mathrm{Q,div} }$ falling, not from the helium divertor density increasing. Therefore, the power exhaust could be of greater concern in high TBE, high $\mathcal{N} A_J$ plasmas due to the low hydrogen divertor density \cite{Wischmeier2015,Asakura2017,Kuang2020}.

A consequence of allowing $f_{\mathrm{He,div}}$ to vary is that the helium particle confinement time also increases. Fortunately, it appears that the increase is not prohibitive according to bounds placed by current experiments. See \Cref{sec:heliumparticleconfinement} for a discussion.

\section{Ignition Stability} \label{sec:ignitionstab}

Operating with spin-polarized fuel could improve the margin of safety between stable and unstable ignited equilibria. Operating with $f_{\mathrm{T}}^{\mathrm{co}} \neq 1/2$ could also form a passive safety mechanism that helps prevent thermal runaway. In this appendix, we perform a simple heuristic analysis. A thorough analysis of the effect of spin polarization on ignition access can be found in \cite{Mitarai1992}.

\subsection{Ignition Stability}

\begin{figure}[!tb]
    \centering
    \begin{subfigure}[t]{0.98\textwidth}
    \centering
    \includegraphics[width=1.0\textwidth]{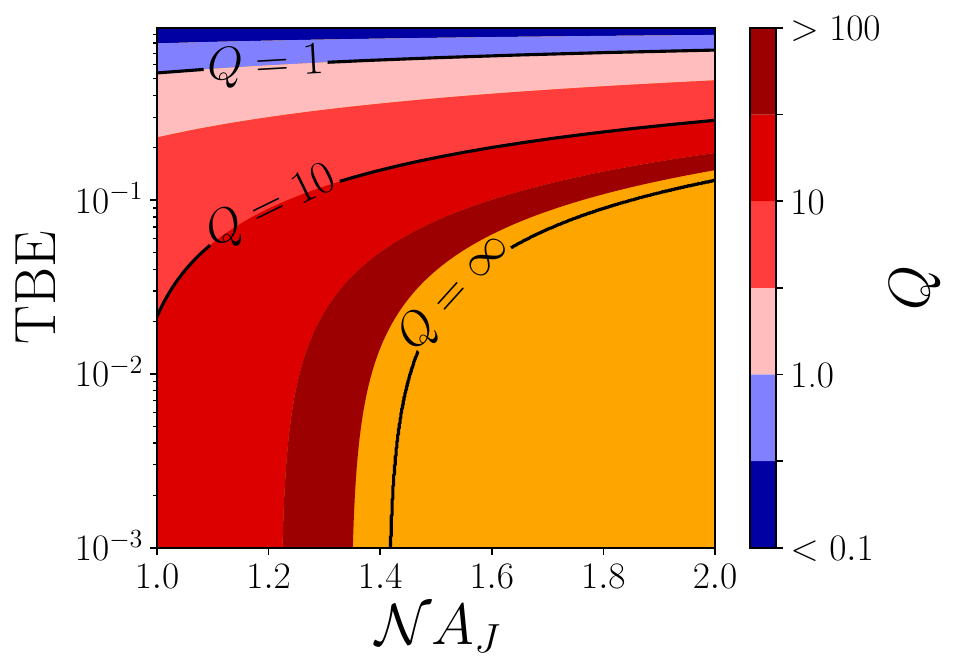}
    \caption{Nominal $Q = 10$.}
    \end{subfigure}
     ~
    \centering
    \begin{subfigure}[t]{0.98\textwidth}
    \centering
    \includegraphics[width=1.0\textwidth]{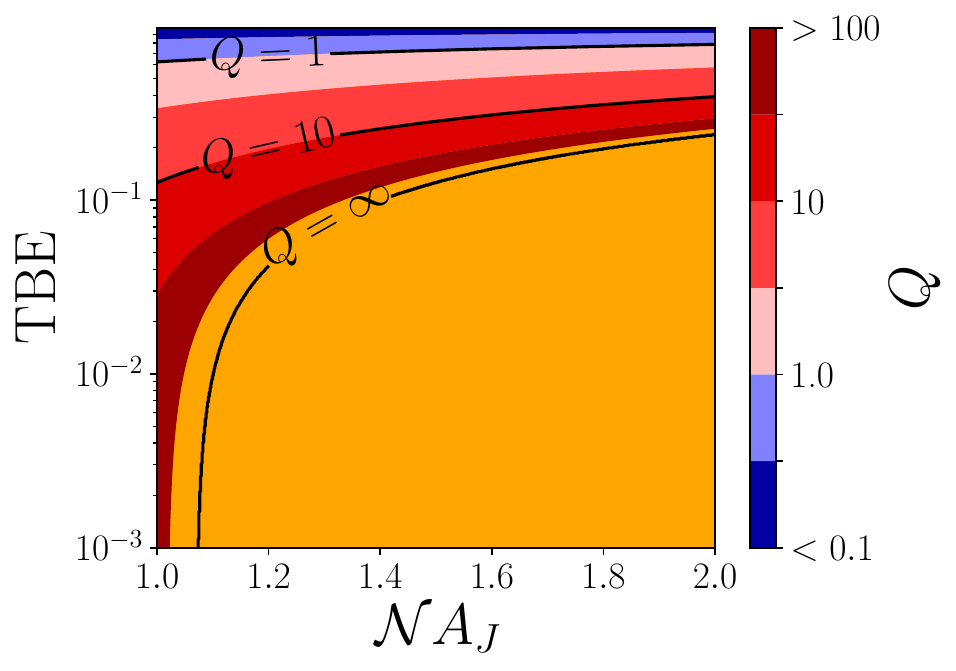}
    \caption{Nominal $Q = 40$.}
    \end{subfigure}
    \caption{Plasma gain $Q$ versus tritium burn efficiency (TBE) and spin-polarization multiplier ($\mathcal{N} A_J$) for a SPARC-like fusion power plant with nominal $Q=10$ (a) and for another hypothetical plant with nominal $Q = 40$. Plasmas below the $Q=\infty$ curves are predicted to ignite. $H_{\mathrm{T}} = \eta_{\mathrm{He} } = 1.0$ and $\Sigma = 0.63$.}
    \label{fig:QplasmaQ10_40}
\end{figure}

The ignition condition for a plasma with constant temperature and density profiles \cite{Wesson2012,Wurzel2022} is modified by tritium fraction and polarization (ignoring helium dilution effects),
\begin{equation}
n_Q \tau_E > \frac{1}{4 \mathcal{N} A_J f_{\mathrm{T}}^{\mathrm{co}}(1-f_{\mathrm{T}}^{\mathrm{co}})} \frac{12}{\langle v \overline{\sigma} \rangle } \frac{T}{E_{\alpha}},
\end{equation}
where $E_{\alpha}$ is the fusion-borne alpha particle energy. Using that the reactivity obeys the following scaling for temperature within $10-20$ keV with at most 10\% error,
\begin{equation}
\langle v \overline{\sigma} \rangle = 1.1 \times 10^{-24} T^2 \mathrm{m}^3 \mathrm{s}^{-1},
\end{equation}
the ignition condition is
\begin{equation}
n_Q T \tau_E > \frac{1}{4 \mathcal{N} A_J f_{\mathrm{T}}^{\mathrm{co}}(1-f_{\mathrm{T}}^{\mathrm{co}})} 3 \times 10^{21} \mathrm{m}^{-3} \mathrm{keV s}.
\label{eq:ignition_condition}
\end{equation}
Operating at $\mathcal{N} A_J = 1.5-1.9$ with spin-polarized fuel could lower the required temperature for ignition in \Cref{eq:ignition_condition} significantly. Alternatively, higher $\mathcal{N} A_J$ could be used to ignite at lower required $n_Q$ or $\tau_E$ values by choosing $T \approx 14$ keV where $n_Q T \tau_E$ is minimized. Using higher $\mathcal{N} A_J$ could also be used to  decrease the temperature required for ignition. This could be useful in operating ignited plasmas in the thermally stable regime, given that the thermal runaway regime occurs above a threshold temperature  $T_{\mathrm{runaway}} \approx 25$ keV \cite{Wesson2012}. For a numerical demonstration of how spin polarization and tritium burn efficiency affects the ignition condition, refer to \Cref{fig:QplasmaQ10_40} in \Cref{sec:Q10_scan} and \Cref{fig:Qplasma} in \Cref{sec:casestudy}.

Stability analysis performed in \cite{Wesson2012} shows that for an ignited plasma's temperature to be stable to thermal runaway,
\begin{equation}
T d \ln \tau_E / dT < 1 - T d \ln \langle v \overline{ \sigma} \rangle  / dT.
\label{eq:stability_condition}
\end{equation}
Runaway occurs when the alpha self-heating power increases faster than the power loss decreases. This condition is not obviously modified by a fuel that is spin-polarized or has $f_{\mathrm{T}}^{\mathrm{co}} \neq 1/2$. Therefore, past results obtained for the maximum D-T temperature before runaway, $T_{\mathrm{runaway}} \approx 25$ keV, may still be accurate. Spin-polarized fuel could allow plasma ignition while ensuring that $T \ll T_{\mathrm{runaway}}$.

\subsection{Passive Stabilization}

If operating with $f_{\mathrm{T}}^{\mathrm{co}} \neq 1/2$ in steady state, transition to a thermal runaway process may be slowed or prevented by the following process: a temperature increase leads to an increase in $\langle v \overline{ \sigma} \rangle $, which in turn increases $|1/2 - f_{\mathrm{T}}^{\mathrm{co}}|$ if the fueling rate is steady, which can decrease the fusion power, even at higher $T$. These effects are not captured by \Cref{eq:stability_condition}, which would need to include a time-dependent $f_{\mathrm{T}}^{\mathrm{co}}$.

\section{Gain for SPARC-like Experiment} \label{sec:Q10_scan}

In this section, we plot the plasma gain $Q$ versus tritium burn efficiency (TBE) and spin-polarization multiplier $\mathcal{N} A_J$ for a SPARC-like fusion power plant. For ease of comparison, we have used exactly the same parameters as the ARC-like fusion power plant in \Cref{sec:casestudy}, except decreasing the nominal plasma gain to $Q = 10$. For completeness, we also perform the same scan but for a power plant with a nominal plasma gain of $Q = 40$. The results are shown in \Cref{fig:QplasmaQ10_40}.

\section{Performance at Lower Tritium Enrichment} \label{sec:tritium_enrichment}

\begin{figure}[t]
    \centering
    \includegraphics[width=0.99\textwidth]{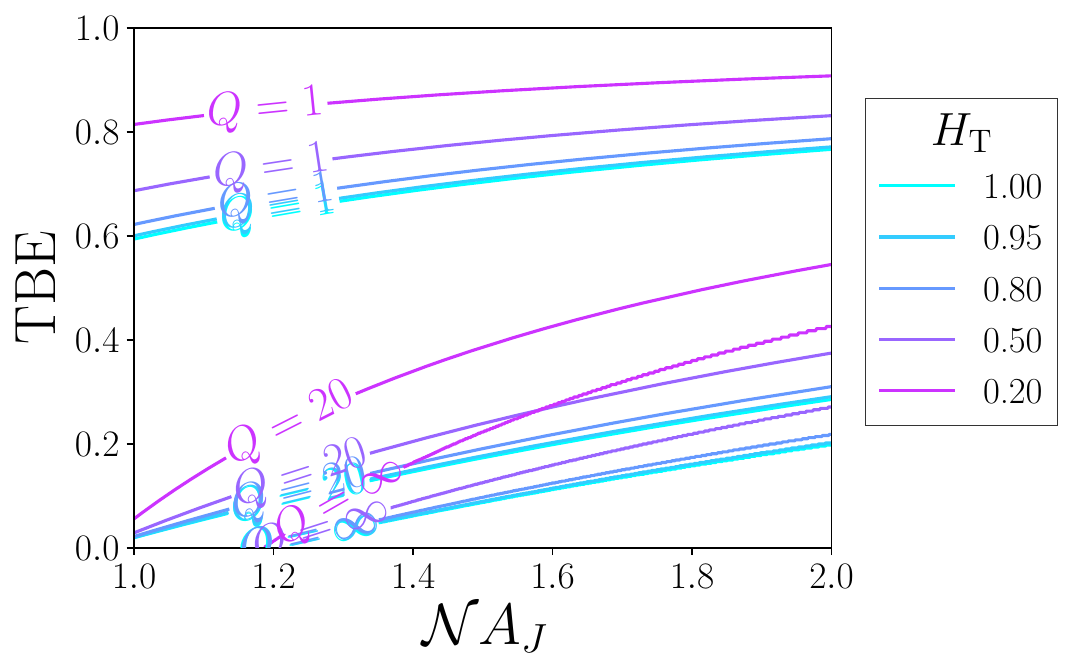}
    \caption{Tritium enrichment scan for the plasma gain $Q$ versus tritium burn efficiency (TBE) and spin-polarization multiplier ($\mathcal{N} A_J$) for an ARC-like fusion power plant with nominal $Q = 20$. Plasmas below the $Q=\infty$ curves are predicted to ignite.}
    \label{fig:QplasmaQ20_HT_scan}
\end{figure}

In this work, we adopted a conservative approach typically reporting results for $H_{\mathrm{T}} = 0.95, 1.00$. However, with lower tritium enrichment values, the fusion power increases at fixed TBE. Here, we report the effects of lower tritium enrichment on the plasma gain. Shown in \Cref{fig:QplasmaQ20_HT_scan}, curves of $Q=1$, $Q=10$, and $Q=\infty$ are plotted against TBE and $\mathcal{N} A_J$ for five enrichment values, $H_{\mathrm{T}} \in [1.00, 0.95, 0.80, 0.50, 0.20]$. Lower $H_{\mathrm{T}}$ always gives a higher TBE at fixed $Q$, but the absolute increase in TBE due to lower $H_{\mathrm{T}}$ improves as $Q$ increases. In the most extreme example in \Cref{fig:QplasmaQ20_HT_scan}, a $Q = \infty$ plasma with $\mathcal{N} A_J = 2$ achieves TBE = 0.42. A plasma with the same polarization and TBE -- $\mathcal{N} A_J = 2$, TBE = 0.42 -- but enrichment $H_{\mathrm{T}} = 0.95$ only achieves $Q \simeq 6$.

\bibliographystyle{apsrev4-2} %
\bibliography{EverythingPlasmaBib} %

\end{document}